\documentclass[11pts]{iopart}

\usepackage{iopams}
\usepackage{graphicx}% Include figure files
\usepackage{appendix}
\usepackage{wrapfig}
\usepackage{braket}
\usepackage{dsfont}

\begin{document}

\title[]{Two-terminal transport measurements with cold atoms}

\author{Sebastian Krinner$^{1}$, Tilman Esslinger$^{1}$, and Jean-Philippe Brantut$^{2,\dag}$\footnote[0]{$^{\dag}$ Author to whom any correspondence should be addressed.}}

\address{$^1$Institute for Quantum Electronics, ETH Zurich, 8093 Zurich, Switzerland}
\address{$^2$Institute of Physics, EPFL, 1015 Lausanne, Switzerland}

\ead{jean-philippe.brantut@epfl.ch}

\begin{abstract}

In the last years, the ability of cold atoms experiments to explore condensed-matter related questions has dramatically progressed. Transport experiments, in particular, have expanded to the point that conductances and other transport coefficients can now be measured in a way directly analogous to solid state physics, extending cold atoms based quantum simulations into the domain of quantum electronic devices. In this topical review, we describe the transport experiments performed with cold gases in the two terminals configuration, with an emphasis on the specific features of cold atomic gases compared to solid state physics. We present the experimental techniques and the main experimental findings, focusing on but not restricted to the recent experiments performed in our group. We eventually discuss the perspectives opened by this approach, the main technical and conceptual challenges for future developments, and the potential applications as a quantum simulator for transport phenomena and mesoscopic physics problems. 

\end{abstract}

\maketitle

\tableofcontents

\section{Introduction}

The measurement of transport between particle reservoirs, or terminals, exposes fundamental properties of quantum matter, such as insulating and superconducting behavior, or the quantization of conductance. These properties are in fact often used to classify solid-state systems. The understanding of the underlying processes is essential for the construction of almost all solid-state devices, where transport properties determine the suitability of a material for a desired application, ranging from energy harvesting, to the conveyance, storage and processing of information. From a more fundamental point of view, transport measurements provide a well-defined setting to study challenging questions in out-of-equilibrium quantum many-body physics.

Indeed, the study of transport in condensed matter physics has been a driving force for advancements in quantum mechanics, since transport entails both static, equilibrium properties of matter, such as the density of states, and its dynamic aspects, such as the nature of excitation above the ground state. In parallel with the understanding of increasingly more complex materials, progress in the control of semiconductor materials has allowed for the fabrication of almost defect-free systems. These led to the development of quantum devices, where the quantum behavior is engineered at the mesoscopic level (see for example \cite{Ihn:2010aa}). Contrary to materials where the quantum nature of the system emerges spontaneously in the ground state of the (possibly many-body) problem, such as in superconductors or quantum hall effects, here the quantum aspect is imposed by a specially designed structure. In most cases the quantum mechanical nature of the effects were revealed by their consequences on transport properties. 

During the last decade, cold atoms have emerged as a powerful tool to investigate models originating from condensed matter physics. The idea is to use a quantum degenerate gas of neutral atoms to simulate the behaviour of electrons in those model systems and to explore regimes not accessible to solid-state systems. Prominent examples include the Hubbard model \cite{Bloch:2008ab,Esslinger:2010aa}, BCS pairing and the BEC-BCS crossover \cite{Zwerger:2011aa}, as well as topological quantum matter \cite{Goldman:2016aa}. By construction, the atomic gas is free of any impurity or disorder, and is almost perfectly isolated from the environment. In addition, the relevant length and energy scales in an atomic cloud differ by orders of magnitude from those of an electron gas in a solid:

\begin{description}
\item{Density scale}. %A typical degenerate Fermi gas has a density of about $10^{12}\,$cm$^{-3}$, which amounts to an inter particle spacing of about $1\,\mu$m. This is comparable to the wavelength of typical near-infrared or visible light. This allows for the direct manipulation and observation of atomic gases by optical techniques: optical lattices made using optical standing waves operate in the single band Hubbard regime \cite{Bloch:2008ab,Esslinger:2010aa}, disorder created by optical speckle patterns leads to Anderson localization \cite{Shapiro:2012aa}, and high resolution microscopes can resolve individual particles in a many body system \cite{Bakr:2009aa,Sherson:2010aa,PhysRevLett.114.213002,PhysRevLett.114.193001,Haller:2015aa,Yamamoto:2015aa,PhysRevA.92.063406,PhysRevLett.115.263001}.
A typical degenerate Fermi gas has a density of about $10^{12}\,$cm$^{-3}$, which amounts to an interparticle spacing of about $1\,\mu$m, comparable to the wavelength of typical near-infrared or visible light. This allows for the creation, manipulation and observation of atomic gases by optical techniques: optical lattices are made from interfering laser beams and often operate in the single band Hubbard regime \cite{Bloch:2008ab,Esslinger:2010aa}; disorder is created by optical speckle patterns leading to Anderson localization \cite{Shapiro:2012aa}; and high resolution microscopes can resolve individual particles in a many-body-system \cite{Bakr:2009aa,Sherson:2010aa,PhysRevLett.114.213002,PhysRevLett.114.193001,Haller:2015aa,Yamamoto:2015aa,PhysRevA.92.063406,PhysRevLett.115.263001}. 

\item{Energy scale}. %Because of the large mass of atoms, and the low densities, cold gases operate with Fermi energies in the micro Kelvin regime, and have to be cooled using a combination of laser cooling and evaporative cooling to reach quantum degeneracy. In turn, these low energy scales translate into long timescales for the microscopic evolution of the system. This allows for a time resolved control over the atoms up the Fermi time through the use of standard radio frequency technologies. 
Because of the large mass of atoms, and the low densities, cold gases operate with Fermi energies in the micro Kelvin regime, and are cooled using a combination of laser cooling and evaporative cooling to reach quantum degeneracy \cite{Ketterle:2008aa}. In turn, these low energy scales translate into long timescales for the microscopic evolution of the system. This allows for a time resolved control over the internal degree of freedom of atoms on time scales shorter than the Fermi time, using standard radio frequency technologies. 

\end{description}

%The question of measuring the transport properties of cold atomic gases has naturally emerged in this context, both as a tool for the quantum simulation of condensed matter models and as a topic of research for quantum gases themselves. First, in the perspective of using cold atomic gases to model solid state systems, it is desirable to extract the same observables from a solid state device and from its cold atoms simulation. This would allow for a direct comparison between the experimental outcomes of both cases, without the need of modeling the system our even the knowledge of the precise Hamiltonian. The typical cases where quantum simulation could be useful are actually cases where the exact Hamiltonian is not known, so it is important for the quantum simulation to reproduce the phenomenology observed in condensed matter. Second, from the point of view of quantum engineering, cold atoms offer a situation where device-like systems can be carved from a fully controlled material. The control extends to parameters such as the strength of interactions between the particles, or the nature and strength of disorder, which are not controlled in quantum devices made of solid-state materials. 

The quest to measure transport properties of cold atomic gases originates in the requirement to further complete the toolbox for quantum simulations of condensed matter models and in the prospect to address a new set of questions in many-body physics. For quantum simulations of solid-state systems it is essential to extract the same observables as in the simulated device and reproduce its phenomenology. This also allows for direct comparison between measurements carried out on the device with those obtained from a quantum simulation, even without the need for modeling the system or having precise knowledge of all aspects of the underlying Hamiltonian.  From the point of view of quantum engineering of new many-body systems, cold atoms offer a situation where device-like systems can be carved from a fully controlled material. The control extends to parameters, such as the strength of interactions between the particles, or the nature and strength of disorder, which are not controlled or out of reach in quantum devices made of solid-state materials. 

The question of transport has been investigated from various points of view in cold atoms physics, but for a long time experiments could not explore it in full generality: in most cases transport processes occur as density redistribution within atomic clouds, where thermal equilibration time and length scales are not separated from the transport scales themselves (see for example \cite{Ott:2004aa,Schneider:2012ys}). This coupled evolution hinders the systematic extraction of transport coefficients independent of the static properties. This limitation was overcome by the two-terminal technique that we applied to cold atomic gas systems in our group during the last years. The principal concept was pioneered in mesoscopic solid-state physics and provides access to transport coefficients. In cold gases, it can be applied to both weakly and strongly interacting systems and it yields observables that are the direct counterparts of the condensed-matter measurements \cite{Brantut:2012aa,Stadler:2012kx,Brantut:2013aa,Krinner:2013aa,Krinner:2015aa,Krinner:2015ac,Husmann:2015ab,Krinner:2016ab}. In particular, a precise meaning is given to conductances and other transport coefficients despite the gases being charge-neutral.

The paper is structured as follows: the remaining of the introduction is devoted to a short description of the various techniques related to transport that were used in cold atoms, and to the conceptual presentation of the two terminal approach providing a precise definition of transport coefficients for general charge neutral systems. The second section describes the experimental techniques, in particular the initialization of the measurements and the interpretation of the results. In the third section, we describe experiments with non-interacting particles, where the conductor is in a multimode regime, illustrating the experimental techniques. The fourth section is dedicated to experiments with a single mode conductor, the observation of quantized conductance and its interpretation using the Landauer formalism. In section five, we describe some transport experiments with Bose-Einstein condensate and discuss the role of superfluidity. In section six, recent experiments with strongly attractive Fermions are described, showing some of the consequences of pairing and superfluidity. The last section discusses the perspectives and challenges opened by these results, both as tools for the study of quantum gases and as a quantum simulator for condensed matter related questions.

\subsection{Transport measurements and cold atomic gases}

To investigate the transport properties of quantum gases, a variety of approaches have been tested. The reason for this diversity in approaches is the isolated nature of quantum gases, which prevents the direct connection of a cold atom device with a battery capable of biasing the system, and injecting and collecting atoms driven through the system. Therefore no strictly DC currents can exist in a cold gas and the response is inherently transient. The main techniques used so far can be roughly classified as follows. 
%The investigations of transport in quantum gases have taken a wide variety of forms, with very different types of experiments and conclusions driven from them. The main reason for this diversity is the isolated nature of quantum gases, which prevents the direct connection of a cold atom device with a battery capable biasing the system, and injecting and collecting atoms driven through the system. Therefore no strictly DC currents can exist in a cold gas and the response is necessarily transient.The main techniques used so far can be roughly classified as follows. 

{ \it Response to an external force}. The cloud is subject to a homogeneous force and the subsequent evolution is observed. In a harmonic trap, this yields dipole oscillations of the center of mass of the cloud, which is fully decoupled form the internal properties of the gas by virtue of Kohn's theorem \cite{Dalfovo:1999aa}. Upon superimposing an extra potential, such as a lattice or a random potential, the oscillations get damped and their relaxation rate is used as a measure of the transport properties \cite{PhysRevA.66.011604,PhysRevA.68.011601,Ott:2004aa,Pezze:2004aa,Fertig:2005aa,Lye:2005aa,Strohmaier:2007aa,Chen:2008aa}.

The isolated nature of the atoms allows for the quantum coherent evolution of atoms under the influence of the homogenous force, leading to counterintuitive results. For example, atoms in an optical lattice show Bloch oscillations in the presence of an external force: atoms are constantly accelerated by the force until they reach the Bragg condition, and are then coherently reflected by the lattice yielding an oscillatory motion \cite{Ben-Dahan:1996aa}. This observation illustrates the difficulties in interpreting the results of such homogenous force response in terms of transport coefficients: one would conclude that the DC conductance of a lattice is always zero as the current response averages over the Bloch cycles. At time scales shorter than the Bloch period, the atoms are continuously accelerated as the quasi-momentum increases. This increase results from the phase coherence of the single particle wave function of the atoms over neighbouring sites of the lattice. In the presence of perturbations, such as interactions between particles or disorder, Bloch oscillations are damped and dephased \cite{Anderson1686,PhysRevLett.87.140402,Roati:2004aa,PhysRevLett.100.080404,PhysRevLett.100.080405,Drenkelforth:2008aa}, and for some specific cases only the transport results of linear response theory can be recovered \cite{Eckstein:2011aa}.

{\it Release and expansion of a cloud}. In this case, a cold atom cloud is prepared in a trap, and the trap, or a specific direction of confinement, is released, letting the atoms freely expand. The initial cloud thus acts as a transient atom source \cite{Fort:2005aa,Clement:2005aa,Roati:2008aa,Billy:2008aa,Robert-de-Saint-Vincent:2010ab,Kondov:2011aa,Schneider:2012ys,Jendrzejewski:2012vn,PhysRevLett.110.205301}. The atom laser, which is generated by continuously out-coupling atoms from a cloud is a particularly refined instance of this situation \cite{Hagley1706,PhysRevLett.82.3008,PhysRevLett.97.200402,Couvert:2008mh,Robins:2008aa}. Typically, the atoms interact with a background potential during the expansion, or among each other. The time evolution of the distribution of atoms following the release reveals informations about the diffusion processes. In the simple case of non-interacting particles, the diffusion coefficient averaged over the initial energy distribution is directly accessible \cite{Robert-de-Saint-Vincent:2010ab,Kondov:2011aa,Jendrzejewski:2012vn}, and was measured to identify Anderson localisation. In periodic structures, the coherent evolution yields ballistic expansion, compatible with the intuitive picture of a perfect lattice being a perfect conductor. In the presence of interactions between the particles the evolution is more intricate as the edges of the cloud behave differently from the center. Nevertheless, some features have been observed as interactions are varied in an optical lattice \cite{Schneider:2012ys,PhysRevLett.110.205301}. For very strongly interacting gases, the evolution can be modeled by a hydrodynamic expansion, and momentum transfer between the different directions during the expansion can be traced back to the shear viscosity of the homogenous gas \cite{Cao:2011aa,Cao:2011ab}.  

{\it Coherent dynamics in superfluids}. 
The long range coherence in a distinctive feature of superfluids. The existence of a complex order parameter is at the root of some of the most intriguing manifestations of superfluidity. As far as transport is concerned, the most important consequence is that the superfluid current is proportional to the gradient of the phase of the order parameter. This implies in particular that the superfluid flow is irrotational, and angular momentum within the fluid is carried by quantized vortices. Such vortices have been extensively studied in the context of Bose-Einstein condensates, and successfully demonstrated for fermionic superfluids as well (see \cite{Fetter:2008aa,Cooper:2008aa} for reviews). The irrotational flow is particularly striking in the connected geometry of a ring trap, where it produces quantized permanent currents robust against weak perturbations \cite{Ryu:2007aa,Ramanathan:2011aa,Beattie:2013aa,Ryu:2013aa}. 

For clouds or systems composed of several parts separated by short barriers or weak links, the long range phase coherence leads to the celebrated Josephson oscillations. These have been extensively studied with Bose-Einstein condensates, both as a manifestation of the superfluid character of the system, and as a tool to realise phase-sensitive measurements \cite{Orzel:2001aa,Cataliotti:2001aa,Albiez:2005aa,Levy:2007fk,LeBlanc:2011aa}. Such weak links have now been extended to Fermionic superfluids \cite{Valtolina:2015ab}.

{\it Spin and heat diffusion}
Beyond the transport of particles, several other conserved quantities can be transported within cold atomic clouds. In particular for Fermionic quantum gases, the internal degree of freedom plays an essential role. For electronic system it is the spin of the particles, while for cold atoms the different hyperfine states i.e. the relative orientation of the nuclear and electronic spin, play an analogous role. In the reminder of this review we refer to internal degrees of freedom as spin in analogy with the electronic case. 
Using spin-selective manipulation tools, spin inhomogeneities can be produced in cold atomic clouds and their evolution monitored \cite{Sommer:2011uq,Fukuhara:2013aa}. For instance, a gradient of spin populations was created in \cite{Sommer:2011uq} by spatially separating the spin components and the relaxation of the system was observed, providing insights in the spin-diffusion coefficient of the strongly interacting Fermi gas. In other sets of experiments, the decoherence of a spin texture was investigated, and the transverse spin diffusion coefficient was observed \cite{Koschorreck:2013aa,Bardon:2014aa,Hild:2014aa,Trotzky:2015aa}. One of the main foci of these experiments is the investigation of strongly interacting Fermi gases, where the scattering mean free path is of the order of the inter particle spacing and the coefficients are believed to take universal values. Heat diffusion has been investigated in the same spirit by locally heating clouds, but is technically more difficult in that it requires local measurements of temperatures. The effects of superfluidity and interactions have been investigated via heat transport \cite{Meppelink:2009ab,Hazlett:2013aa}, and the second sound was observed in strongly interacting Fermi superfluids\cite{Sidorenkov:2013aa}.

{\it Response to particle decay due to natural or induced losses}
The existence of particle loss processes can be used to trigger transport across the system. Intrinsic losses such as three-body collisions have a natural density dependence, and tend to create chemical potential and temperature gradients that can be used to extract transport coefficients \cite{Hazlett:2013aa}. Local addressing techniques allowed it to induce losses at very short length scales within cold atomic clouds, creating a well defined non-equilibrium situation, which triggers a time evolution that provides informations on particle flow within the cloud \cite{PhysRevLett.110.035302,Labouvie:2015aa}. 

Even though these techniques have all intimate links with transport, their relation to the concepts of conductivity or conductance, as understood in condensed matter physics, is sometimes very indirect. Questions of timescales, length scales, trap averaging have to be investigated in a case-by-case fashion. 

This review is devoted to the implementation of the Landauer two-terminal configuration for cold gases. We have developed this technique with the aim of providing a generic method to extract the equivalent of the DC conductance for a quantum gas system, in particular for Fermionic quantum gases where the analogy with condensed matter systems is most relevant. The route that was followed is inspired from the Laudauer approach to transport, which (i) explicitly considers terminals that introduce the bias in the system and (ii) applies equally well to charged and charge neutral particles, since it does not refer in any respect to electric fields or electrical potentials.% \footnote{Interestingly, the motivation for this formulation seems to have been that the actual field distribution within a conductor is a very complex problem, and supposing that the bias simply adds a uniform electric field seemed unsatisfactory. It was rather assumed that the injected current is the main driving, and the field and charge distribution within the conductor adapts to the driving without having to be dealt with explicitly.}.

\subsection{Two terminal transport of particles}

\begin{figure}[htbp]
\begin{center}
\includegraphics[width=0.65\textwidth]{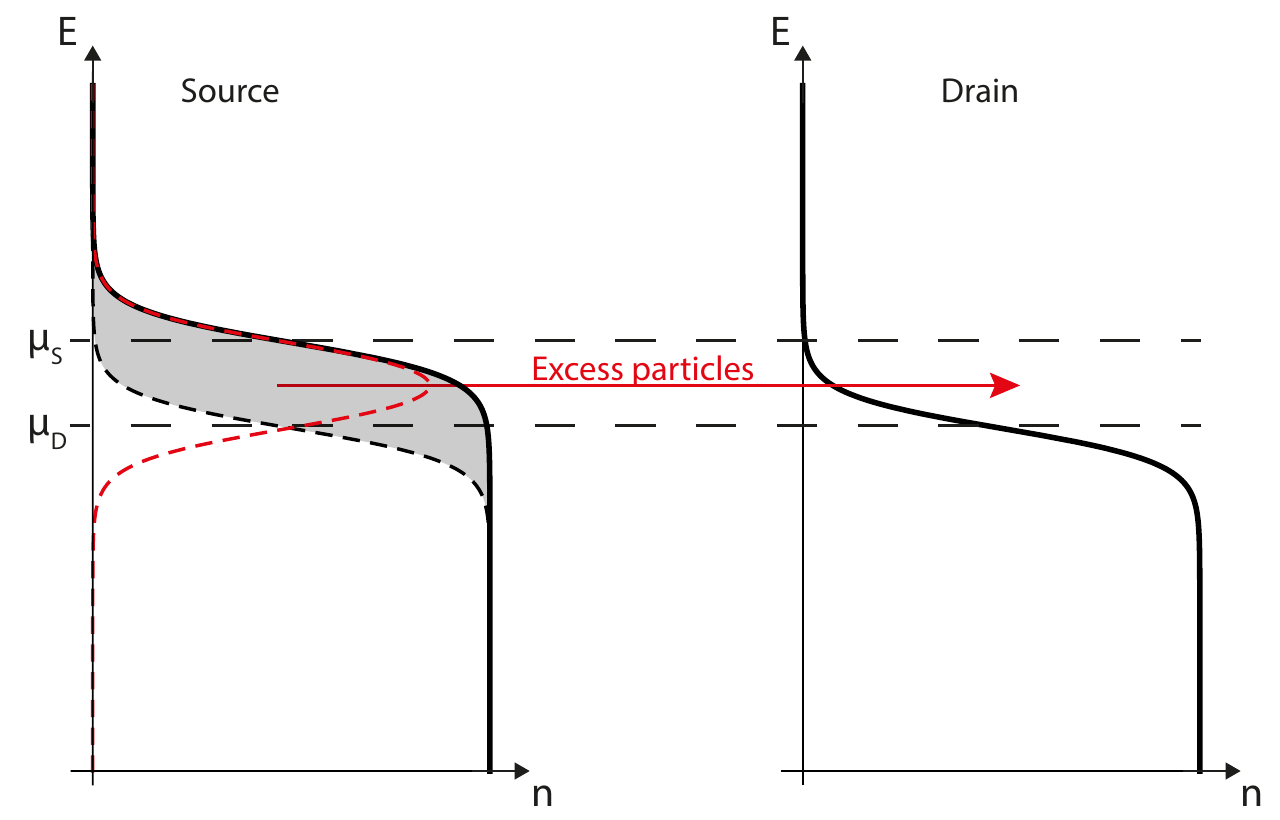}
\caption{Concept of a two-terminal transport measurement. Source and drain reservoirs are contacted to the conductor of interest. A chemical potential difference is introduced between them, leading to different energy distributions of particles in the source and drain. A net current of particles is driven from the source, as a result of the excess of particles above the Fermi level of the drain. The red dashed line indicates the energy distribution of the particles giving rise to the net current.}
\label{fig:twoTermConcept}
\end{center}
\end{figure}

The generic concept of a two terminal transport measurement is presented in figure \ref{fig:twoTermConcept}. The system of interest is connected on two sides to particle reservoirs, which supply and collect particles to and from the system. A bias is then introduced between the two reservoirs, in the form of a chemical potential difference, and the resulting current through the system is observed. Conceptually, the chemical potential bias can be thought of as a thermodynamic force, driving a current of the extensive, thermodynamically conjugate quantity, namely atom number.

By specifying the system of interest, the process by which the system is attached to the reservoirs is also specified. The nature of the interfaces with the reservoirs may have a large influence on the results, for example when their geometry leads to particle reflections. We also suppose that transport inside the conductor is fully elastic, i.e. energy is strictly conserved, which is realized in cold atoms experiments and in many condensed matter physics situations as well. 

These reservoirs play a central role in the concept of two- or multiple terminal transport measurements. It is assumed that they contain fast relaxation processes that ensure a quick thermalization of the incoming particles. By analogy with optics, reservoirs act like {\it perfect Fermionic black-bodies}. Two important characteristics follow from these requirements:
\begin{description}
\item[Phase incoherence] No quantum interference processes take place inside the reservoirs at the time and length scales relevant for the transport. This implies that particles emitted from different reservoirs have no phase relation. This also implies that energy is the only relevant character of the particles emerging out of one reservoir. In particular all energetically available momenta are populated irrespective of the possible directions. 
\item[Energy relaxation] Incident particles from the conductor are drawn from the energy distribution of another reservoir, possibly modified by interactions with the conductor. Fast scattering processes serve to erase the memory of the incident particles such that, should it be reemitted at later times, it will appear as drawn from the energy distribution of the reservoir it is emitted from.
\end{description}
In addition, the reservoirs are assumed to have standard reservoir properties from the point of view of statistical mechanics: they have a very large heat capacity, compressibility, spin susceptibility, etc, such that their chemical potential, temperature or spin potential is not affected by the emission and absorption of particles into and from the conductor. 

In condensed matter physics, these hypothesis are natural and follow from the experimental situation, where leads are essentially infinite, and the intrinsic complexity of the materials ensure efficient inelastic processes in the leads. The strength of the Landauer approach is that it does not require a precise description of the physics of the reservoirs. The main disadvantage is that the hypothesis are very hard to check independently of the transport measurements themselves. 
As will be explained in the next sections, cold atoms offer a situation where the reservoirs can be experimentally observed and controlled, providing perspectives for novel tests of the validity and breakdown of the hypothesis underlying the Landauer approach. 

The process by which the bias is introduced is usually not specified, and it is assumed that the system as a whole reaches a steady state in which a DC current establishes, and its magnitude only depends on the bias rather than the precise history of the setup operation. While this is very likely to be the case in most experimental situations, it is worth noting that in some important cases such as glasses this may not be the case, or the establishment of the steady state may take exponentially long times. There again, the cold atom implementation offers a renewed perspective on this problem, owing to the long timescales at play, opening the possibility to observe the establishment of a steady state. 

The transport process can be understood from the point of view of the energy distribution in the reservoirs, as presented in figure \ref{fig:twoTermConcept}. The two Fermi distributions describing the left and right reservoirs differ by their chemical potential, the difference being the bias. Since transport is elastic in the conductor, the current can be considered separately for each energy. This leads to the fundamental conclusion that {\it the net current driven by the bias through the conductor originates from a narrow energy window around the Fermi level}, the width of which is set by the bias or the temperature. In essence, it is a differential measurement where the contributions of the low energy levels cancel between the source and the drain. The detailed calculation will be presented later in the specific context of the Landauer formula applied to a quantum point contact. 

The fact that transport is a property of the Fermi surface is well known in condensed matter physics. The two terminal configuration provides a concrete implementation of a transport measurement that fulfills this property. Interestingly, this approach directly applies to charge-neutral particles without having to exhibit a counterpart to electric fields and electric potentials. In cold atoms, the typical experimental observable is the density distribution measured destructively at a given time. It is not energy resolved, leading for example to well known problems of thermometry in the degenerate regime where most of the atoms reside in the Fermi sea and carry no information about temperature. Energy resolution is restored using for example Radio-frequency spectroscopy \cite{Chin:2004aa,Torma:2014aa}, sensitive to the density of states, or Bragg spectroscopy measuring the density fluctuation spectrum \cite{Stenger:1999aa,Veeravalli:2008aa}. These techniques do not directly probe the localized or extended nature of the excitations at the Fermi level, which is an essential ingredient for the classification of phases of matter into insulating or metallic. The two terminal transport measurements fills this gap in the cold atoms toolbox. In addition, it gives a precise meaning to the conductance of a cold atomic gas.

By putting forward a description in terms of system and reservoirs connected together but spatially separated, the Landauer approach leads naturally to a description of the system in terms of {\it device}, connected to each other in a circuit. Indeed, figure \ref{fig:twoTermConcept} provides a minimal circuit with two leads and a conductor. While this is natural for electronic systems, this approach is new in the field of cold atoms, where most of the systems considered are "bulk". The engineering of devices for atoms has been coined atomtronics, with an emphasis on reproducing the functionality of electronic devices. Here the device approach to cold atoms provides the natural way to perform and interpret transport measurements.

\subsection{Heat and spin transport}
The two-terminal approach can be directly generalized to the transport of other conserved quantities such as energy, spin etc. The cases of spin and heat (or energy) are relevant for some experiments and theoretical proposals in the context of cold atomic gases.

\begin{figure}[htbp]
\begin{center}
\includegraphics[width=0.65\textwidth]{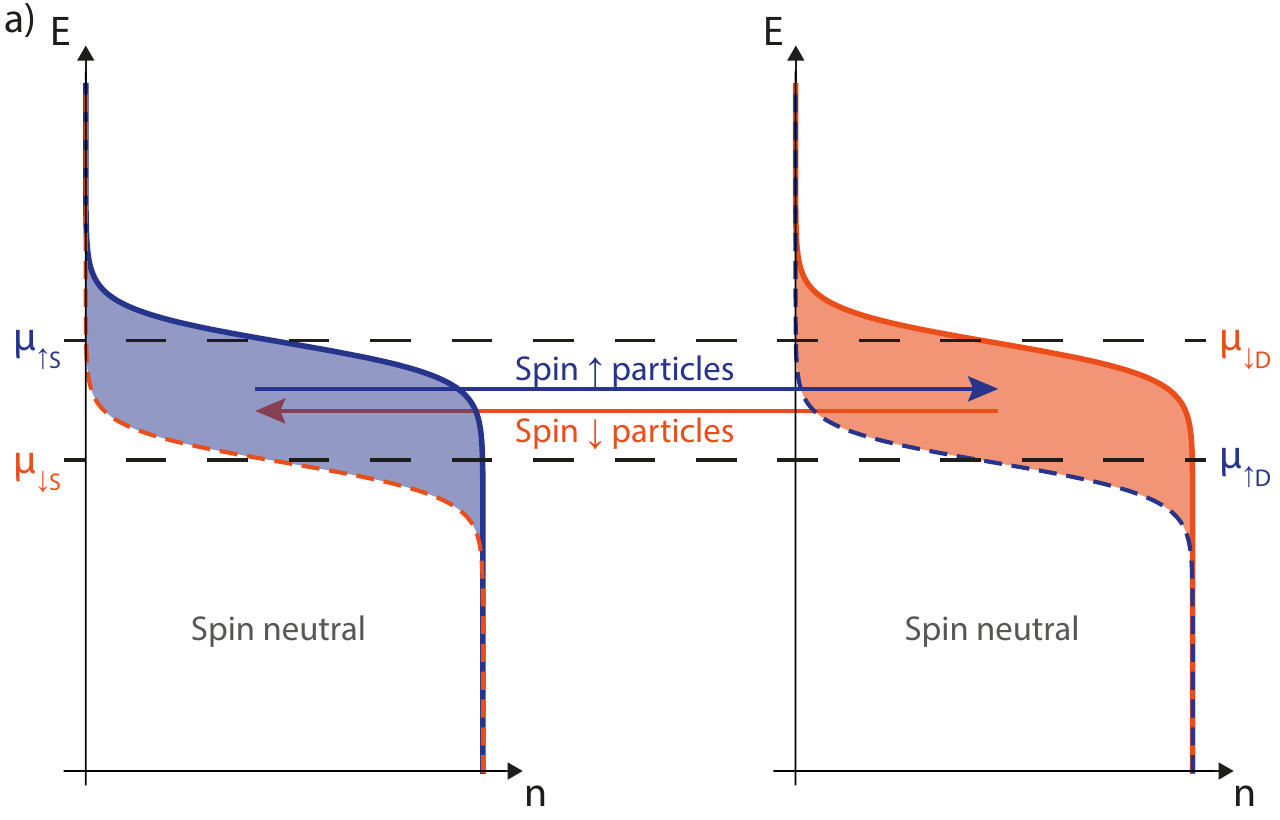}
\includegraphics[width=0.65\textwidth]{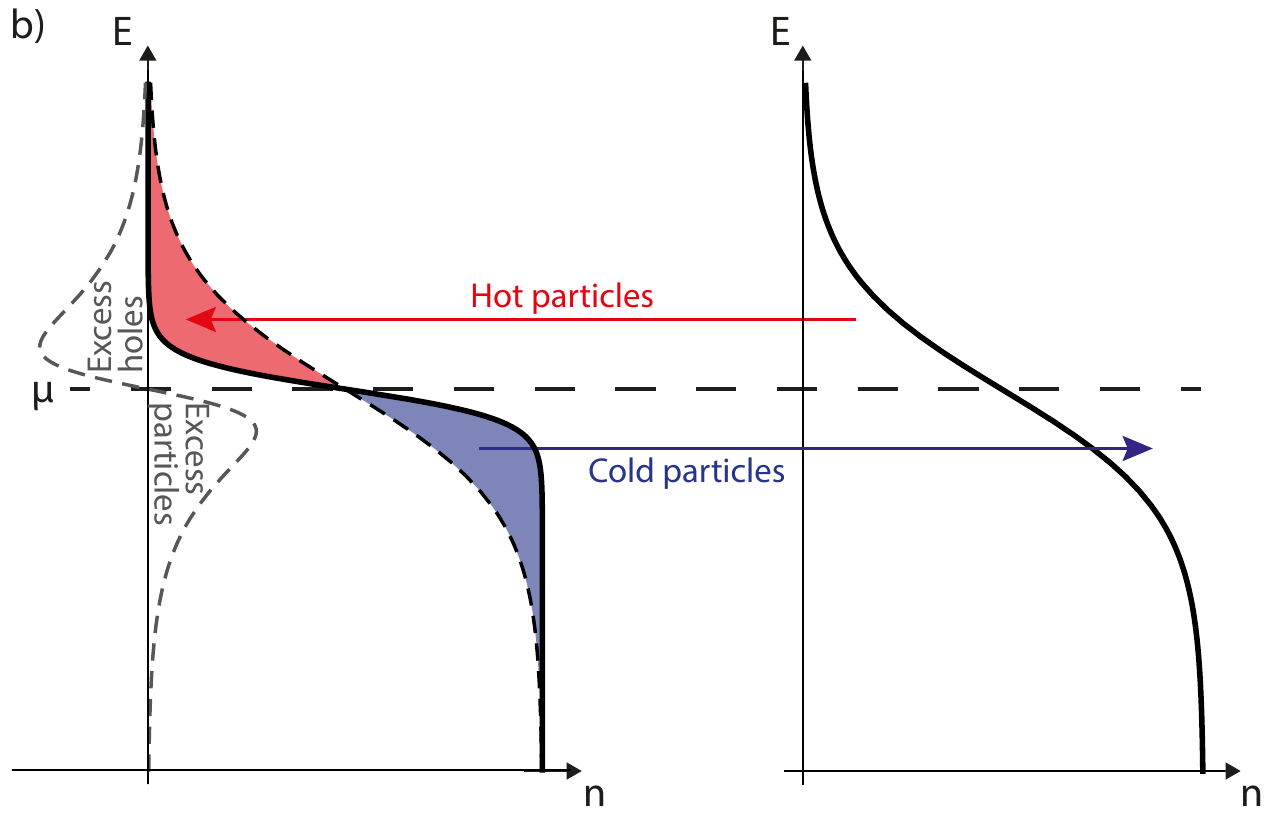}

\caption{(a) Two terminal transport of spin. Opposite chemical potential differences for the two spin components are introduced between the source and bias, driving currents of spin up from source to drain and vice versa for spin down. (b) Two terminal transport of heat. A temperature difference between the source and drain drives a current of high energy particles from hot to cold, and a current of low energy particles from cold to hot. }
\label{fig:twoTermSpinHeat}
\end{center}
\end{figure}

Figure \ref{fig:twoTermSpinHeat}a illustrates the case of spin transport. Here we suppose in addition that there is no interactions between the particles in different spin states within the reservoirs, or at least that these can be captured within a Fermi liquid picture. The spin bias, the thermodynamic force conugate to magnetization, can be thought of as equal in magnitude with opposite chemical potential biases applied to the two spin components across the conductor. Following the reasoning above, in the narrow energy window around the Fermi energy, there is an excess of spin up particle incident from the source, and an excess of spin down particles incident from the drain. As a result, the net particle current is zero. 

Such a configuration is particularly sensitive to scattering between the two spin components inside the conductor. Indeed, scattering between identical spin particles is forbidden, and scattering between different spin components is elastic, and thus conserves the total current. Thus, we expect particle current to be weakly affected by interactions. In contrast, the spin current can be strongly damped by scattering since momentum is not conserved within a single spin component. In the extreme case of a one dimensional system, this yields the famous spin-charge separation. In conductors where interactions are important, the conductances and spin currents yield fundamentally different informations compared to particle conductance alone. 

In a situation where no net spin polarisation is present, the spin and particle conductance fully describe the transport properties at the linear response level, and these are not coupled to each other. When a net spin polarisation is present, spin and particle transport are no longer independent from each other, and couplings appear already at the level of linear response. 

Heat transport can be envisioned within the same framework, as illustrated in figure \ref{fig:twoTermSpinHeat}b. Here we consider one spin component, and the two reservoirs have identical chemical potentials but different temperatures, apparent in figure \ref{fig:twoTermSpinHeat}b from the different broadenings of the Fermi distributions. Again reasoning energy by energy, we observe that a net current of particles is driven from the cold to the hot reservoirs for energies below the Fermi level, and vice-versa above the Fermi level. Since transport is elastic in the conductor, these processes are responsible for carrying a net energy or heat current from the hot to the cold reservoir.

There is a fundamental coupling of heat and particle current at the linear response level \cite{Onsager:1931aa,Callen:1948aa}: following the sketch of figure \ref{fig:twoTermSpinHeat}b, there is no reason a priori for the high energy particle current above the Fermi level and low energy particle currents below to cancel each other. When the rate of transport in the conductor depends on energy, which is typically the case, then a {\it thermoelectric} coupling arises, proportional to the derivative of the transport coefficients with energy \footnote{Even though we are dealing with neutral particles, we still use the word thermoelectricity to describe the coupling between particle and heat transport}. 

In general, when two possible transport processes can take place simultaneously within one conductor, they are coupled to each other. Thermoelectricity is only one example of such couplings. The linear response takes the form of an Onsager matrix with non-zero off-diagonal coefficients relating the currents of extensive quantities to the generalized thermodynamic forces. Like for the simple case of particle transport, the two terminal configuration directly implements these concepts, and suggests a practical route for experimental investigations.

\subsection{Bosons and classical particles}

Even though the concepts were so far illustrated on the basis of Fermi gases, the general ideas of the two terminal setup can be extended to other types of particles. The case of particles following Boltzman statistics such as high temperature cold atoms gases is very similar to that of Fermions. The range of energies to be considered in the transport process is determined by the mean internal energy of the particles and the temperature. 

The case of low temperature bosons is much more delicate, because of the emergence of the Bose-Einstein condensate. There, the reservoirs acquire a superfluid character, with properties that are not universal but depend for example on the interactions between the particles. Contrary to Fermions, the case of non-interacting Bosons is a pathological case with little experimental relevance. The long range phase coherence produces macroscopic interferences, that renders the coupling from the reservoirs to the conductor sensitive of details of the geometry. The cases of superfluid reservoirs is of particular relevance to cold atoms and will be treated in details in section \ref{section:BEC}. 

The particular case of the atom laser, derived from a Bose-Einstein condensate but constituting a strongly non-equilibrium case with a highly monochromatic beam of particles strongly resembles the case of Fermions with a low temperature but a large bias (one reservoir is the vacuum) \cite{Hagley1706,PhysRevLett.82.3008,Robins:2008aa}. Guided atom lasers are actually being used as sources for atom optics, giving access to quantities such as the energy dependent transmission coefficient for atoms in structures, that are closely related to the transport coefficients in the Landauer configuration \cite{PhysRevLett.97.200402,Couvert:2008mh}, as explained in section \ref{section:QPC}.

\section{Two-terminal configuration for cold atoms}
\label{section:RCmodel}

The two-terminal formulation provides a natural framework for the study of transport processes in charge neutral systems such as cold atoms. We now describe the two-terminal transport setup that we operate in our group, that implements the principles exposed above. For concreteness, we will give precise numbers from our experimental setup, but most of the experimental techniques can be generalized for example to different atomic species or different platforms such as atom chips \cite{Reichel:2011aa}. 

The concept shares some similarities with previously implemented techniques such as double-wells or ring traps in the context of Bose-Einstein condensates. These experiments will be discussed in section \ref{section:BEC} in the wider context of superfluid transport with cold atoms. We only briefly describe the aspects of cold atomic gases that are not directly related to the transport experiments, referring the readers to general textbooks about this topic (for example \cite{Torma:2014aa}).

\subsection{Cold Lithium gas with tunable interactions}
The preparation of quantum degenerate Fermi gases in our setup follows standard procedures developed in the field over the last years. Our apparatus is described in \cite{Zimmermann:2011aa}. It produces a quantum degenerate Fermi gas consisting of about $N_\sigma = 10^5$ atoms in each of two different hyperfine states labeled by $\sigma$, which play the role of a spin degree of freedom. We reach temperatures in the range of $T = 0.1 T_F$ to $0.3 T_F$, where $T_F = \hbar \bar{\omega} \left( 6 N_\sigma \right)^{1/3}$ is the Fermi temperature of the trapped gas. Here $\bar{\omega} = \left( \omega_x \omega_y \omega_z \right)^{1/3} $, where $\omega_i$ is the angular trap frequency in direction $i$. 

In cases where there are no interactions, $T_F$ coincides with the Fermi temperature of a homogenous system $\hbar^2/2 m \left( 6 \pi ^2 n_\sigma \right)^{2/3}$ with a density per spin component $n_\sigma$ equal to the density of the trapped gas at the center. Importantly, in the presence of interactions, the total atom number is constant, hence $T_F$ is constant even though the actual density of atoms varies. There is thus in general no direct relation between $T_F$ and the actual density of particles anywhere in the trap. Complementarily, certain situations require a description of the gas within the local density approximation where the local density of the gas at one point $\mathbf{r}$ in the cloud is associated with a local Fermi temperature $T_0(\mathbf{r}) = \hbar^2/2m \left(6 \pi^2 n_\sigma(\mathbf{r}) \right)^\frac{2}{3}$. $T_0$ and $T_F$ can be deduced from each other using the equation of state of the gas. 

The interactions between atoms in the same spin component are suppressed by Fermi statistics, and s-wave scattering occurs for pairs of atoms in different hyperfine states. It stems from the van der Waals force between atoms, and at the density and energy scales of cold atoms experiments, the only relevant parameter is the scattering length $a$. It depends on the precise molecular structure of the two atoms problem. The coupling between two atoms colliding in free space and a molecular channel leads to Feshbach resonances: the scattering length diverges upon tuning the energy of the molecular channel with respect to the scattering channel using a homogenous magnetic field. As a result, the scattering length is a control parameter that can be varied, kepping everything else equal over a wide range, up the unitary limit reached on the resonance itself. More details on this phenomenon and the associated physics can be found in dedicated reviews, for example \cite{Kohler:2006aa,Chin:2010aa}.

\begin{figure}[htbp]
\begin{center}
\includegraphics[width=0.65\textwidth]{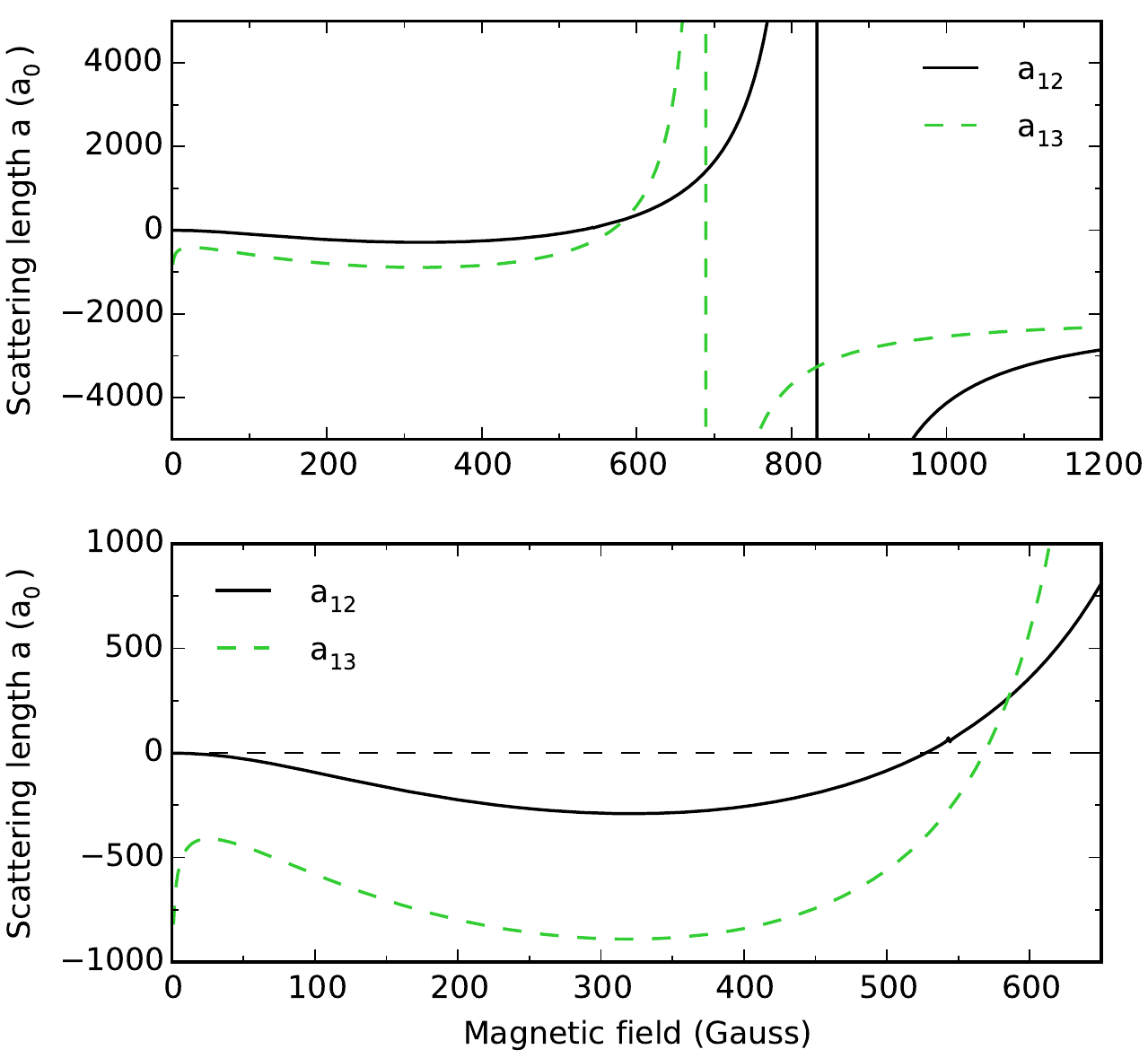}
\caption{ Feshbach resonances between two pairs of the $|1\rangle$, $|2\rangle$, $|3\rangle$ states of $^6\textup{Li}$, used in the experiments. \textbf{a}, Scattering length as a function of magnetic field. \textbf{b}, Zoom of a, onto the low magnetic field region to the left of the resonances.}
\label{fig:scattLength}
\end{center}
\end{figure}

The evolution of the scattering length of two Lithium atoms as a function of magnetic field is presented in figure \ref{fig:scattLength}, based on the data of \cite{Zurn:2013aa}. Depending on the situations, different mixtures of hyperfine states are used. In general, we found that the mixture of the lowest and third lowest hyperfine states (labeled $1$ and $3$) offers more flexibility in tuning the interactions while showing negligible losses, and our most recent measurements operate with this mixture. Otherwise the mixture of the first and second states (labeled $1$ and $2$) is used. Two operation regimes are used for the experiments: (i) the low or moderate interaction regime, at low magnetic field, where the scattering length presents a local minimum and can be continuously tuned to zero, and (ii) the BEC-BCS crossover region which contains the unitary regime, where the system explores the strongly attractive regime, and becomes an s-wave superfluid at low temperatures \cite{Bloch:2008ab,Zwerger:2011aa}.

\subsection{Trapping configuration}

\begin{figure}[htbp]
\begin{center}
\includegraphics[width=0.65\textwidth]{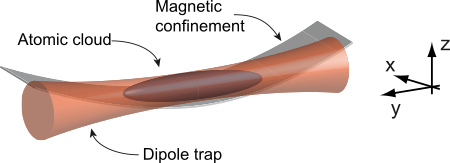}
\caption{Trapping configuration. The dipole trap (red focused laser beam) creates the radial confinement, whereas the axial confinement ($y$ direction) is created by the curvature of the Feshbach magnetic field. }
\label{fig:trapConf}
\end{center}
\end{figure}

The configuration of traps that is used in the setup is presented in figure \ref{fig:trapConf}. It consists in (i) an optical dipole trap confining the atoms along the $x$ and $z$ directions and (ii) a magnetic field with positive curvature providing a weak confinement along the $y$ direction. The trapping frequency along $y$ depends on the magnetic field and thus is slightly different between the weakly and strongly interacting regimes. The position of the minimum of the magnetic field is controlled by the addition of a magnetic field gradient. 

\begin{figure}[htbp]
\begin{center}
\includegraphics[width=0.65\textwidth]{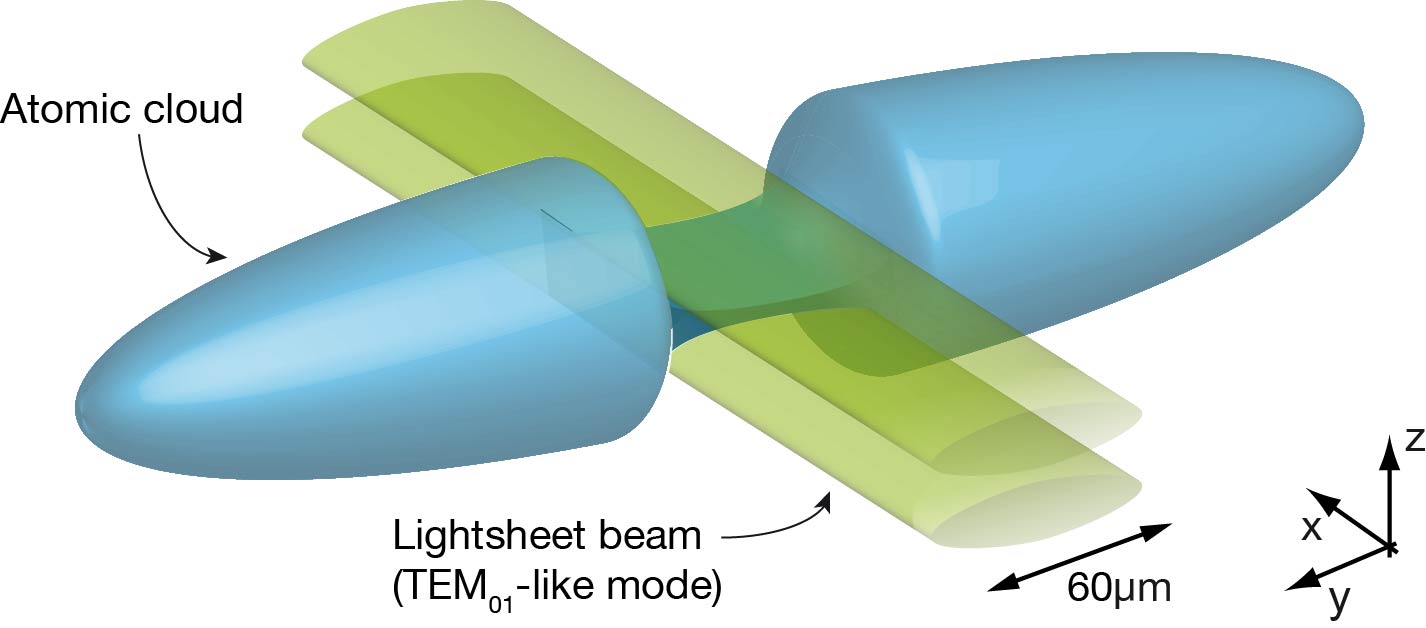}
\caption{A blue-detuned $\textup{TEM}_{01}$-like laser mode propagates along the $x$ direction and is focused on the center of the cigar-shaped cloud. This defines a quasi-2D channel and two smoothly connected reservoirs.}
\label{fig:lightsheet}
\end{center}
\end{figure}

This elongated cloud is the basis in which the two-terminal system is carved. To this end a beam propagating along the $x$ direction, having a nodal line oriented along $y$ at its center, intersects the cloud, as presented in figure \ref{fig:lightsheet}. This beam is produced using holographic techniques \cite{Smith:2005aa,Meyrath:2005aa} and approximates a TEM$_{01}$ mode close to the center. Its wavelength is $532$\,nm, so that it expels atoms from the high intensity regions, thus creating a quasi-two-dimensional region at the center of the cloud. It has a waist of $30\,\mu$m in the $y$ direction, much shorter than the size of the cloud (about $300\,\mu$m). The Gaussian nature of the beam ensures a smooth connection between the tightly confined region and the large, unperturbed regions at both sides. The number of atoms in the confined region is of the order of $10^3$, very low compared to the population in the unperturbed regions at both sides. The separation of scales between the mesoscopic, tightly confined region and the large reservoirs consisting in the large, unperturbed harmonic traps at both ends allows to consider this system as a realisation of the two terminal configuration for transport. 

The tightly confined region resides in the focal plane of two confocal microscope objectives \cite{Zimmermann:2011aa}. This arrangement allows for the observation of the details of the atomic distributions. Furthermore, a wide range of potential landscapes can be projected onto the confined region, by sending light with a tailored intensity distribution through the microscopes. The high numerical aperture of the microscopes leads to potential variations at the scale of micro-meter, the scale of the Fermi wavelength. 

\begin{figure}[htbp]
\begin{center}
\includegraphics[width=0.25\textwidth]{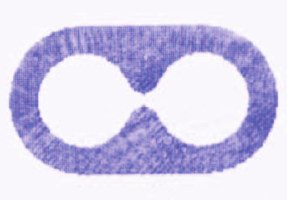}
\caption{Potential pattern used to create a atomic capacitor, with two reservoirs connected by a channel. The transverse size of the channel is $>240\,\mu$m. The roughness in the potential distribution comes from optical imperfections. Figure extracted from \cite{Lee:2013fk}.}
\label{fig:NISTlasercooledAtoms}
\end{center}
\end{figure}

A conceptually similar system was realized at the same time for classical atomic gases of laser cooled atoms \cite{Lee:2013fk}. There, a cold atom cloud of $^87$Rb was trapped in a two reservoirs configuration produced using a spatial light modulator (SLM) producing in the region of the atoms a binary structure as shown in figure \ref{fig:NISTlasercooledAtoms}. The use of a SLM provides a wide tenability for the structure, and the transverse size of the channel connecting the reservoirs was varied between $240$ and $576$\,$\mu$m. The optical imperfections rendered the walls of the structure rough at the scale of $10$\,$\mu$m. %This experiment was recently pushed into the regime of microscopic potentials with transverse sizes of the channel in the range of $10\,\mu$m. 

\subsection{Initialisation}

\begin{figure}[htbp]
\begin{center}
\includegraphics[width=0.85\textwidth]{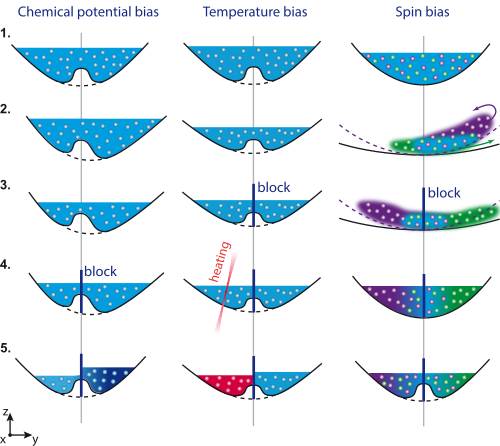}
\caption{Biasing the reservoirs. The solid line represents to potential landscape combining a large parabolic trap and an extra confinement at the center, the level of the blue color represents the chemical potential. Left: Creation of a chemical potential bias. Middle: Creation of a temperature bias. Right: Creation of a spin bias.}
\label{fig:biases}
\end{center}
\end{figure}

In order to induce currents through the mesoscopic region, we introduce a bias between the two reservoirs. Three different procedures are used for the three kinds of biases used throughout our past experiments. These are summarized in figure \ref{fig:biases}.

\paragraph{Particle number imbalance}
A particle number difference between two reversoirs having the same geometry allows for the introduction of a chemical potential bias. Such a difference is produced by shifting the position of the center of the magnetic confinement along the transport direction (y), before the evaporation process, by applying an extra magnetic field gradient \footnote{Note that in the relevant regimes where a large offset magnetic field is present, all hyperfine states have the same magnetic moment. Thus the magnetic field gradient only couples to density.}. The subsequent evaporation produces a cold cloud with unequal populations, but with the same chemical potential, due to thermal equilibration during the evaporation process. Afterwards, a "wall" beam is introduced, consisting in a blue detuned, elliptic beam creating a large repulsive potential barrier over the conductor, preventing any exchange between the reservoirs. The magnetic field gradient is then adiabatically reduced, restoring the symmetry between the geometry of the two reservoirs, but maintaining the population difference constant. The chemical potential bias can be finely adjusted by controlling the strength of the magnetic field gradient. 

\paragraph{Temperature imbalance}
To introduce a temperature imbalance, a balanced situation with the same atom number and temperature is first produced, then a "wall" beam produced using a blue detuned laser beam focused on the channel is turned on, separating the two reservoirs. Another laser beam, red detuned from resonance, is then focused on one reservoir creating a perturbation. The power of this beam is then modulated at the frequency of the radial breathing mode of the cloud, thereby depositing energy into one reservoir by parametric heating. Controlling the duration and amplitude of the excitation allows for a precise adjustment of the heating. 

\paragraph{Spin bias}
To create a spin bias, we build on a technique developed in \cite{Sommer:2011uq}. A difficulty is that for the relevant cases where a high offset magnetic field is present, the different hyperfine components have the same magnetic moment. For weak magnetic field offsets, the densities and scattering lengths do not permit efficient evaporative cooling. We thus first produce a cold Fermi gas in a balanced mixture of states $1$ and $2$ at an offset field of $302\,$G, then ramp down the magnetic field to $52\,$G, where the spin components have different magnetic moments. A pulse of magnetic field gradient initiates dipole oscillations, which have different frequencies due to the magnetic nature of the confinement along the transport direction. The dipole oscillations for the two components dephase with time, producing transiently a cloud centered in the trap but having finite displacement between the spin component. The oscillations are then abruptly interrupted by turning on a strong "wall" beam, disconnecting the two sides of the cloud. The confinement at the center of the cloud is then ramped up together with the other beams producing the potential landscape in the conductor, and the offset field is ramped up to high values again.  Atoms in state $2$ are transferred to state $3$ using a high efficiency Landau-Zener radiofrequency sweep. Finally, an extra step of evaporative cooling brings the reservoirs down to low temperatures allowing for superfluidity \cite{Krinner:2016ab}.

%This relatively complex procedure renders fine tuning of the spin bias more difficult. In particular, achieving a pure spin bias without extra particle imbalance usually requires a fine optimisation of each step. 

\subsection{Transport measurements}

Transport measurements are initiated by removing the "wall" beam separating the reservoirs, which allows the reservoirs to exchange particles through the conductor. After an evolution time $t$, the wall beam is abruptly turned back on, and the laser beams producing the confinement in the conductor and the potential structures are adiabatically turned down. This turns the reservoirs into half-harmonic traps, where atom number and temperatures can be estimated using the time-of-flight technique applied to the transverse direction. 

In order to extract the conductance from the time evolution of the atom number in the two reservoirs, we make use of a simple linear resistor-capacitor (RC) model. This model captures the essential features of the measurements, and is very helpful to disentangle the effects genuinely originating from transport from those reflecting the thermodynamics of the reservoirs. We introduce it here for the simple case of particle transport. The cases of heat and spin transport will be covered in dedicated sections below.

\begin{figure}
    \includegraphics[width=0.65\textwidth]{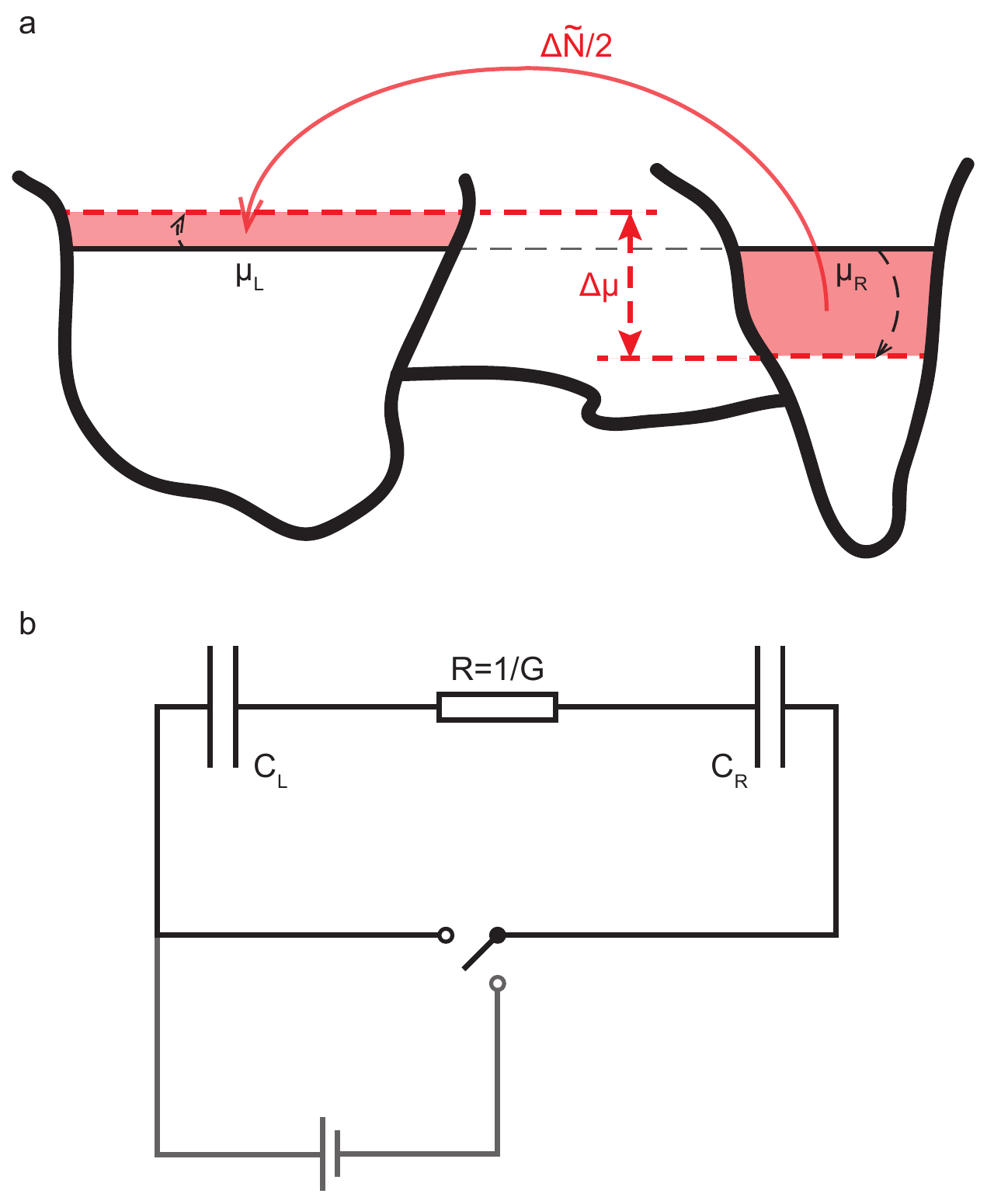}
    \centering
    \caption{%
    {\bf Generic two-terminal model.} \textbf{a,} Two reservoirs of different size and shape are connected by a transport channel. Upon transfer of $\Delta \tilde{N}/2$ particles from the right to the left reservoirs a chemical potential bias $\Delta\mu$ arises. \textbf{b,} Corresponding RC circuit: left and right reservoir are represented each by a capitor, having a capacitance $C_{\textup{L}}$ and $C_{\textup{L}}$. They can be charged by moving the switch in the vertical position. A discharge through the resistor $R$ is initiated by moving the switch in the horizontal position.}
    \label{fig:rc-model}
\end{figure}

Consider two arbitrarily shaped atomic reservoirs connected by a channel characterized solely by its conductance $G$. It may have any dimension and an arbitrary shape, the only requirement should be that it contains a negligible atom number compared to the reservoirs. This condition implies that $G$ is low enough to not place the reservoirs out of equilibrium by a too fast emptying of the reservoirs. At equilibrium the reservoirs have the same chemical potential $\mu_0$, but may contain in general a different number of atoms, $N_{\textup{L},0}$ and $N_{\textup{R},0}$, as sketched in Fig. \ref{fig:rc-model}a. 
Imagine that we shuffle $\Delta \tilde{N}/2$ atoms from reservoir R to reservoir L, i.e.
\begin{eqnarray}
N_{\textup{L}} &=& N_{\textup{L},0} + \Delta \tilde{N}/2 \\
N_{\textup{R}} &=& N_{\textup{R},0} - \Delta \tilde{N}/2.
\end{eqnarray}
This will change the chemical potential in each of the reservoirs to
\begin{equation}
\mu_{\textup{L(R)}} = \mu_0 + \Delta\mu_{\textup{L(R)}}.
\end{equation}
The changes in chemical potential can be expressed in linear response in terms of the compressibilities $C_{\textup{L(R)}}=\partial N_{\textup{L(R)}}/\partial \mu_{\textup{L(R)}}$ of the reservoirs:
\begin{eqnarray}
\Delta\mu_{\textup{L}} &=& \frac{1}{C_{\textup{L}}}\frac{\Delta \tilde{N}}{2}\\
\Delta\mu_{\textup{R}} &=& -\frac{1}{C_{\textup{R}}}\frac{\Delta \tilde{N}}{2}.
\end{eqnarray}
The difference in chemical potential or the chemical potential bias $\Delta\mu = \mu_{\textup{L}} - \mu_{\textup{R}}$ is thus given by
\begin{equation}
\Delta\mu = \left(\frac{1}{C_{\textup{L}}}+\frac{1}{C_{\textup{R}}}\right)\frac{\Delta \tilde{N}}{2} = \frac{1}{C_{\textup{eff}}}\frac{\Delta \tilde{N}}{2}, \label{eqn:DeltaMu}
\end{equation}
where we have introduced an effective compressibility $C_{\textup{eff}}$. To get the dynamics of the system we differentiate Eqn. (\ref{eqn:DeltaMu}) with respect to time obtaining
\begin{equation}
\frac{\textup{d}}{\textup{d}t}\Delta\mu = \frac{1}{C_{\textup{eff}}}\frac{\textup{d}}{\textup{d}t}\frac{\Delta \tilde{N}}{2} = -\frac{1}{C_{\textup{eff}}}I = -\frac{G}{C_{\textup{eff}}} \Delta\mu, \label{eqn:diffEqn_DeltaMu}
\end{equation}
where we used of the definition of the current $I=-\frac{\textup{d}}{\textup{d}t}\frac{\Delta \tilde{N}}{2}$, which is positive when the particle flow goes from reservoir L to reservoir R. Since we assumed linear response, $\Delta\mu$ is linearly related to $\Delta \tilde{N}$ by Eqn. (\ref{eqn:DeltaMu}) and we obtain the same differential equation for $\Delta \tilde{N}$:
\begin{equation}
\frac{\textup{d}}{\textup{d}t}\Delta\tilde{N} = -\frac{G}{C_{\textup{eff}}} \Delta\tilde{N}. \label{eqn:diffEqn_DeltaNtilde}
\end{equation}
The solution of Eqn. (\ref{eqn:diffEqn_DeltaMu}) and (\ref{eqn:diffEqn_DeltaNtilde}) with the initial condition $\Delta\mu(t=0)=\Delta\mu_{\textup{ini}}$ and $\Delta \tilde{N}(t=0)=\Delta \tilde{N}_{\textup{ini}}\,\,(=2C_{\textup{eff}}\Delta \mu_{\textup{ini}})$ is an exponential decay of the chemical potential bias or particle number imbalance as a function of time,
\begin{eqnarray}
\Delta \mu(t) &=& \Delta\mu_{\textup{ini}} \, \textup{exp}(-t/\tau) \\
\Delta\tilde{N}(t) &=& \Delta\tilde{N}_{\textup{ini}} \, \textup{exp}(-t/\tau), \label{eqn:imbalance-fct-of-time}
\end{eqnarray}
with a time constant $\tau=C_{\textup{eff}}/G$. This is analogous to the discharge of a capacitor, having a capacitance $C_{\textup{eff}}=\partial Q/\partial U$, through a resistor $R=1/G$: the chemical potential difference $\Delta\mu$ is the analogue of the voltage $U$ across the capacitor and $\Delta\tilde{N}/2$ corresponds to the charge $Q$ on the capacitor. More precisely, the capacitor corresponding to $C_{\textup{eff}}$ can be thought of as two capacitors in series, one corresponding to $C_\textup{L}$ and the other to $C_\textup{R}$. The corresponding $RC$ circuit is drawn in Fig. \ref{fig:rc-model}b.
In the experiment we rather measure the total atom number difference between both reservoirs
\begin{equation}
\Delta N = N_{\textup{L}} - N_{\textup{R}} = (N_{\textup{L},0}-N_{\textup{R},0}) + \Delta \tilde{N} = \Delta N_{0} + \Delta \tilde{N}, \label{eqn:deltaN}
\end{equation}
where $\Delta N_{0}$ is the difference in atom number between the two reservoirs at equilibrium. Typically, $\Delta N_0/(N_{\textup{L}} + N_{\textup{R}})$ is on the few per cent level and depends on the precise alignment of the channel with respect to the cigar-shaped cloud. 

Using expression (\ref{eqn:deltaN}) in Eqn. (\ref{eqn:imbalance-fct-of-time}) we obtain the time evolution of $\Delta N$,
\begin{equation}
\Delta N (t) = \Delta N_0 + \Delta\tilde{N}_{\textup{ini}} \, \textup{exp}(-t/\tau). \label{eqn:rc-model-exp-decay}
\end{equation}

This simple reasoning requires the following important remarks. First, the hypothesis underlying the reasoning is that the reservoirs maintain their internal equilibrium at each point in time, as necessary for the validity of the Landauer approach. In our experiment, this is ensured by the separation of length scales between large reservoirs and a small conductor. Currents through the conductor are weak enough that the corresponding perturbation on the reservoirs is small. In particular, we operate in the regime where $\omega \tau \gg 1$, where $\omega$ is the typical trap frequency of the reservoirs, such that the current does not couple to the internal excitation modes of the reservoirs. 

Second, the quantity which is experimentally accessible without any further assumption is the time scale $\tau$, corresponding to the exponential decay of the initial perturbation. Such an exponential decay is typical for many cold atom experiments where dynamics following a quench is measured. The important difference introduced by operating with reservoirs connected to a mesoscopic system is that $\tau$ can now be decomposed into two contributions: a thermodynamic contribution, the compressibility, and a genuine transport coefficient $G$. In situations where a global relaxation of the system is observed, transport and thermodynamics occur simultaneously at the same time and length scales, preventing the study of transport independently of the static properties.

Third, in our mesoscopic approach, the compressibility is a property of the reservoirs only. Thus, the conductance can be directly studied, either by modifying the conductor only and comparing the subsequent evolution of $\tau$, or by directly modeling the reservoirs and calculating $C_{\textup{eff}}$, allowing for the extraction of $G$. For non interacting gases, accurate modeling of the reservoirs is available for any temperature and spin polarisation, provided the shape of the potentials is known. For interacting gases, recent experiments have provided high precision measurements of the equation of state of the homogenous Fermi gas \cite{Luo:2009aa,Horikoshi:2010aa,Nascimbene:2010ys,Navon:2010ab,Ku:2012aa}, that can be used as an input to estimate $G$. For cases not covered by these measurements, one can rely on theoretical calculations, but without experimental benchmark these can have systematic errors that may be hard to estimate, in particular with strong interactions. 

Fourth, the exponential decay of the initial imbalance is a direct consequence of the linear response hypothesis. Deviations from an exponential decay directly indicate a breakdown of linear response, which can be tracked down to a non linear current-bias relation. This is in particular expected for superfluids \cite{Husmann:2015ab,Krinner:2016ab}, see section \ref{section:superflow}.

In this reasoning, it was supposed that the currents are weak enough that no inductive effects have to be taken into account. These typically arise due to the finite mass of the particles: the potential or interaction energy stored in the capacitive element can also be converted and stored in the form of kinetic energy of the atoms. This is best illustrated by the trivial limiting case where the channel is absent. One is then left with a harmonic trap where an initial imbalance triggers undamped dipole oscillations. The parasitic inductance gives a significant contribution only when the relaxation timescale is comparable or lower than the trap frequency along the transport direction. A possible model in this case is the RLC circuit, which was put forward in \cite{Lee:2013fk}, where explicit expressions were derived for inductive terms. These effects are important in the case of superfluids where large currents can be driven even through narrow channels.

The period of dipole oscillations is also a lower bound on the thermalisation time scale of the reservoirs. As a result in the presence of large currents the reservoirs cannot be considered at equilibrium at each point in time: among other issues, the instantaneous chemical potential cease to have a meaning and the capacitive effect of the reservoirs is not anymore given by the equilibrium compressibility.

\section{Transport in the multimode regime}
\label{section:multimodeWIF}
We now examine transport experiments with weakly interacting Fermions, where the concepts described in the previous section are used to investigate  ballistic and diffusive transport \cite{Brantut:2012aa,Brantut:2013aa}. This section is dedicated to experiments where transport takes place in a regime of weak transverse confinement in the channel, such that the transverse size of the channel is larger than the Fermi wavelength, or equivalently the Fermi energy is larger than the trap frequencies. In this case, the quantization of the motion of atoms along the transverse direction of the conductor do not play a strong role. These experiments have provided the first proof of principle for the two-terminal approach in cold atomic gases, demonstrating the potential of the technique. They are also used as a benchmark for the more intricate experimental situations explored in later sections. 

\subsection{Conduction}
The two-terminal transport experiments reported in \cite{Brantut:2012aa} were performed with Fermi gases at a scattering length of -145$\,a_0$, in the weakly interacting regime and at a temperature of $0.36(18)$\,$T_F$. The estimated mean free path for inter particle collisions was $\simeq1.3$\,mm, larger than the typical size of the cloud ($\sim 300\,\mu$m). The collision rate in the reservoirs was estimated to be $\simeq 30$\,ms, ensuring thermalisation at this time scale. \footnote{In this experiment, no "wall" beam was used to separate the clouds, but the slow timescale of transport compared to the switch off time of the magnetic field gradient was used instead to initialise the transport.}

In these experiments, the channel is defined solely by the tight vertical confinement of the TEM$_{01}$-like mode. The frequency along this direction was at most $3.9\,$kHz, lower than temperature and Fermi temperature such that even along the tightly confined direction, several quantum mechanical modes are populated. The resulting transport is thus taking place in a quasi-classical regime. 

\begin{figure}
\includegraphics[width=0.65\textwidth]{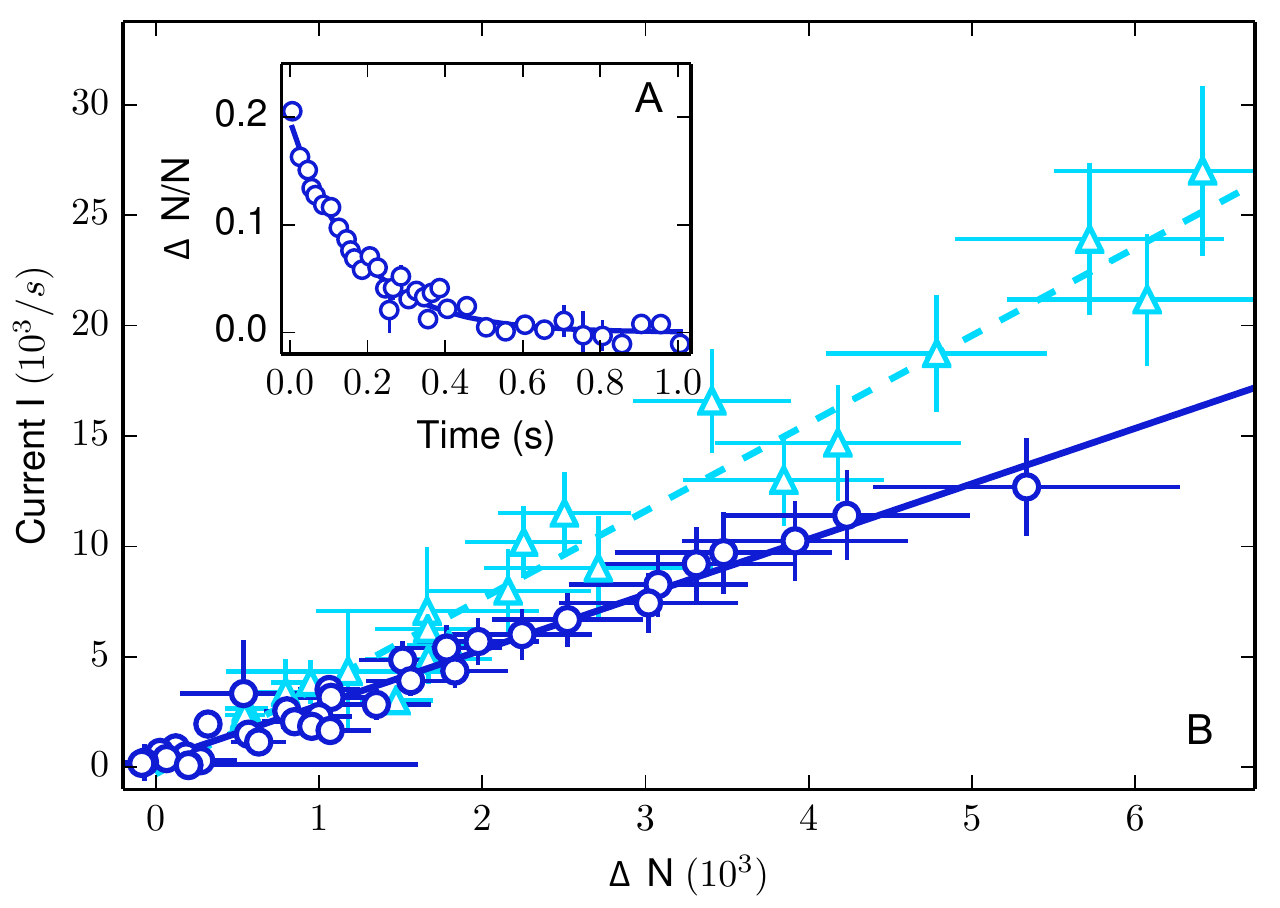}
\centering
\caption{%
{\bf Observation of ohmic conduction.} (A) Measured number difference between the two reservoirs as a function of time. The solid line is an exponential fit to the data. (B) Current as a function of number difference between the two reservoirs, measured from the exponential fit of panel A, for two different confinements in the channel. A small offset obtained from the fits in panel A, which is due to a slight misalignment of the channel with respect to the center of the trap, has been subtracted. Circles: maximum center frequency along $z$ set to 3.9\,kHz, triangles : 3.2\,kHz. The lines are linear fits to the data. Figure adapted from \cite{Brantut:2012aa}.}
\label{fig:conduction}
\end{figure}

The decay of the initially prepared imbalance was observed to be exponential, as shown in figure \ref{fig:conduction}A, in agreement with the linear response hypothesis. The decay time $\tau$ is $170$(14)\,ms, larger than the collision rate in the reservoirs, and much larger than the trap frequencies in the reservoirs, such that the decomposition of $\tau$ into a reservoir compressibility and a conductance for the channel is justified. Extracting the derivative of the decay curve at each point in time, using a linear fit to successive points, yields a set of measured currents, with the corresponding average particle number difference. The resulting data are presented in figure \ref{fig:conduction}B, representing the first observation of Ohm's law with cold atomic gases. 

The microscopic origins of the resistance in the case of ballistic transport is the highly restricted phase space available in the contact compared to the reservoirs, such that most of the atoms impinging onto the channel are actually reflected elastically. This can be put on formal ground using the Landauer formalism, which will be presented in the next section. A consequence of the ballistic nature of transport in the channel is that the density inside the channel in the presence of a DC current is uniform. Thus, the variations of densities in the presence of the current are concentrated at the contacts between the channel and the reservoirs \cite{datta_electronic_1995}.

\begin{figure}
\includegraphics[width=0.65\textwidth]{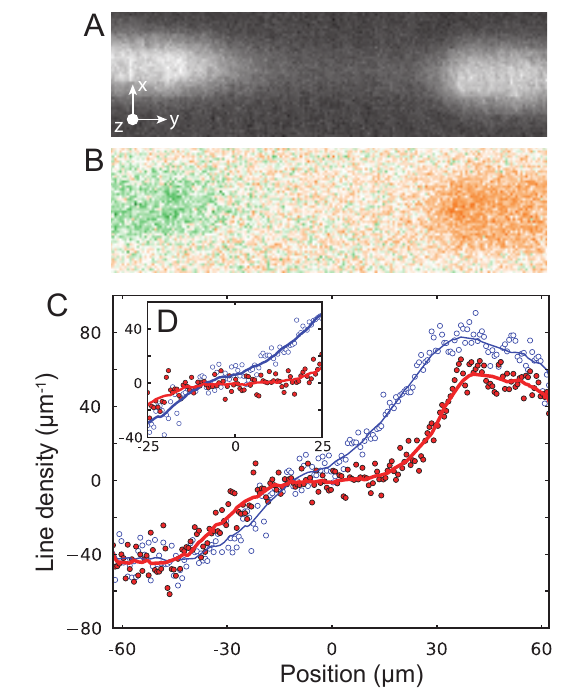}
\centering
\caption{%
{\bf Investigation of ballistic and diffusive conduction using high-resolution imaging.} (A) Absorption picture of the density in the channel, for a cloud at equilibrium (no current). (B) Difference between two pictures taken at equilibrium and with a current of $10^4\,s^{-1}$. The color is orange for positive difference and green for negative. (C) Line-density difference obtained by accumulating B along the $x$-axis, for a ballistic channel (red full circles) and for a diffusive channel having the same conductance (blue open circles), in the presence of the same current. The solid lines are smoothed data to guide the eye. (D) Focus on the central part of the line density difference. Figure adapted from \cite{Brantut:2012aa}.}
\label{fig:conductionAbs}
\end{figure}

Absorption imaging was used in order to observe the density distribution across the channel in the presence of current. Such a picture is shown in figure \ref{fig:conductionAbs}A, where the two reservoirs can be seen on both sides of a low density region, the channel. In the presence of a chemical potential difference between the reservoirs driving a current though the channel, the density distribution is slightly modified. The difference between the density distributions in the presence and in the absence of current is shown in figure \ref{fig:conductionAbs}B. Integration along the direction transverse to the channel yields a line density difference shown in figure \ref{fig:conductionAbs}C. The plateau at zero centered on the channel demonstrates the ballistic character of the channel, while the drop of density is concentrated at the contacts. 

These observations were repeated with a channel on which a strong disorder was imposed, using the potential induced by a laser speckle pattern \cite{Shapiro:2012aa}. The confinement along the vertical direction was reduced in order for the disordered channel to exhibit the same conductance as the ballistic one (the same bias applied to the reservoirs was producing the same current). The corresponding line-density difference along the channel is shown in blue on figure \ref{fig:conductionAbs}C, showing a linear decrease of the density across the channel, as expected for a diffusive system. 

The distribution of density difference across the channel contains informations about the local transport properties, such as mobility or resistivity. An important difficulty however lies in the non-equilibrium nature of the transport phenomena, preventing the unambiguous attribution of a chemical potential to a region of space from the knowledge of the density. In the case of charged particles this can be circumvented by referring to a local value of the electrical potential, which can always be well defined by solving Maxwell's equations, and connects with the chemical potential at equilibrium. For neutral particles, the distribution is not coupled to any field and thus cannot be related to a potential unless local equilibrium is achieved. A pragmatic solution suggested in \cite{datta_electronic_1995} and used in \cite{Brantut:2012aa} is to attribute to a point a chemical potential that would be required to achieve the corresponding density at equilibrium. This gives a meaning to the question "were does the chemical potential drop ?". However one should keep in mind that this refers to a fictitious equilibrium situation, which is not the one actually realised in the experiments. Measurements in condensed matter systems have actually shown that the distribution is very far from thermal inside a driven conductor \cite{Pothier:1997aa,Pothier:1997ab}. 

\subsection{Thermoelectric effects}
%This describes the results of Brantut et al, Science 342 713 (2013). Also a short description of the most important findings. Possible extensions to cooling are presented. 

As the cold atoms are isolated from the environment, energy is strictly conserved throughout the entire transport measurement. This gives an opportunity to investigate heat transport in a situation where energy can only be carried from one reservoir to the other via particle exchange, in contrast to solid materials where phonons and other excitations can contribute. In addition, the direct imaging of the momentum distribution in the reservoirs gives a direct access to the total entropy during the transport process and permits to keep track of entropy exchange and creation, thus allowing to isolate the reversible processes from irreversible ones. These rather unique conditions are ideal test beds for the fundamental principles of thermoelectricity, and more generally heat engines, with a quantitative comparison possible with a microscopic theory.

In the quasi-classical regime with weakly interacting particles, measurements of heat transport and thermoelectric power were performed and reported in \cite{Brantut:2013aa} using the two terminal configuration. The theoretical description parallels that of particle transport, except that linear response relations take matrix forms with off-diagonal coefficients describing coupling between heat and particle currents and the temperature and chemical potential bias. 

The linear response transport is described within linear response by the "Onsager" matrix connecting the currents of extensive quantities to the gradients of the thermodynamically conjugated quantities:
\begin{equation}\label{eq:lin_response}
\left(
\begin{array}{c}
 I_N\\
 I_S
\end{array}
\right)
=
-G\left(\begin{array}{cc}
		1&\alpha_{ch}\\
	      \alpha_{ch}&L+\alpha_{ch}^2
             \end{array}
\right)
\left(
\begin{array}{c}
 \mu_c-\mu_h\\
 T_c-T_h
\end{array}
\right)\,.
\end{equation}
where $I_N$, $I_S$ are the particle and entropy currents, $T_h, \mu_h$ and $T_c, \mu_c$ refer to temperature and chemical potentials of the hot and cold reservoirs respectively, and $\bar{T}=(T_h + T_c)/2$. $G$ is the particle conductance, $L=\frac{G_T}{\bar{T} G}$ is the Lorenz number with $G_T$ the heat conductance and $\alpha_{ch}$ is the thermopower of the channel. The symmetric nature of the matrix is a consequence of Onsager's relations \cite{Onsager:1931aa}. 

Note that we formulate the problem in terms of the entropy currents rather than energy currents. The final results are the same but the use of entropy provides a more direct distinction between reversible and irreversible processes, which are of fundamental interest in assessing the thermodynamic efficiency \cite{antoineCollegeThermoelec}. 

Similarly, the reservoirs are characterised by a "Maxwell" matrix, symmetric by virtue of the Maxwell relations
\begin{equation}\label{eqn:thermodynamics-reservoirs}
\left(
\begin{array}{c}
 \Delta N\\
 \Delta S
\end{array}
\right)
=
\left(\begin{array}{cc}
		\kappa&\kappa \alpha_{r}\\
	      \kappa \alpha_{r}& \frac{C_{\mu}}{T}
             \end{array}
\right)
\left(
\begin{array}{c}
 \Delta \mu \\
 \Delta T
\end{array}
\right), 
\end{equation}
with $\kappa=\left. \frac{\partial N}{\partial \mu}\right|_{T}$, $\alpha_r = \left. \frac{\partial S}{\partial N}\right|_{T} = -\left. \frac{\partial \mu}{\partial T}\right|_{N}$, $C_{\mu}=\left. T\,\frac{\partial S}{\partial T}\right|_{\mu}$ the compressibility, dilatation coefficient, and specific heat of each reservoir respectively. Further we have written $\Delta\mu=\mu_c-\mu_h$ and $\Delta T = T_c-T_h$. All thermodynamic quantities are evaluated at the average temperature $\bar{T}$ and particle number $(N_c+N_h)/2$.

A reasoning similar to that of equation \ref{eqn:diffEqn_DeltaNtilde} yields the time evolution of the populations and temperatures of the reservoirs: 
\begin{equation}\label{eq:temp_imb}
 \tau_0
\frac{d}{dt}\left(
\begin{array}{c}
 \Delta N\\
 \Delta T
\end{array}
\right)
=
-\left(\begin{array}{cc}
                                 1 &-\kappa(\alpha_r-\alpha_{ch})\\
                                 -\frac{\alpha_r-\alpha_{ch}}{\ell\kappa} & 
\frac{L+(\alpha_r-\alpha_{ch})^2}{\ell}
                                \end{array}\right)
\left(
\begin{array}{c}
 \Delta N\\
 \Delta T
\end{array}
\right)\,.
\end{equation}

Here, $\ell = \frac{C_N}{\kappa \bar{T}}$ is an analogue of the Lorenz number for the reservoirs, measuring the relative magnitude of thermal fluctuations of entropy and atom number, and $\tau_0 = \kappa G^{-1}$ is the particle transport timescale identical to the decay constant for particle transport. 

\begin{figure}
\includegraphics[width=0.55\textwidth]{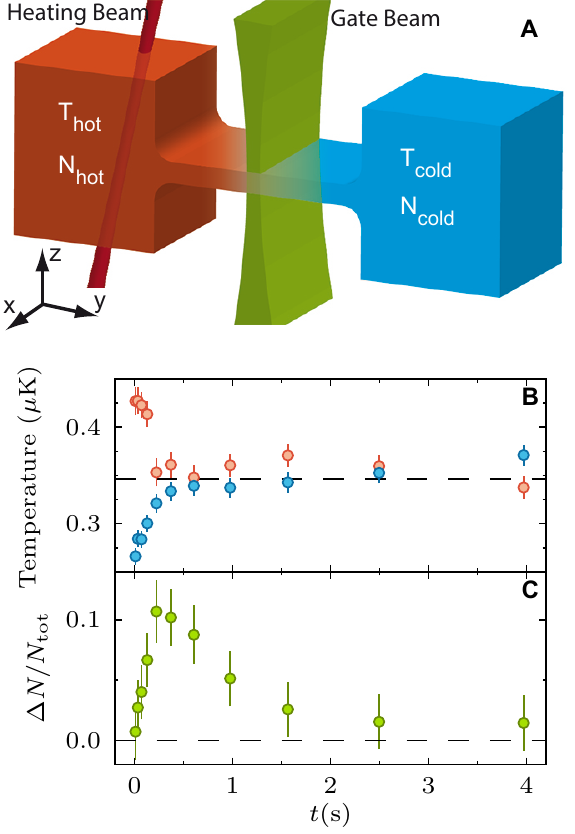}
\centering
\caption{%
{\bf Thermoelectricc transport experiment.} A: A quasi-two dimensional channel connects two atomic reservoirs. A gate beam intersects with the channel and blocks particle and heat transport. A heating beam traverses the left reservoir and heats it in a controlled way. B: $T_h$ (red) and $T_c$ (blue) as a function of time. Dashed line~: $\bar{T}$ at the initial time. C: $\Delta N/N_{tot}$ as a function of time. In the channel $\nu_z$ was set to $3.5\,$kHz with a disorder of average strength $542\,$nK~(see text). Figure adapted from \cite{Brantut:2013aa}}
\label{fig:thermo1}
\end{figure}

The experiments where performed in conditions similar to that of the previous section. Transport is initialised by introducing a temperature bias between the reservoirs, using a laser focused in one of the reservoirs which is modulated to induce parametric heating, as depicted in figure \ref{fig:thermo1}A. The subsequent evolution of particle number and temperature in the reservoirs after various evolution times is presented in figure \ref{fig:thermo1}B and C. It should be noted that measuring temperatures separately in the two reservoirs is much more challenging than evaluating the atom numbers, and these experiments could only be performed at moderate temperatures where the atom numbers are large in each reservoir and the signal to noise ratio for fitting the cloud profiles is not strongly reduced by Fermi statistics \cite{Demarco:2001ab}.

Temperatures in there reservoirs converge exponentially, reflecting heat transport between the reservoirs in a manner similar to particle transport in the simple RC model. That the convergence temperature is equal to the average of the initial temperatures of the reservoirs is a further indication that despite the large temperature difference the evolution is still within liner response. The atom number imbalance, initially zero, first increases linearly and then decreases back to zero. This transient effect represents the thermoelectric response of the system. 

%A fundamental difference with respect to condensed matter situations is the contribution of the thermodynamic of the reservoirs to the response. Indeed, like with particle transport, the observations result from the combination of a thermodynamic response from the reservoirs and a transport response of the channel. 
As highlighted in the previous sections, the response is the combination of two factors, a thermodynamic response of the reservoirs and a transport response from the channel. This is crucial in the context of thermoelectricity. Here, the thermodynamic part leads to a decrease of the chemical potential with increasing temperature at fixed particle number, as is well known in cold atoms physics. This leads to a temperature induces chemical potential bias, that induce a particle current. This effect has nothing to do with thermoelectric effects, it would be present even with a purely diagonal Onsager matrix. 

The genuine thermoelectric response of the system, contained in the off-diagonal elements of the Onsager matrix, creates an additional contribution to transport which depends on the details of the channel geometry. In the experiments of \cite{Brantut:2013aa}, it turns out to be positive, i.e. directed from high temperature to low temperature. As a result the thermoelectric and thermodynamic contributions compete with each other. The positive current in figure \ref{fig:thermo1}C indicates that the thermoelectric effects dominate. 

An important consequence of this particular combination is that the current is driven from the low chemical potential reservoir to the high chemical potential reservoir, i.e. against the bias. In other words, the channel operates like a heat engine, generating work in the form of current flowing against the bias out of the connection of a hot and cold reservoir. Because of its isolated nature, the engine is not able to perform cycles since the heat inserted in the reservoir remains in the system. However, the cycle is closed up to a global increase of temperature and was analysed in \cite{Brantut:2013aa}. 

The strength of the thermoelectric response was quantified experimentally using a figure of merit $\mathcal{R} = (\Delta N /N_{\mathrm{tot}}) / (\Delta T_0/T_F)$, with $\Delta T_0$ the initial temperature imbalance. With increasing the confinement in the channel, the thermoelectric response and the time scale $\tau$ characterising particle transport increase. 

\begin{figure}
\includegraphics[width=0.55\textwidth]{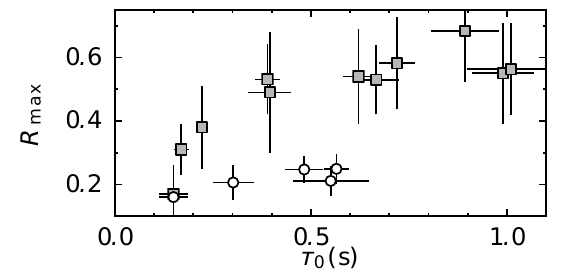}
\centering
\caption{%
{\bf Thermoelectricity in the ballistic to diffusive crossover.} $\mathcal{R}_{\mathrm{max}}$ versus timescale $\tau_0$ for the diffusive (gray squares) and ballistic case (open circles). Figure adapted from \cite{Brantut:2013aa}.
}
\label{fig:thermoRVStau}
\end{figure}

The experiments were repeated in the presence of disorder in the channel, where transport crosses over from ballistic to diffusive. Also in this case, transport time scale and thermoelectric response were seen to increase together. The thermoelectric response is presented in figure \ref{fig:thermoRVStau} as a function of the measured $\tau$ for both the ballistic and diffusive cases. Interestingly, for a given $\tau$, i.e. for a given resistance of the channel, the disordered channel is seen to have a much higher thermoelectric response, up to three times higher than the ballistic channel. 

The physical reason is the much stronger dependence on energy of the rate of particle transport across the channel, compared to ballistic transport. In the ballistic case, the dependence in energy comes from the variations of the density of modes with energy. In the case of disordered channels, transport typically proceeds by random walks, modified by the wave nature of matter. Typical distributions of the random potential, like the speckle patterns used in most experiments with cold atoms, yield mean free paths that have very strong energy dependance. These variations lead to a strong asymmetry around the Fermi level between high energy particles predominantly emitted by the hot reservoir and low energy particles emitted by the cold one (or equivalently between particles and holes), leading to enhanced thermoelectric response \cite{Grenier:2012tg}. 

A theoretical model for the reservoirs and channel was used in \cite{Brantut:2013aa} in order to extract transport coefficients from the time evolution of temperatures and atom number differences. For the case of ballistic transport, good agreement was found between the predications, based on the Landauer theory and ideal gas thermodynmics and the observations. The case of a disordered channel was modeled using a heuristic interpolation formula allowing to describe the crossover from ballistic to diffusive \cite{datta_electronic_1995}. To model the diffusive transport, a power law dependance of the scattering time on the disorder strength was fitted to the data, showing good agreement. 

The microscopic understanding of transport of cold atoms in realistic speckle potentials is a topic on its own, far beyond the specific case of two-terminal transport (see \cite{Kuhn:2007aa} for a systematic treatment). We only point out the fact that the energy dependence of the transport properties is a general feature of disordered systems, which is likely to be a key for the interpretation of the new generation of experiments studying high temperature transport in closed systems \cite{Choi:2016ab}.

In the same spirit, the coupling of spin and heat transport in Fermi gases was analyzed theoretically \cite{Wong:2012ab,Kim:2012aa} following the spin diffusion experiments \cite{1367-2630-13-5-055009,Sommer:2011uq}. By introducing a spin polarization in a Fermi gas, one creates an energy mismatch between the Fermi surfaces of the minority and majority populations. As polarization gradients or temperature differences are introduced, currents of heat or spin arise as a result of the differential relaxation of the high energy atoms with respect to the low energy ones, similarly to the thermoelectric effects described above. As interactions are varied in the vicinity of the Feshbach resonance, the interaction cross section between the two spin components changes from hard-core, energy insensitive in the weakly interacting limit to unitary with an inverse square dependence on momentum. This yields qualitative changes in the spin-Seebeck coefficients. \cite{Wong:2012ab,Kim:2012aa}.

The operation of the channel as a heat engine evidenced by the driving of currents against the chemical potential bias provides perpectives for the use of transport in order to cool down Fermi gases, using the Peltier effect. This was theoretically investigated in details in \cite{Grenier:2014aa}. The idea is to design a channel that can filter out some classes of energy, i.e. having a peaked transmission as a function of energy. By biasing the channel with a well chosen energy transmission band it is possible to inject low entropy atoms into one of the reservoirs in order to 'rectify' the energy distribution and lower the entropy per particle. Importantly, it was shown that this process is intrinsically more efficient than plain evaporative cooling, specially in the degenerate regime, raising exciting perspectives for quantum simulation in the low temperature regime with Fermi gases. More details on the possible applications are described in the recent review \cite{Grenier:aa}.  A strong energy dependence of the density of states is also a feature of BCS superfluids. The thermoelectric effects in these systems is topic of general interest in condensed matter, and the case of superfluid Fermi gases in the two terminal configuration was recently investigated theoretically \cite{PhysRevA.94.033618}.

The operation of complete heat engines with cold atoms, and more recently with individual trapped ions \cite{Rosnagel:2016aa}, opens perspective for the fundamental understanding of thermodynamics at the quantum level, in particular the exploration of the microscopic origins of irreversibility and the operation of quantum-enhanced heat engines \cite{PhysRevLett.112.030602}.

\section{Single mode regime and quantum point contacts}
\label{section:QPC}
%In the experiments presented in the previous section, the energy scales of temperature and bias are large compared to the spacing between the modes in the channel, such that the quantization of motion was not manifest. Such a situation is very common in cold atomic gases, where harmonic traps are used in most experiments. There, a trap with frequencies much smaller than the chemical potential is usually treated in the local density approximation, corresponding to a situation where the de Broglie wavelength of particles, or healing length for superfluids, is much smaller than the length scales of the potential variations \cite{Giorgini:2008ab}. Conversely, low dimensional quantum gases have been investigated for more than a decade with cold atoms. They are produced by strongly confining atoms such that the trap frequencies exceed the chemical potential and temperature, and their properties are seen to strongly deviate from their three dimensional counterparts \cite{Bloch:2008ab}. %However, to our knowledge, clouds of atoms with coexisting regions of different dimensionality, like a channel and reservoirs, had not been investigated before.
In the experiments presented in the previous section, the energy scales of temperature and bias are large compared to the spacing between the transverse modes in the channel, such that the quantization of motion was not relevant. Most traps for atomic gases operate in this regime, where the trapping frequencies are much smaller than the chemical potential, so that the chemical potential can be treated in the local density approximation. This corresponds to a situation in which the de Broglie wavelength of particles, or healing length for superfluids, is much smaller than the length scales of the potential variations \cite{Giorgini:2008ab}. To study quantum gases in low dimensions atomic ensembles were strongly confined along one or two directions, such that the respective trapping frequencies exceeded the chemical potential and temperature. The properties of these low dimensional gases were indeed found to strongly deviate from their three dimensional counterparts  \cite{Petrov:2004ab,Bloch:2008ab}. 

%As far as transport is concerned, the regime where confinement in the channel is large enough that a single vibrational state is accessible is of fundamental importance, both as a conceptually simple case, but also because of the experimental relevance to many solid state physics experiments, in particular the quantum Hall effect where the one dimensional character arises from the edge state \cite{Ihn:2010aa}. 
As far as transport is concerned, the regime, where confinement in the channel is large enough to separate transverse vibrational states on the relevant energy scale, is of fundamental importance, both as a conceptually simple case, and also because of phenomena observed in various solid-state physics experiments. Prominent examples are the quantum Hall effect, where the one dimensional character arises from the edge state \cite{Ihn:2010aa}, and the quantized conductance observed in quantum point contacts. 
However, the concept of two-terminal transport still requires reservoirs, that in most cases are two or three dimensional\footnote{While this is de-facto the case for condensed matter systems connected to leads, cold gases could in principle realize purely one dimensional reservoirs without connection to higher dimensional systems. This case would deserve a particular treatment as most of the properties expected from reservoirs in the Landauer formulation are questionable in one dimension, in particular the ability to efficiently reach thermal equilibrium \cite{Kinoshita:2006aa,Gring:2012aa}. }. The investigation of transport in low dimensional conductors leads thus naturally to the realization of a {\it hybrid} dimensional system, with different parts of the system having different dimensionality. Such systems had to our knowledge not been considered in cold atoms. More generally, this provides an example of the type of physics that emerges from a device approach to quantum gases, where interfaces are a central topic. The particular case of the dimensional crossover at the reservoir-conductor interface in cold atoms has been very recently theoretically analyzed \cite{PhysRevLett.117.255302}.

In this section we introduce the Landauer theory of conductance quantization in tightly confined conductors, with an emphasis on the physical hypothesis and their relation to cold atoms experiments, and derive the quantization of conductance. We then describe in details the experimental observation of this phenomenon with cold gases and the perspectives for future studies. 

\subsection{Landauer formula}
\begin{figure}[htbp]
\begin{center}
\includegraphics[width=0.65\textwidth]{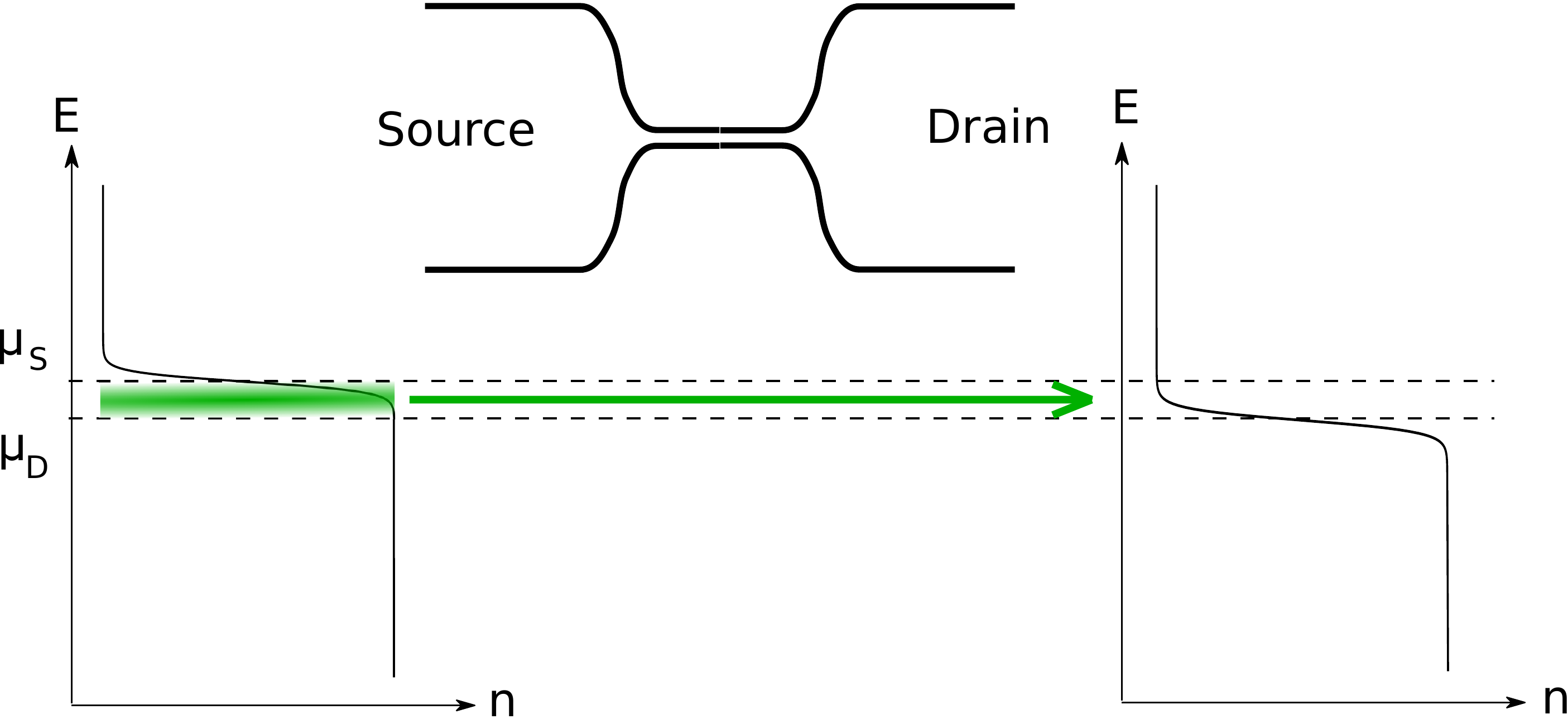}
\caption{Two-terminal Landauer configuration. Source and drain reservoirs are contacted by a ballistic one dimensional channel. A chemical potential difference is introduced between them, driving a net current of particles is driven from the source, as a result of the excess of particles above the Fermi level of the drain.}
\label{figLandauerSetup}
\end{center}
\end{figure}

%This describes the physical content of the Landauer formula. This is mainly directed at cold atoms physicists who are usually not familiar with this. Relations with laser spectroscopy and atom lasers are discussed. 
%The Landauer formula for the two terminal conductance of a ballistic system is derived in many standard textbooks and reviews of mesoscopic physics \cite{datta_electronic_1995,Ihn:2010aa,Imry:1999aa}. 
The derivation of the Landauer formula, as presented in many standard textbooks and reviews of mesoscopic physics \cite{datta_electronic_1995,Ihn:2010aa,Imry:1999aa}, can directly be adapted to atomic systems. 
In short, consider the situation depicted in figure \ref{figLandauerSetup}. A one dimensional system is connected smoothly to large particle reservoirs described by the Fermi-Dirac distributions $f(\mu_i,T)$, where $\mu_i$ is the chemical potential of the reservoir $i$, $i=\mathrm{L,R}$ and $T$ is temperature. 

Because the channel is ballistic, atoms traveling through it with positive (resp. negative) momentum follow the energy distribution in the left (resp. right) reservoir. The total current inside the channel thus reads
\begin{equation}
I = \int d\epsilon v(\epsilon) g(\epsilon) \left( f(\epsilon;\mu_L,T) - f(\epsilon;\mu_R,T) \right)
\end{equation}
with $v(\epsilon)$ being the group velocity and $g(\epsilon)$ the density of states. In one dimension, $g(\epsilon) = 1/hv(\epsilon)$ and the two terms cancel, leaving
\begin{equation}
I = \frac 1 h \int \left(f(\epsilon;\mu_L,T) - f(\epsilon;\mu_R,T) \right)d\epsilon
\end{equation}
where $h$ is Planck's constant. The formula reduces to $I = \Delta \mu /h$ with $\Delta \mu = \mu_L - \mu_R$ in the low temperature, linear response limit. Thus, the conductance is equal to one over Planck's constant. 

It may seem at first sight that the exact cancellation of the group velocity with the density of states in one dimension is a fortuitous effect that could be disturbed by many experimental imperfections. This is not the case, as can be seen from the following argument \cite{Martin:1992aa,Batra:1998aa}. Consider the Fermi distributions in the two reservoirs, as depicted in figure \ref{figLandauerSetup}. As mentioned in the first section, at low temperature the net current originates from the particles populating the energy states between $\mu_L$ and $\mu_R$. Let us operate a unitary change of basis for these states and consider their representation in the time domain. Fourier transforming the square window in energy leads to a discrete set of wavelets with a $sinc$ shape, and separated in time by $\Delta t=h/(\mu_L - \mu_R)$. This is nothing but a statement of the Heisenberg principle. Consider now the zero temperature limit, where all these energy states in the left reservoirs are populated by one and only one particle, as required by Fermi-Dirac statistics. Then after the unitary transformation to the time domain, each wavelet carries one and only one particles. Counting now the number of particles per unit time gives rise to the net current and we directly obtain the Landauer result $I=1/\Delta t=(\mu_L - \mu_R)/h$, the Landauer result. 

The case where several modes can independently contribute to transport is a straightforward extension, where a sum over the transverse modes has to be added to the reasoning. Each mode contributes to transport by the same amount, provided it is energetically accessible. As a result the expected conductance at low temperature is $N / h$, where $N=\sum_n \theta( \mu - E_n )$ is the number of transverse modes energetically accessible: $\mu$ is the average chemical potential of the reservoirs, $\theta$ is the step function and $E_n$ is the energy of the mode $n$, obtained by solving Schr\"odinger's equation in the transverse directions in the channel. 

In the presence of an obstacle in the channel capable of reflecting particles, the formula for the current includes the energy dependent, intensity transmission coefficient $t(\epsilon)$:
\begin{equation}
I = \frac 1 h \int t(\epsilon) \left(  f(\epsilon;\mu_L,T) - f(\epsilon;\mu_R,T) \right) d\epsilon
\end{equation}

For charge currents measured in Ampere and bias measured in Volt, an overall charge factor $e$ in front of the current appears and the bias reads $\Delta \mu = -e V$, where V is the voltage. This represents merely a choice of units (Voltage for the bias and Amperes for the current), which yields a conductance quantum of $e^2/h$. A factor of two for spin degeneracy is usually also included, since conductance measurements in condensed matter physics are not spin resolved. In contrast, measurements based on optical imaging with atomic gases are spin-resolved, thus all currents are to be understood per spin states and the factor for spin degeneracy is left out. 

\subsection{Reservoirs}
As already mentioned earlier, a crucial advantage of the Landauer approach is that it separates the channel, which is a quantum coherent and elastic system, from the reservoirs where all the inelastic processes are taking place, without the need to describe these precisely. Cold atomic gases offer a situation where the physics of the reservoirs can be investigated directly, and their properties can be tuned. The validity of the Landauer approach as a function of the size of the reservoirs has been investigated theoretically and deviations at short time scales were predicted, that could be tested in the future \cite{Chien:2014aa,Chien:2015ab}. In particular, the dynamics associated with the onset of a quasi-steady state where the transport is described by the Landauer is expected to be observable in future experiments. 

\subsection{Experimental observation in cold atoms}

\begin{figure}
\includegraphics[width=0.25\textwidth]{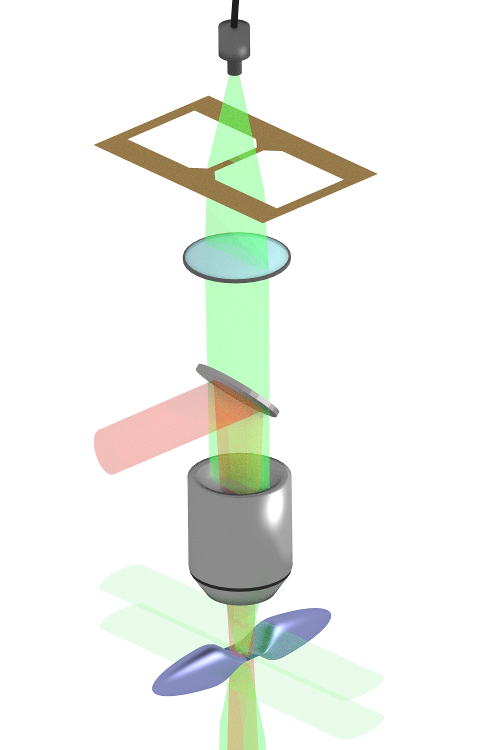}
\includegraphics[width=0.45\textwidth]{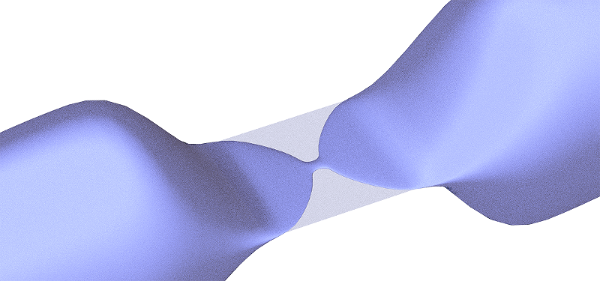}
\centering
\caption{%
{\bf Quantum point contact}. Left: schematic of the experimental setup. A binary mask is imaged onto the channel using a microscope objective. Right: artist's view of the contact region connecting the two reservoirs. Figure adapted from \cite{Krinner:2015aa}.}
\label{fig:QPCSetup}
\end{figure}

We now come to the experimental observations reported in \cite{Krinner:2015aa} using the two terminal configuration. There, a binary mask featuring a narrow opaque line was imaged onto the two dimensional channel through a high resolution microscope objective, using blue-detuned light, generating a repulsive potential. A schematic view of the setup and the resulting structure is presented in figure \ref{fig:QPCSetup}. This produces a split gate structure similar to that used in mesoscopic electronics. 

Along the transverse direction ($x$), the high resolution of the microscope allows for very tight confinements, with typical transverse sizes of $1.5\,\mu$m. With the laser power available, the oscillation frequency reaches up to $50\,$kHz at the center of the structure. Along the direction of transport ($y$), the envelope of the structure is Gaussian, corresponding to the shape of the laser beam used for the imaging. The confinement therefore smoothly evolves from very weak in the wings of the beam, i.e. at the entrance and exit of the QPC, to very strong at the center. Geometrically, this length scale is determined by the beam waist which is $5.5\,\mu$m in this experiment. 

Additionally, a red detuned laser beam with a larger gaussian envelope is superimposed with the structure. This beam attracts atoms towards the center of the QPC, hence controlling the density in the QPC and its entrance and exit regions. The potential created by this beam is called gate potential, by analogy with top gates in semiconductor structures and its strength at the center of the QPC is denoted $V_g$. This beam only affects the channel, and unless its power is very large it does not deform the reservoirs significantly.

\begin{figure}
\centering
\includegraphics[width=0.6\textwidth]{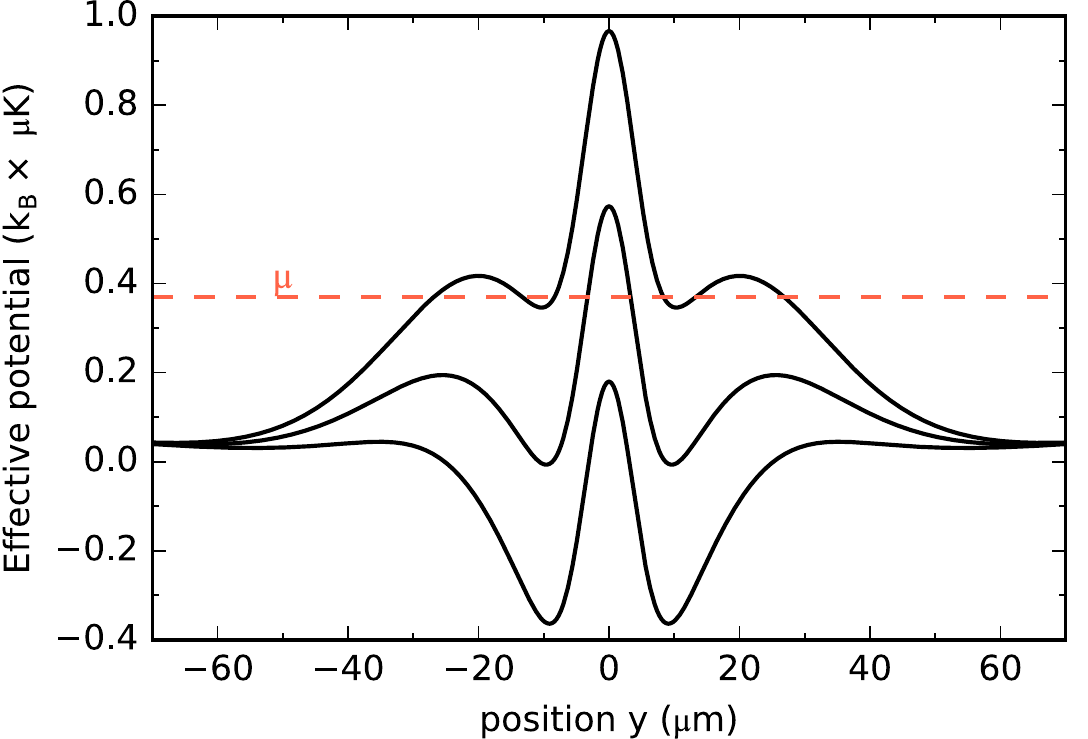}
\caption{%
{\bf Semi classical potential landscape}, for the three lowest transverse modes with a gate potential of $0.78\,\mu$K. Here only the lowest transverse mode is below the chemical potential (dashed orange line).}
\label{fig:potentialLandscape}
\end{figure}

The combination of laser beams creating the potential landscape is quite complex, but allows to fulfill the apparently contradictory constraints for the observation of quantized conductance: (i) tight confinement, with frequencies much larger than temperature, (ii) a high degree of quantum degeneracy and low temperatures and (iii) a controllable chemical potential, independent of temperature. Criteria (i) is met through the use of tightly focused beams for the lateral confinements, and criteria (ii) is met through evaporative cooling and adiabatic decompression of the trap, yielding temperatures and chemical potentials lower than the transverse mode spacing: the QPC is empty due to the zero point motion. Criteria (iii) is met using the gate potential, that forces atoms in the QPC region, while keeping the temperature imposed by the reservoirs unchanged.

In order to understand transport through the structure, consider a semi classical picture where motion along the transverse direction is quantized, leading to a longitudinal potential landscape for each transverse mode where atoms propagate quasi-classically. Such a quasi-classical potential landscape is presented in figure \ref{fig:potentialLandscape}. This results in a one-dimensional problem where the energy resulting from the transverse confinement $E_\perp(y)=\frac{1}{2}h\nu_xf_x(y)+\frac{1}{2}h\nu_zf_z(y)$ acts as an additional potential, with $f_{x,z}(y)$ describing the spatial variation of the trapping frequencies of the QPC. Further contributions arise from the spatial profile of the gate potential $V_g(y)=-V_g f_g(y)$, and the underlying harmonic trapping potential extending over the entire cloud $V_{\textup{trap}}(y)=\frac{1}{2}m\omega_y^2 y^2$.

\begin{equation}
V_{\textup{eff}}= E_\perp + V_g+ V_{\textup{trap}} + E_{\textup{resid}}.
\end{equation}

Extra contributions resulting form experimental imperfections, such as the finite contrast of the projected structures can also be included and calibrated by independent transport measurements. The envelope functions are listed in Table \ref{envelopefunctions}.

\begin{table}[htp]
\centering
\caption{Envelope functions determining the effective potential.}
\begin{tabular}{ccc}
Envelope function & Waist & Description\\
\hline
$f_x(y) = \mathrm{exp}(-y^2/w_x^2)$ & $w_x=5.6(6)\,\mu$m & QPC, $x$ conf. \\
$f_z(y) = \mathrm{exp}(-y^2/w_z^2)$ & $w_z=30(1)\,\mu$m & QPC, $z$ conf. \\
$f_g(y) = \mathrm{exp}(-2y^2/w_g^2)$ & $w_g=25(1)\,\mu$m & Gate potential \\
\hline
\end{tabular}
\label{envelopefunctions}
\end{table}

The finite length of the QPC is controlled by the curvature of the semi classical potential around the center of the contact, modeled as an inverse harmonic oscillator. The typical parameters of the QPC are presented in Table \ref{tab:QPC}. Importantly, temperature is typically an order of magnitude lower than the lowest trap frequency at the center of the QPC, allowing for the resolution of individual transverse modes in the transport experiment.

\begin{table}[h]
\centering
\begin{tabular}{ccc}
\hline\hline
Parameter & Typical value & Description\\
\hline
$\omega_x$ & $2\pi\times30\textup{kHz}$ & Horizontal QPC confinement\\
$\omega_z$ & $2\pi\times10\textup{kHz}$ & Vertical QPC confinement\\
$\Omega_{y,n_x=0}$ & $2\pi\times1.3\textup{kHz}$ & QPC curvature arising from split gate\\
$\Omega_{y,n_z=0}$ & $2\pi\times0.1\textup{kHz}$ & QPC curvature arising from 2D confinement\\
$E_{\textup{F}}/\hbar$ & $2\pi\times8.0\textup{kHz}$ & Fermi energy\\
$\mu/\hbar$ & $2\pi\times7.7\textup{kHz}$ & Chemical potential\\
$\Delta\mu/\hbar$ & $2\pi\times2.0\textup{kHz}$ & Chemical potential bias\\
$\textup{k}_{\textup{B}}T/\hbar$ & $2\pi\times0.9\textup{kHz}$ & Temperature\\
\hline\hline
\end{tabular}
\caption{Energy scales of interest for transport through our QPC.}
\label{tab:QPC}
\end{table}

\begin{figure}
\includegraphics[width=0.65\textwidth]{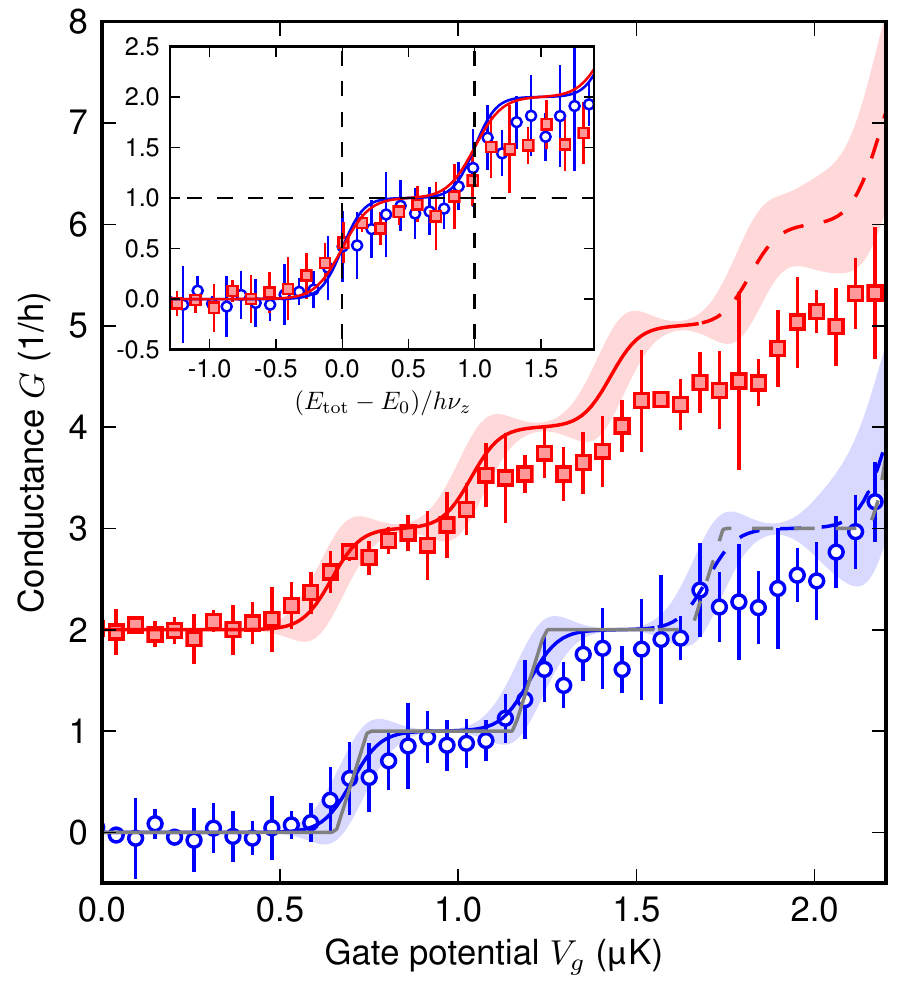}
\centering
\caption{%
{\bf Conductance as a function of gate potential}. Open blue circles correspond to a vertical confinement of $\nu_z=10.4\,\rm{kHz}$. Filled red squares correspond to $\nu_z=8.2\,\rm{kHz}$ and are vertically shifted by two units for clarity. Each data point represents the mean of six measurements and error bars indicate one standard deviation. Solid lines are theoretical predictions based on the Landauer formula of conductance. The shaded regions reflect the uncertainties in the input parameters (see text). Dashed lines are continuations of the solid lines and correspond to a change in the effective potential. Inset: first conductance plateau as a function of reduced energy, showing universal scaling. Vertical dashed lines indicate the width of the first plateau, whereas the horizontal dashed line indicates the universal conductance value $1/h$. Extracted from \cite{Krinner:2015aa}.}
\label{fig:gateScan}
\end{figure}

\begin{figure}
\centering
\includegraphics[width=0.65\textwidth]{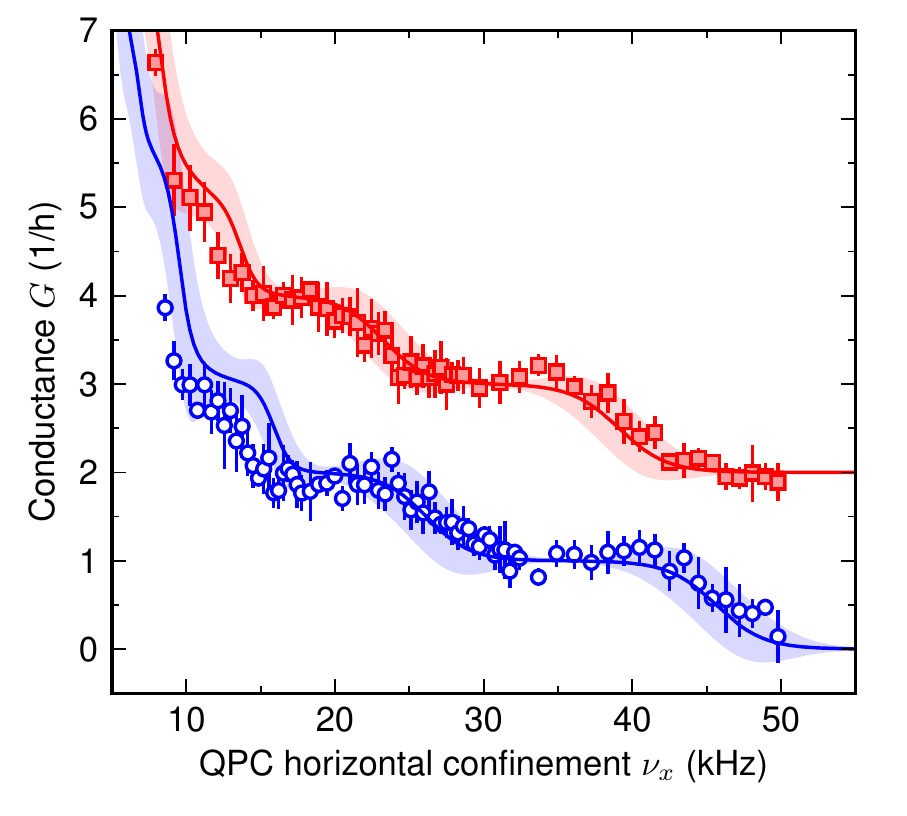}
\caption{%
{\bf Conductance as a function of horizontal confinement.} Open blue circles correspond to a vertical confinement of $\nu_z=10.9\,\rm{kHz}$ and a gate potential of $V_g=1.0(1)\,\rm{\mu K}$. Filled red squares correspond to $\nu_z=9.2\,\rm{kHz}$ and $V_g=0.8(1)\,\rm{\mu K}$, and are vertically shifted by two units for clarity. Solid lines are theoretical predictions based on the Landauer formula of conductance. The shaded regions reflect the uncertainties in the input parameters (see text). Error bars are the same as in Fig. \ref{fig:gateScan}. Extracted from \cite{Krinner:2015aa}.}
\label{fig:confinementScan}
\end{figure}

The experimental measurement of the conductance of the QPC as a function of gate potential and horizontal confinement are presented in figures \ref{fig:gateScan} and \ref{fig:confinementScan} respectively. For the largest confinement at fixed gate potential or the lowest gate potential at fixed confinement, the chemical potential in the channel is lower than the energy of the lowest mode of the QPC, hence the QPC is empty and the conductance is zero. As the gate potential is increased, or confinement decreased, the conductance increases up to $1/h$ and saturates, forming a conductance plateau, until the second mode becomes available leading to another plateau of conductance. 

These observations are compared with the predictions of the Landauer formula accounting for the semi classical potential landscape, finite temperature and the finite bias, as shown with the solid lines in figures \ref{fig:gateScan} and \ref{fig:confinementScan}. The agreement is very good, in particular the width and height of the plateaus agree very well, as well as the width of the transition region between plateaus which is well accounted for by temperature. For the largest gate potentials, the deformations of the potential landscape are significant and transverse modes do not open at the center of the structure anymore but rather at the entrance and exit, leading to strong deviations compared to the calculations. The quantization is robust, and can be observed for reduced confinements along the vertical direction, as shown in red in the figures. While the data are compared here to independently measured quantities such as the trap frequencies, these measurements being absolute provide a useful calibration for the geometry of the QPC and is used as such for the most recent experiments.

Compared to its condensed matter counterpart, the quantum point contact for atoms is rather long. The geometric length ($1/e^2$ radius of the beam) is about $5.6\,\mu$m, compared to the length scale associated with tunneling through the structure $\sqrt{\hbar/m \Omega} \sim 1.1\,\mu$m. Hence, the width of the transitions between successive plateaus is entirely attributed to finite temperature, rather than tunneling as is the case for mesoscopic structures in electronic systems.

Another important difference is the nature of the reservoirs. These are strictly isolated, and actually no dissipative process is taking place inside them: the potential is strictly conservative, spontaneous emission from the dipole trap is negligible as well as atom losses. Inter particle scattering allows for the thermalisation of the incoming particles, but the scattering mean free path for collisions in the reservoirs is about $12\,$mm, much larger than the reservoirs size. Thus energy relaxation is entirely non local. The effects of collisions in the reservoirs were further investigated by reducing the scattering length in order to entirely suppress energy relaxation on the time scale of transport. The conductance measured with the very same protocol was found to be identical within experimental uncertainties \cite{Krinner:2015aa}. This counterintuitive result can be explained by the fact that even though particles incident from the channel into the reservoirs do not release their energy, the three dimensional, non-harmonic shape of the reservoirs makes them most likely chaotic. Thus the probability that the incident atom can come back to the channel and reduce the measured current is very small, even in the absence of collisions. 

\subsection{Quantum interferences in transport}

Even though the conductance is quantized in units of $1/h$, the motion of atoms along the transport direction remains semi-classical. In the presence of more complex structures in the channel, such as fine-grained disorder or any set of partially transmitting obstacles, transport is strongly modified by quantum interferences. The description of these phenomena in the condensed matter context is a major topic of research and goes far beyond the present paper. We refer the readers to standard textbooks for the description of these effects in condensed matter physics \cite{Ihn:2010aa}. 

In the context of cold atoms, theoretical studies have been conducted considering canonical examples of one dimensional Hubbard chains or single state quantum dots weakly coupled to reservoirs \cite{Bruderer:2012aa,Nietner:2014aa}. In these cases the genuinely quantum properties of transport could be predicted, including the transient establishment of the steady state or the current noise. The effects of fixed atom number or finite extensions of the reservoirs could be modeled, and shown to influence many transport properties including the noise spectrum of the current or the thermoelectric response.

\subsection{Role of quantum statistics}

In the derivation of the Landauer formula, the role of quantum statistics in the conductance quantisation appears in setting the occupation of the energetically available modes. This raises the question of the observations of quantized conductance for cold Bosonic atoms. The fact that the group velocity and the density of states cancel in the derivation of the Landauer formula is a single particle property. A first proposal was actually written for the transmission of a beam of cold atoms through a tight magnetic waveguide \cite{Thywissen:1999aa} regardless of the statistics of the atoms.

\begin{figure}
\centering
\includegraphics[width=0.65\textwidth]{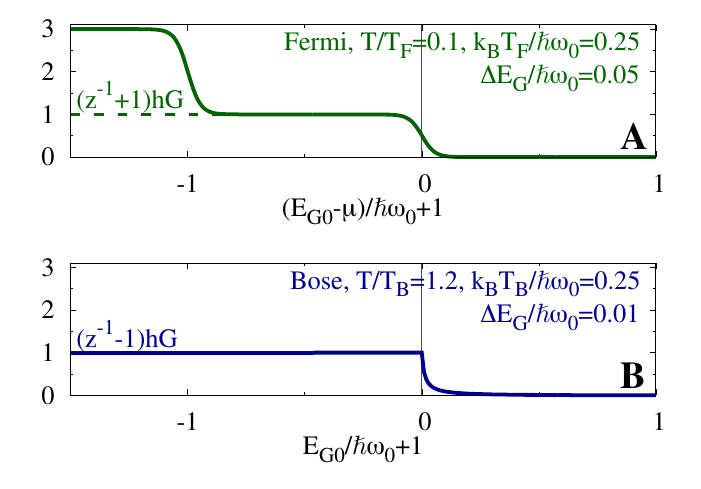}
\caption{%
{\bf Theoretical comparison of quantized conductance for Fermions (A) and non-condensed Bosons (B)}. Conductance is presented in units of $1/h$ as a function of gate potential. $\omega_0$ is the transverse trap frequency, $\Delta E_G$ denotes the width of the transition region and $T_B$ the critical temperature for Bose condensation in the reservoirs. Extracted from \cite{PhysRevA.94.023622}. }
\label{fig:boseVSfermi}
\end{figure}

In the case of actual particle reservoirs containing bosons at thermal equilibrium, the situation is more complex since at low temperature Bose Einstein condensation renders the reservoirs phase coherent and superfluid (see section \ref{section:BEC}). The situation of reservoirs above the critical temperature was analyzed in details in \cite{PhysRevA.94.023622}. Because of Bosonic stimulation for the low energy states in the channel, the height of the conductance plateau is strongly enhanced even with non-condensed reservoirs, up to a non-universal, temperature dependant value. The shape of the transition from insulating to conducting as the first channel opens is also very different for Bosons, again due to Bosonic stimulation effects leading to sharper transitions. The predictions are presented in figure \ref{fig:boseVSfermi}.

\subsection{Other neutral particles}

The first generalisation of the concepts underlying the Landauer approach to neutral particles was the case of photons. In this situation, there is no counterpart to chemical potential and thus no chemical potential bias as the number of photons is not conserved. Nevertheless the transmission probability through narrow apertures also increases stepwise as the aperture is opened, as was observed in \cite{Montie:1991aa}. The crucial ingredient in observing this effect is to mimic incoherent reservoirs which populate every energetically accessible modes while keeping monochromatic light to mimic the low temperature case, and symmetrically detectors should integrate over all possible momenta for the photons emerging from the channel. This was achieved using diffuse illumination and an integrating sphere for the detection. Jumps in the transmission occur with a spacing of $\lambda/2$. No quantum effects are involved in the experiment, highlighting the fundamental role of the wave character of matter for massive particles.

The case of phonons in solid state systems is conceptually similar, but of much more fundamental interest due to its connection with heat transport. For dielectric narrow wires where heat is only carried by phonons the Landauer approach yields a universal heat conductance quantum equal to $k_B^2 \pi^2 T/3h$, where $T$ is the average temperature of the system, in the low temperature limit \cite{Rego:1998aa}. Experiments conducted on suspended nano-structures have shown saturations of the heat conductance at low temperature compatible with this bound \cite{Schwab:2000aa}. Quite remarkably, the quantum of heat conductance is actually a universal bound, not restricted to phonon transmission . It is rooted in Shannon's theorem stating the upper bound to the amount of information that a single channel can carry \cite{Pendry:1983aa}. Recently, a new set of experiments have measured the quantum of heat conductance associated with electrons in a mesoscopic structure in the quantum Hall regime, showing good agreement with the universal limit \cite{Jezouin601}.

Quantized transport measurements were also proposed for dilute mixtures of $^3$He in $^4$He. For very dilute mixtures at the lowest temperatures, $^3$He behaves like an ideal Fermi gas with a very long mean free path of several tens of micrometers. An array of nano-pores was then proposed as a connection between two tanks of Helium, with holes of the order of 10\,nm, compatible with the Fermi wavelength of $^3$He \cite{Sato:2005aa,Lambert:2008aa}.

\section{Superfluid transport in Bose-Einstein condensates}
\label{section:BEC}

The results that were presented in the previous sections concerned weakly interacting Fermi gases, and the physics was that of ideal gases. In the presence of interactions between particles, the gases can turn superfluid in part if not all the system. This is the case for low temperature Bosons, and low temperature Fermions in the presence of attractive interactions. Both cases are realised in cold atoms experiments, and transport in the two terminal setup has been studied under these conditions. 

In contrast to normal currents, which are driven by chemical potential gradients, superfluid currents are driven by gradients in the phase $\phi$ of the superfluid order parameter $\psi = |\psi| e^{i\phi}$. More precisely, superfluid velocity is given by $v_s = \frac \hbar m \nabla \phi$.  Most of the dynamical properties of superfluids can be understood based on this principle contained in the Josephson-Anderson equation \cite{Josephson:1962aa,ANDERSON:1966aa,Packard:1998aa,Leggett:2006aa,RevModPhys.87.803}: 
\begin{equation}
\hbar \frac{d \phi}{dt} = \mu
\end{equation}
where $\mu$ is the chemical potential. In the two terminal configuration, this reduces to the usual Josephson relation, but it has a very general validity for both bosonic and fermionic superfluids, weak and strongly interacting, and explains in particular a wide range of phenomena such as vortex motion and nucleation. 

For a general superfluid system such as Helium, the relation between the superfluid order parameter and the microscopic parameters of the system is a very involved problem, since densities are large and interactions are strong \cite{Leggett:2006aa}. In contrast, cold atoms are often weakly interacting systems for which a mean field theory based on the microscopic Hamiltonian provides an accurate description. The Gross-Pitaevskii equation for Bose-Einstein condensates identifies the macroscopically populated single particle wave function with the superfluid order parameter, allowing for simple and transparent derivation of the Anderson equation from the microscopic model \cite{Dalfovo:1999aa,Leggett:2006aa}. In the case of attractively interacting Fermions, the BCS mean-field model provides a similarly simple description, identifying the pairing gap with the amplitude of the order parameter \cite{Leggett:2006aa}. However, in contrast to the Gross Pitaevskii equation for Bosons, this description of Fermionic superfluids is usually only qualitatively correct for dilute atomic gases, as the strength of interactions necessary to produce superfluid Fermi gases at the temperatures achievable in the experiments is usually too strong for mean field theory to be quantitatively correct. Nevertheless, in both cases, most of the phenomenology can be captured by the Anderson equation, which provides a unified picture of superfluidity. 

Cold atoms experiments have provided a wide playground for superfluid transport measurements where the phase coherence plays a key role \cite{Madison:2000aa,Zwierlein:2005ab}. Like in the previous section, coupling a device with universal transport properties to a reservoir allows for driving the system and measuring the transport coefficients. A general feature throughout these experiments is also the combination of the universal Josephson-Anderson equation, with a reservoir dynamics that depends on the geometry and the details of the setup. 
%Cold atoms experiments have provided a wide playground for superfluid transport measurements where the phase coherence plays a key role. Like in the previous section where the device having universal transport properties was coupled to a reservoir allowing for driving and measurement, a general feature throughout these experiments is also the combination of the universal Josephson-Anderson equation, with a reservoir dynamics that depends on the geometry and the details of the setup. 

\subsection{Superfluid flow}
The question of the superfluid nature of Bose-Einstein condensates has been thoroughly investigated in experiments over decades following the realization of Bose-Einstein condensation in dilute gases. Contrary to long range order, which is directly probed at equilibrium using time-of-flight or more elaborated interferometric techniques, superfluidity is a dynamic property that requires the system to be set in motion. The existence of a critical velocity for the creation of excitations in a cloud is usually regarded as demonstrating superfluidity. Such experiments have been realized by steering an obstacle inside a condensate and probing the heating rates \cite{Onofrio:2000aa,Desbuquois:2012cr,PhysRevLett.114.095301}, or imprinting a moving optical lattice and measuring heating as a function of velocity \cite{Miller:2007aa}.

\begin{figure}
\centering
\includegraphics[width=0.65\textwidth]{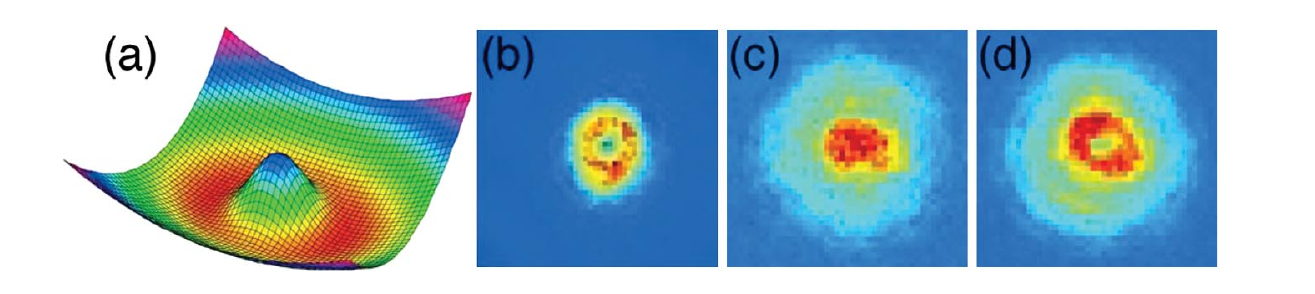}
\caption{%
{\bf Bose-Einstein condensate in a ring trap}. (a) potential landscape produced by a combination of magnetic trap and repulsive optical plug at the center. (b) In-situ absorption image of the condensate in the ring trap. (c) and (d) time of light images of the condensate released from the ring trap, in the absence (c) and presence (d) of rotation. Figure extracted from \cite{Ryu:2007aa}.}
\label{fig:ring}
\end{figure}

A set of experiments have been performed using connected geometries, created by placing a Bose-Einstein condensate in a ring shaped trap \cite{Ryu:2007aa,Ramanathan:2011aa,Beattie:2013aa,Ryu:2013aa}. An example of such experiment is presented in figure \ref{fig:ring}, extracted from \cite{Ryu:2007aa}. Such a configuration realises a minimal version of a circuit for atoms, and has a lot of similarities with superconducting quantum interference devices \cite{:/content/aip/journal/rsi/77/10/10.1063/1.2354545} and superfluid helium gyroscopes \cite{Packard:1998aa,RevModPhys.87.803}. 

The dynamics in such systems is probed by injecting angular momentum in the system. This can be achieved using Raman transitions, by which a cloud initially at rest in one internal state is coherently transferred into a different internal state, using a pair of laser beams with a frequency difference matching the energy difference between internal states \cite{Ryu:2007aa}. During this process the relative phase of the two laser beams is imprinted onto the atoms. Choosing for one of the laser beams a Laguerre-Gaussian profile, a phase circulation is coherently imprinted, effectively transferring one quanta of angular momentum from the beam to each atom. The persistence of the rotation is a direct consequence of the existence of the macroscopic order parameter $\Psi$ whose phase gradient gives the superfluid current. Indeed, for $\Psi$ to be single valued, the number of phase windings around the ring can only vary by integer units. Any unwinding event has to occur via the creation of vortices within the ring, which represents an energy cost whenever $|\Psi|$ takes a macroscopic value. This can be seen as a macroscopic free energy barrier that prevents the relaxation of the current to the ground state. 

Currents in the ring are inferred from time-of-flight measurements: the trap is turned off, allowing the ring to be filled due to expansion of the condensate in the radial direction. In the presence of a net circulation in the ring, corresponding to a phase winding of $2\pi$ around the ring, the hole is seen to persist after expansion. This can be understood by considering that atoms emerging from opposite sides of the ring interfere destructively at the center, leaving a hole in the far field. Equivalently, one can  considering that the circulation in the ring is equivalent to a vortex trapped in the zero density region inside the ring. Upon changing the trap configuration, this vortex will remain at the center of the cloud and be revealed upon increasing the density. 

The stability of the persistent currents depends in practice on many parameters of the experiments, such as the smoothness and stability of the potentials used to trap the atoms. It was observed that currents injected in these rings persisted for several tens of seconds or even minutes . Remarkably, the lifetime of these currents was longer than the lifetime of the cold gas itself, such that even condensates having lost a large fraction of their initial atom number would maintain their rotation, a striking demonstration of persistent currents \cite{Beattie:2013aa}.

The decay of super currents was investigated thoroughly in the presence of a controlled weak link \cite{Ramanathan:2011aa}. An elliptic laser beam, producing a repulsive barrier, was focused onto a small portion of the ring trap. The length of the weak link was larger than the healing length of the condensate, such that its main effect was to reduce the local density, hence the speed of sound and the condensate fraction. It was observed that the current was stable until the height of barrier reached a certain fraction of the chemical potential of the condensate. This constituted a direct evidence for the existence of a critical velocity. However, the critical velocity was different from the speed of sound, as predicted by the Landau theory, showing that the currents where not decaying via the emission of single particle excitations. The creation of vortices within the ring was identified as the decaying mechanism, following Feynman's original arguments \cite{Feynman:1955aa}, and confirmed using direct numerical simulations. 

The reverse operation of creating rotation using a rotating barrier was later investigated \cite{PhysRevLett.110.025302}. This is qualitatively equivalent to the decay of rotation up to a change of reference frame from the lab frame to that moving with the barrier. Accordingly, stepwise changes in the winding were observed upon increasing the rotation velocity. This technique however is not restricted to imprinting one quantum of angular momentum but can be used to introduce larger angular momentum in the system. Because of the large free energy barriers that need to be crossed for varying the angular momentum in the ring, the circulation presents hysteresic behavior when trying to increase and decrease the rotation, which was directly demonstrated in \cite{Eckel:2014aa}. This is reminiscent of the persistent character of the currents and represents a hallmark of superfluid circuits. 

\subsection{Two reservoirs experiments with Bose-Einstein condensates}

The first experimental realization of a weak connection between two cold atoms clouds was realised in Heidelberg in 2004 \cite{Albiez:2005aa}. Earlier experiments had been performed by placing a condensate in a one dimensional optical lattice, so as to form a chain of coupled condensates \cite{Cataliotti:2001aa,Anderson1686}. The evolution of phase coherence and number fluctuations in particular was measured in \cite{Orzel:2001aa}, a precursor of the superfluid to Mott insulator transition in 3D optical lattices \cite{Greiner:2002aa}.

\begin{figure}
\centering
\includegraphics[width=0.65\textwidth]{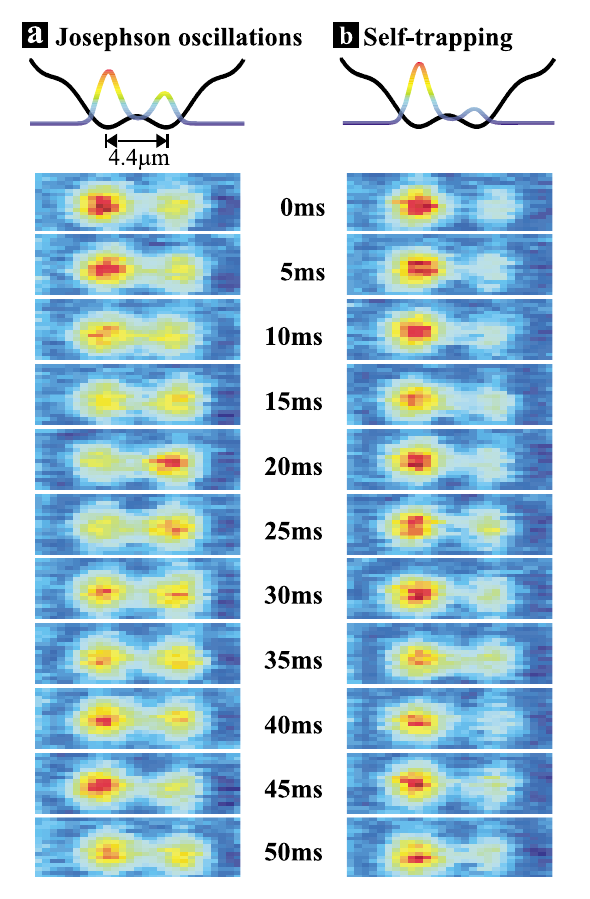}
\caption{%
{\bf Tunneling in Josephson coupled Bose-Einstein condensates}. A Bose-Einstein condensate is placed in a symetric double well potential with an initial population imbalance, ans left to evolve freely, either with a weak initial population imbalance (left) or a strong imbalance (right). The resulting dynamics shows either plasma oscillations or self trapping, respectively. Extracted from \cite{Albiez:2005aa}. }
\label{fig:Oberthaler1}
\end{figure}

In the Heidelberg experiment, a condensate loaded in a small harmonic trap was split in two in a controlled way, allowing for the preparation of initial population imbalances. The key to the dynamics of this system is the phase coherence between the two reservoirs \footnote{Note that this is in direct contradiction with the prescription of the Landauer description, showing the need to go beyond this approach to describe superfluid systems.}. 

The outcomes of such an experiment were first predicted using a two mode model by which the internal dynamics within the two coupled condensates is neglected \cite{Smerzi:1997aa}. The evolution is captured by a two component Gross-Pitaevskii equation, describing the condensate in each well and a coherent tunneling term coupling the condensate wave functions of the two condensates. Introducing canonically conjugated variables for the phase difference between the two condensates $\phi$ and atom number difference between the two wells $z=\frac{N_L-N_R}{N_L+N_R}$, one obtains two coupled non linear equations \cite{Smerzi:1997aa}

\begin{eqnarray}
\dot{z}&=& -\sqrt{1-z^2} \sin(\phi) \\
\dot{\phi} &=& \Lambda z + \Delta E + \frac{z}{\sqrt{1-z^2}} \cos(\phi)
\end{eqnarray}

where $\Delta E$ controls the asymmetry in the trap, and $\Lambda$ controls the interaction strength between the particles, the equivalent of a charging energy for capacitive Josephson junctions. These equations predict a regime of "Josephson" oscillations, where a small initial population imbalance oscillates between the two wells at the plasma frequency $\omega_J = \sqrt{E_C E_J}$ determined by both the coupling between the wells, controlled by the coupling strength $E_J$ and the strength of the repulsive interactions yielding an effective inverse-compressibility measured by $E_C$. The phase difference oscillates in quadrature. These oscillations are reminiscent of the AC Josephson effect in superconductors, but the trapped nature of the system sets them at a non universal frequency. This regime is observed in the experiments, see figure \ref{fig:Oberthaler1}. A direct observation on the Josephson oscillations at a frequency equal to the bias was reported in a later experiment \cite{Levy:2007fk}.

With stronger initial imbalance between the two wells, the equations predict a "self-trapping" regime, where the populations in the two wells show weak oscillations around their initial value, and the phase difference increases linearly. This regime is reached when the initial particle number difference, or more generally the initial internal energy difference between the two wells is large compared to the coupling between the condensates. Except for the weak oscillatory behaviour, the dynamics is that of decoupled condensates with different chemical potentials and follow directly from the Josephson-Anderson equation. This dynamical decoupling was also observed in the experiment, as shown in figure \ref{fig:Oberthaler1}. 

Rather than starting from a two-modes model, adapted to large barriers or weak connection, one can start from a single condensate and study the evolution of the dynamics as a tunnel barrier is raised. Theory and experiments in this regime have been reported in \cite{LeBlanc:2011aa}. In the absence of the tunnel barrier, the cloud performs dipole oscillations. The uniform velocity within the cloud translates to a uniform phase gradient: the phase difference across the cloud oscillates in quadrature with the population. As the barrier is raised, the oscillation frequency decreases and smoothly connects to the plasma frequency in the regime of large barriers. Qualitatively, the phase gradient which was linear in the absence of the barrier steepens in the low density region of the barrier, leading to flatter phase profiles inside the wells and a steep phase drop in the junction. A theoretical model explicitly accounting for the coupling between the oscillating mode and other excitation modes in the condensates was found to successfully reproduce the data. In particular, the beating of several frequencies was observed experimentally in the population dynamics as a result of this coupling.

\begin{figure}
\centering
\includegraphics[width=0.65\textwidth]{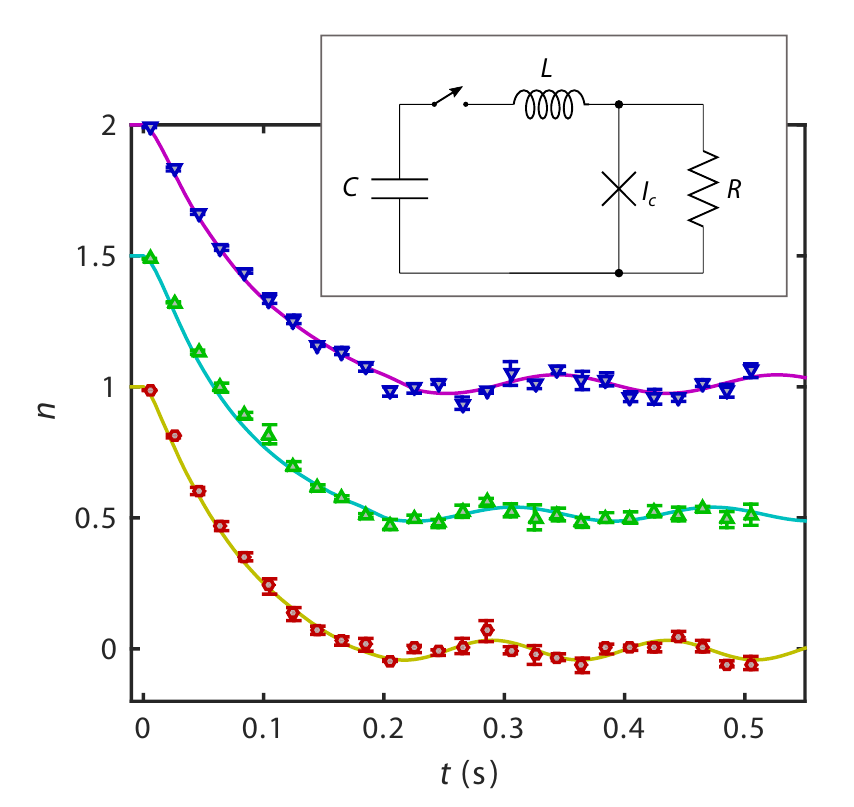}
\caption{%
{\bf Evolution of populations in a strongly imbalanced two-terminal setup with Bose-Einstein condensates connected by a few-modes quantum wire}. After the initial exponential decay, the imbalance oscillates. The three colors represent increasing densities in the channel $599(17)\,\mu$m$^{-1}$ (blue), $665(16)\,\mu$m$^{-1}$ (green), $790(25)\,\mu$m$^{-1}$ (red). The inset presents a lump element model for the dynamics capturing the dissipative, reactive and Josephson dynamics expected in this circuit. Extracted from \cite{Eckel:2016aa}. }
\label{fig:HillDecays}
\end{figure}

A qualitative interpretation for the role of the harmonic trap is a harmonic oscillator, such as an LC circuit, coupled in series to the Josephson junction. In the weak barrier regime, the junction merely renormalizes the resonance frequency by enhancing the inductive effect. In the strong barrier regime with weak population imbalance, the low currents lead to weak effects of the inductance L, such that the dynamics is that of a capacitive Josephson junction with its plasma mode. Eventually, the self trapping regime is reached for stronger imbalances, where the high frequency Josephson oscillations essentially decouple from the harmonic trap: at the time scale of the LC oscillations, the current in the junction averages out to zero and the population imbalance is constant.

Recently, a long quantum wire connecting two Bose-Einstein condensates was realised \cite{Eckel:2016aa} following the experiments with thermal Bosons in \cite{Lee:2013fk}. A very large population imbalance was imposed and the evolution was observed to cross over from an initial exponential decay for large imbalance to an oscillating behaviour, interpreted as LC oscillations of the populations, as shown in figure \ref{fig:HillDecays}. This dynamics was successfully reproduced using an LC circuit coupled to a resistively shunted Josephson junction, where the initial decay of the population imbalance results from the resistive element while the oscillating behaviour represents LC oscillations. Interestingly, the magnitude of the resistance was found to be extremely low, two orders of magnitude lower than expected from the Landauer formula for free particles, highlighting the key role played by superfluidity. The microscopic mechanism for resistive flow was identified to be vortex nucleation at the exit of the channel, and vortex-like excitations were directly identified in the low density reservoir. 

%The case of an array of Bose-Einstein condensate coupled by tunnel junctions is realized in several experiments where atoms are loaded in an optical lattice. The phase coherence featured by low temperature gases ensures that neighboring sites of the lattice are phase-locked. The dynamics of atoms under various external drivings has been investigated in early experiments.
Rather than using two traps as reservoirs, it is possible to locally deplete one well in an optical lattice in order to mimic the drain reservoir, as performed in \cite{Labouvie:2015aa}. There, one site of the lattice was depleted by more than $90\%$, and the subsequent refilling of the site was measured and a direct analogy with a two-reservoirs configuration was drawn. An interesting feature of this experiment is the existence of a regime of large chemical potential bias where the current decreases with increasing bias, hence denoting negative differential conductivity. This phenomenon was attributed to a density-dependent tunneling: the transverse profile of the condensate depends on its density, so do matrix elements for hoping between filled and empty sites. %(TO BE DISCUSSED) This illustrates again the rich interplay between the universal transport law given by the Josephson-Anderson equation and the reservoir properties that can be tuned experimentally.

\section{Two-terminals transport in superfluid Fermi gases}

Using the control offered by Feshbach resonances (see section \ref{section:RCmodel}), it is possible to create attractive interactions between Fermions in different spin (hyperfine) states strong enough that Cooper pairing and superfluidity appears. A Fermi gas in the vicinity of a broad Feshbach resonance realizes the crossover from Bose-Einstein condensate of molecules to BCS superfluidity, as the magnetic field is varied across the resonance. Across the resonance, the system crosses the strongly interacting regime where the scattering length is larger than inter-particle spacing. Superfluidity was demonstrated at low temperature in the entire crossover, and the thermodynamic properties of the system are now experimentally known with great accuracy (see \cite{Zwerger:2011aa} for a recent review). The body of knowledge gathered on such a strongly correlated superfluid is indeed used as a benchmark for numerical techniques aimed at treating strongly correlated Fermions. 

Superfluids are ideal test beds for transport measurements, since variations of conductances upon crossing the transition are expected to be very large. Since superfluidity mainly occurs on the Fermi surface, transport directly adresses the fraction of atoms which experience strong quantum correlations in the superfluid phase. This contrasts with density distribution observed with absorption imaging, which observes the entire Fermi gas, where most of the signal comes from atoms deep in the Fermi see where no correlations appear at the phase transition \footnote{In the case of spin imbalanced Fermi gases, a first order transition between paired and polarized superfluid takes place, and the density profile acquires a distinctive kink that was used to identify the superfluid transition \cite{Partridge:2006fk,Zwierlein:2006fk}}. High precision measurements of the density profiles can be used to reveal a jump of the compressibility of the gas characteristic of the superfluid transition \cite{Ku:2012aa}. In the case of BCS superfluids, well known in condensed matter physics, the actual equation of state is actually unchanged, hence no density variations are expected, even though the resistance is reduced to zero. This was a major motivation for the investigation of transport in the two terminal configuration in strongly attractive Fermi gases.

\subsection{Superfluid Fermions in the multimode regime}
\label{section:superflow}
Following the realization of the two terminal system for weakly interacting Fermions, the very same configuration was used to investigate unitary fermionic superfluids produced in the vicinity of the Feshbach resonance \cite{Stadler:2012kx}. A measurement protocol identical to that described in section \ref{section:multimodeWIF} was applied, allowing for a direct comparison between weakly interacting gases and the unitary Fermi gas in the same experimental conditions. 

As a control parameter, a repulsive gate potential was applied on the two dimensional channel, in the form of a Gaussian beam propagating along the vertical direction. This beam has a waist of $18\,\mu$m, covering the center of the transport channel and allowing for a local reduction of the chemical potential (or equivalently the local density). The main advantage of this control parameter is that it only affects the channel and leaves the reservoirs unchanged. Therefore the thermodynamic properties of the reservoirs do not need to be accounted for precisely in order to interpret the evolution of transport. Later experiments with a similar setup have used the detailed knowledge of the equation of state to infer the precise thermodynamics of the reservoirs even in the strongly interacting state. 

\begin{figure}
\centering
\includegraphics[width=0.65\textwidth]{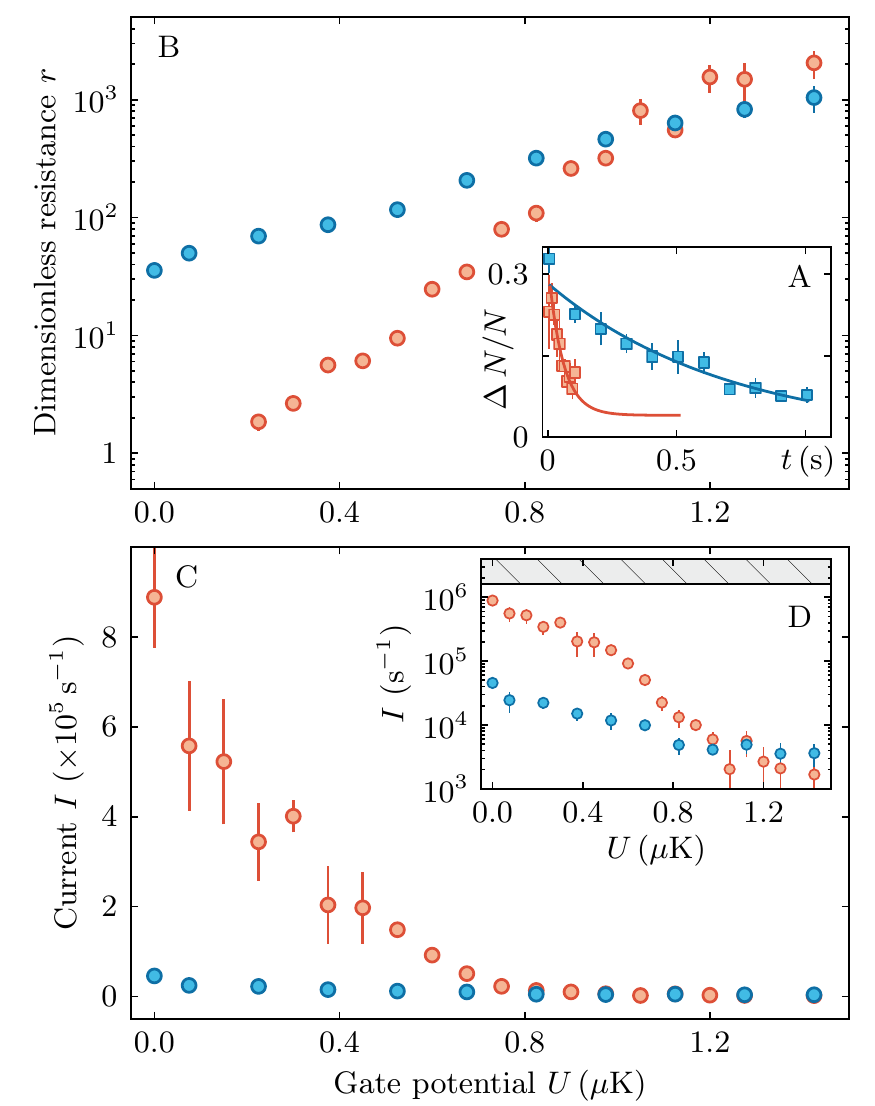}
\caption{%
{\bf Conduction properties through the channel.} Red and blue data points correspond to the strongly and weakly interacting gas, respectively. A: Decay of the relative atom number imbalance between source and drain as a function of time with a gate potential $U = 525(50) \,\mathrm{nK}$. The solid lines are exponential fits with fixed offset of 0.04 for the red curve to account for a small remaining imbalance in the reservoirs. B: Dimensionless resistance $r$ as a function of gate potential. The data points that are shown are those for which the decay is exponential. C: Atom current as a function of the gate potential $U$. A large increase of the current appears for the strongly interacting gas below $U \approx 0.7 \, \mu \mathrm{K}$. D: Atom current in logarithmic scale. The dashed region indicates the maximum current allowed by the internal dynamics of the reservoirs (see text). The error bars show the statistical errors (one standard deviation). Figure extracted from \cite{Stadler:2012kx}.}
\label{fig:Superflow}
\end{figure}

The decay of the initially prepared bias $\Delta N$ was observed for both the unitary Fermi gas and the weakly interacting Fermi gas at low temperature as a function of the gate potential. It was found that this decay follows an exponential behaviour (see figure \ref{fig:Superflow}A), at least down to low values of the gate potential, justifying an $RC$ model description similar to that introduced in section \ref{section:RCmodel}. The decay constant $\tau$ was normalised to the underlying trap frequency $\omega_y$ along the transport direction, measured by dipole oscillations in the absence of any channel or gate potentials at the center, yielding a dimensionless resistance $r=\tau \omega_y$.

The variations of $r$ with the gate potential are presented in figure \ref{fig:Superflow}B. For the largest gate potentials, or lowest densities, $r$ is of the same order for both weakly and strongly interacting systems, but as the gate is reduced, $r$ drops very quickly and reaches values about $50$ times lower than for the weakly interacting system exposed to the same gate potential strength. The lowest measurable resistance is also one order of magnitude lower than that of the weakly interacting gas without any gate potential.

For gate potentials below $0.2\,\mu$K, the decay of $\Delta N$ was found to deviate from exponential behaviour, in the form of residual oscillations of the populations, at frequencies close to the underlying dipole mode along the transport direction. In this regime the assumption of thermal equilibrium of the reservoirs appears no longer to be valid. For such low gate potentials, an observable independent of the reservoirs equilibration is the current through the channel observed during the early stage of the decay. This initial current keeps increasing down to zero gate potential, as can be seen in figure \ref{fig:Superflow}C. Interestingly, the order of magnitude of the current is similar to that expected for free dipole oscillations in the absence of any channel or gate, which represents the upper limit on currents due to energy conservation: indeed, the dipole mode represents the full, periodic conversion of potential energy into kinetic energy. Figure \ref{fig:Superflow}D shows the measured currents together with this limit, where the measurements for the unitary gas approach the limit within a factor of two.

These observations are in line with the qualitative expectations for a superfluid system, i.e. the ability to flow with very little resistance through narrow apertures which efficiently stop any other normal fluids that was discovered with superfluid Helium by Allen and Miesner, and Kapitza in 1938 \cite{Allen:1938aa,Kapitza:1938aa}, and for which the word superfluid was introduced \cite{Balibar:2007ab}. 

Beyond these qualitative observations concerning the drop of resistance in the case of unitary superfluids, more insight can be extracted by combining the current measurements with local observations of the distribution of particles in the channel, using high-resolution absorption imaging. Two main informations are extracted
\begin{itemize}
\item the density distribution at equilibrium, i.e. in the absence of current, can be used to extract the thermodynamic potential $\Omega$ at the center of the channel based on the local density approximation and the Gibbs-Duhem relation, following the procedure described in \cite{Ku:2012aa}. Comparing this with the thermodynamic potential that an ideal Fermi gas at the same density, we form a quantity $\Omega/\Omega_0$ that can be used to quantify the degree of quantum degeneracy and relate it the transport properties.
\item the line density in the channel obtained by integration of the measured density along the transverse direction. Normalising the current measured using the two-reservoirs configuration by the measured line density, we obtain a drift velocity that measures the transport properties independent of the density of particles.
\end{itemize}

\begin{figure}
\centering
\includegraphics[width=0.65\textwidth]{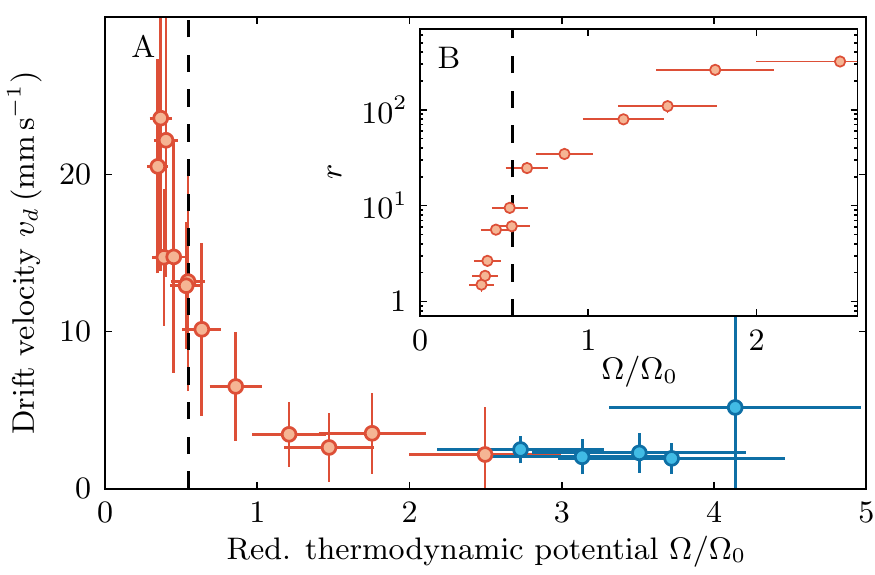}
\caption{%
{\bf Conduction properties as a function of thermodynamic potential.} A: Drift velocity as a function of the reduced thermodynamic potential $\Omega / \Omega_0$ for the strongly interacting (red) and weakly interacting (blue) Fermi gas. B: Dimensionless resistance as a function of $\Omega / \Omega_0$ in logarithmic scale for the strongly interacting gas, showing the drop of resistance. The dashed black lines at $\Omega / \Omega_0 = 0.55$ indicate the position where the superfluid transition in three dimensions occurs. Error bars represent statistical errors (one standard deviation).  Figure extracted from \cite{Stadler:2012kx}.
}
\label{fig:Superflow2}
\end{figure}

The refined observations are presented in figure \ref{fig:Superflow2}, and confirm the order of magnitude increase in velocity for the unitary gas in the quantum degenerate regime. Since in the case of superfluid reservoirs, current is driven by the phase gradient, owing to the Josephson-Anderson equation, the observation of a resistive behaviour implies the onset of a dissipative mechanism allowing for the decay of superfluid flow. Vortex nucleation is compatible with the observations, in contrast with the simple estimates for single particle excitations provided by the Landau argument. This situation is similar to that of smooth weak links in BECs in the ring geometry, where such mechanism is back by strong numerical and experimental evidence. 

That such a mechanism leads to resistive behaviour for sufficiently large velocities was explicitly demonstrated by separating a ring trap in two reservoirs using two distinct links. Moving both links with opposite directions at a fixed velocity amounts to imposing a constant flow, and the onset of dissipation is observed as a density difference builds up between the two parts. 

\subsection{Interplay with disorder}

%The two terminal configuration with superfluid gases, using the simple two-dimensional transport channel described in the previous section can be used to investigate the effects of disorder on superfluidity. Disorder was actually introduced in the investigation of weakly interacting gases in order to cross over from ballistic to diffusive in section \ref{section:multimodeWIF}. The combination of the thin transport channel with tunable disorder allows to directly observe the competition between superfluidity and disorder. 

A very attractive aspect of cold atomic gases is the ability to introduce fully characterized, controlled disorder. The effects of disorder on superfluidity have been investigated in the context of cold atomic gases for about a decade \cite{Sanchez-Palencia:2010aa,Shapiro:2012aa}. The main focus so far has been the investigation of the phase diagram of disordered system, in particular the observation of the elusive Bose-Glass phase\cite{Fallani:2007aa,Chen:2008aa,White:2009aa,Deissler:2010ys,Pasienski:2010vn,Beeler:2012aa,Allard:2012aa,DErrico:2014aa}, predicted by theory \cite{Giamarchi:1988aa,Fisher:1989aa,Pollet:2009aa} and for which clear experimental demonstrations have proven difficult. Several experiments have used phase coherence measurements and transport generated by a homogeneous magnetic field gradient as a evidences for disorder induced localisation, finding in general that a sufficiently strong disorder is capable of destroying superfluidity. The main motivation for extending these studies to transport measurements was to use the high sensitivity of transport to superfluidity in order to provide further insights in the mechanisms destroying superfluidity in disordered systems. This problem is to a wide extent still open in the condensed matter context.

In the two terminal configuration, the effects of disorder were investigated by replacing the homogeneous gate potential used to tune locally the density of atoms by a the disordered potential produced using a speckle pattern \cite{Krinner:2013aa}. The potential was repulsive, consisting in a random distribution of hills with an average strength $\bar{V}$. In addition to its average strength, disorder is also characterised by its correlation length $\sigma = $290(90)$\,nm$, which is the half width at half maximum of the power spectral density of the potential, directly measured in-situ. The correlation length plays a key role in the physics of Anderson localisation for single particles, since it determines the scattering mean free paths and localisation lengths. 

The simple situation of a Bose-Einstein condensate was first investigated, using a gas of tightly bound molecules produced in the molecular regime of a broad Feshbach resonance. The magnetic field was chosen such that the s-wave scattering length is $3545\,a_0$, where $a_0$ is Bohr's radius, and the binding energy of the molecules is $E_b = 2.3\,\mu$K. In comparison all the other energy scales such as the correlation energy $E_\sigma=\hbar^2/2m\sigma^2$ and $\bar{V}$ are weak. As a consequence, a description of the system in terms of Bose-Einstein condensate is adequate. In that framework, the relevant length scale for the effects of disorder on superfluidity is the healing length, which in this experiment is typically $\xi \simeq 230$\,nm, of the order of the correlation length of the disorder, such that disorder does not simply reduce to a smooth modulation of the condensate profile but can deplete the condensate directly.

Similar to the experiments with a smooth gate potential, the measurements where repeated using a weakly interacting Fermi gas under the same conditions, in order to provide a comparison point where superfluidity is absent. This is of particular interest in the case of disordered systems since the single particle problem is already a pretty involved problem in the regime where disorder is not perturbative. In particular, the transport properties change dramatically with disorder strength without the need to invoke localisation or many-body properties, which makes it hard to identify the genuine effects of superfluidity.

\begin{figure}
    \includegraphics[width=0.65\textwidth]{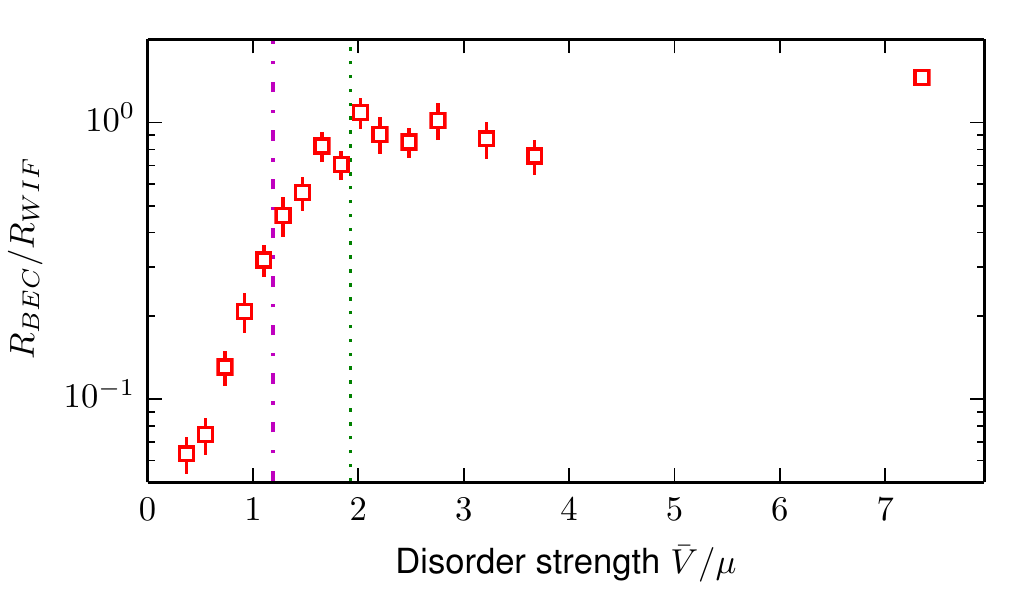}
        \centering
    \caption{{\bf Ratio of resistances of the BEC and the weakly interacting Fermi gas (WIF) as a function of disorder strength}. The disorder strength $V_0$ is normalised to the chemical potential of the BEC $\mu$. The green dotted line represents the percolation threshold of the speckle potential, and coincides with the change of behavior from superfluid to single-particle transport. The dash-dotted line shows the correlation energy $E_{\sigma}$. Error bars represent statistical errors. Figure extracted from \cite{PhysRevLett.112.119901}. }
    \label{fig:resistanceRatio}
\end{figure}

Figure \ref{fig:resistanceRatio} presents the ratio of resistances measured for the BEC and the weakly interacting system. We identify a strongly disordered regime where the ratio is constant and of order one. In this case, a simple interpretation is that disorder dominates over all the other processes and no phase coherence or superfluid effects contribute to transport. Upon reducing the disorder strength a strong reduction of resistance for the BEC shows up and for the weakest disorder, the large currents involved prevent thermalization in the reservoirs and no resistance can actually be attributed, as observed already in the clean case \cite{Stadler:2012kx}.

The transition between the two regimes takes place for disorder strength of the order of twice the chemical potential of the BEC. This is actually the position of the expected percolation threshold of the potential \cite{Weinrib:1982aa,Pezze:2011ab}: a classical particle with energy below $V_{\mathrm{th}} = 0.52 \bar{V}$ is typically trapped in a finite region of the disordered landscape, with a probability approaching one for large systems, while a particle with energy higher than $V_{\mathrm{th}}$ can diffuse. This corresponds to a classical phase transition from metal to insulator. For the case of a BEC, the condition $\mu>V_{\mathrm{th}}$ corresponds to having a connected superfluid pierced by disorder hills, which we expect to support a finite superfluid flow. Conversely, $\mu>V_{\mathrm{th}}$ corresponds to disconnected superfluid islands. 

This picture agrees with the observation for the position of the observed threshold. However, we only expect it to hold qualitatively as it disregards several important ingredients: (i) tunnelling through the disorder barriers can be significant, since the correlation energy is rather large compared to $\mu$. Since such processes can maintain phase coherence, the picture of disconnected superfluid islands does not prevent macroscopic phase coherence and superfluidity. (ii) Anderson localisation due to wave interference is certainly strongly modifying the percolation picture, by further localising particles that classically would diffuse. The interplay between localised single particle states and  interaction induced hybridation has been shown to lead to delocalisation \cite{Deissler:2010ys}. 

Regarding the nature of the strongly disordered state, further informations were gathered using in-situ absorption imaging, allowing for the estimate of the compressibility of the system. This is an important observable as finite compressibility together with an insulating character is a feature of the Bose-Glass phase. the compressibility was indeed observed to remain finite over the whole range of disorder strength explored, providing further hints that the strongly disordered regime has a lot in common with the Bose-Glass.

These qualitative findings were found to be very robust when varying the interaction strength in the BEC-BCS crossover, namely the existence of the two regimes of low disorder and high disorder characterized by their resistance compared to the weakly interacting Fermi gas, and the fact that the compressibility remains finite. Very similar findings were also reported using the unitary Fermi gas and a disorder with longer correlation length \cite{Krinner:2015ac}. In the latter case, the correlation length of the disorder was $\sigma = 0.72(5)\,\mu$m, large enough that individual modulations of the disordered potential could be resolved by direct optical imaging. 

\begin{figure}
    \centering
    \includegraphics[width=0.85\textwidth]{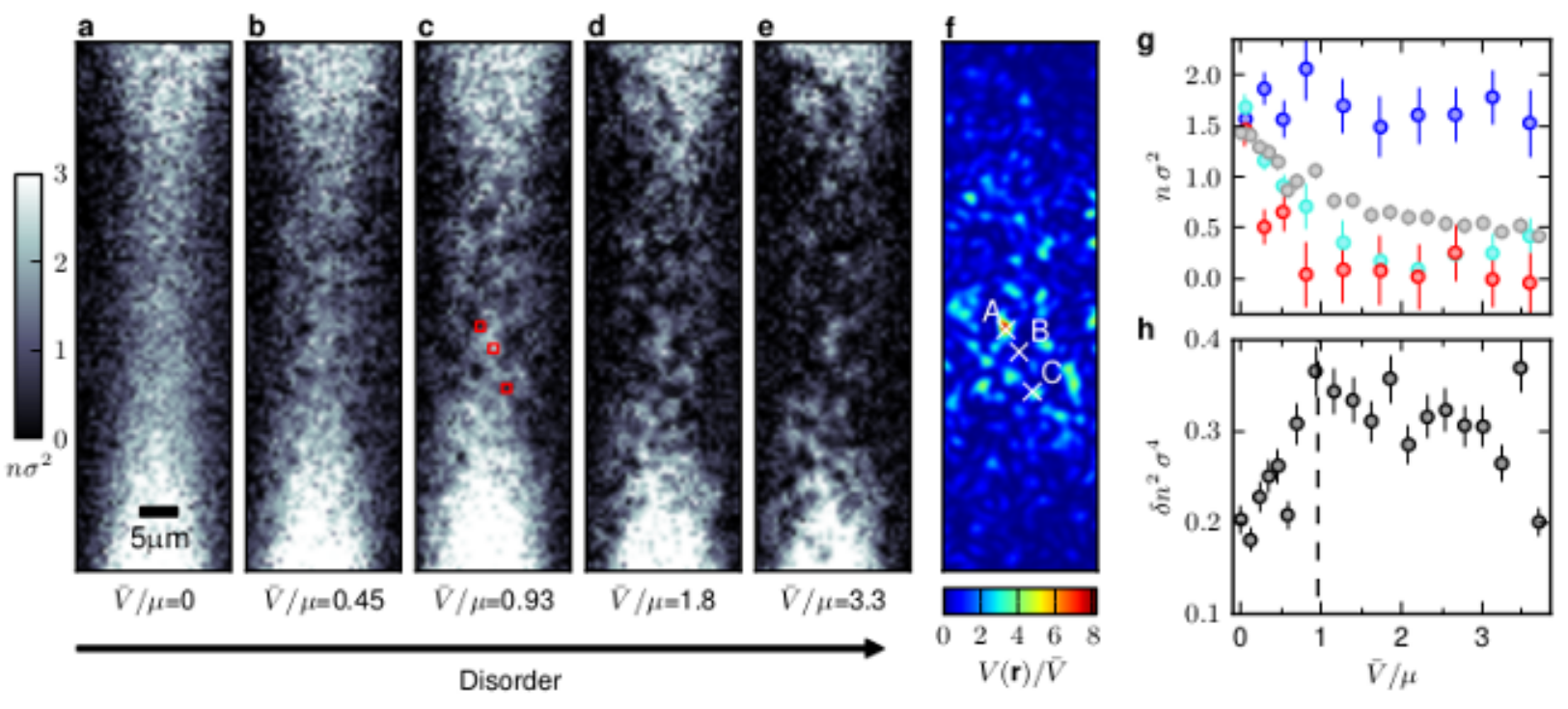}
    \caption{{\bf Evolution of the column density $n$ (in units of $\sigma^{-2}$) as the disorder strength is increased.} \textbf{a-e,} High-resolution images of size $21\,\mu m \times 72\,\mu$m of the in-situ density distribution in the channel for increasing $\bar{V}/\mu$. The saturated column density on top and bottom marks the beginning of the reservoirs, which extend far beyond the field of view. The systematic uncertainty in $\bar{V}/\mu$ is estimated to be 25\,\%. \textbf{f,} Image of the projected speckle pattern. The density ripples, gradually appearing from panel a to e, can be matched one to one to bright (potential hills) and dark spots (potential valleys) in the image. 
    \textbf{g,} Local column density as a function of disorder strength for three specific points indicated in the potential landscape of panel f (point A: red, point B: blue, point C: cyan), each computed within a region of size $1.2\,\mu m \times 1.2\,\mu$m marked as red squares in image c. The grey data points are the mean column density in the channel, computed in a central region of size $18\,\mu m \times 7\,\mu$m. \textbf{h,} Variance of the density computed in the same central region. The dashed line represents the theoretical percolation threshold for the potential seen by point-like pairs. Figure extracted from \cite{Krinner:2015ac}. }
    \label{fig:densityPics}
\end{figure}

Such images are shown in figure \ref{fig:densityPics}(a-e) for increasing disorder strength, and \ref{fig:densityPics}(f) shows the underlying disorder potential landscape. The fragmentation of the density profile is obvious from the evolution with increasing disorder. The amount of modulations in the density profile was measured and seen to saturate (\ref{fig:densityPics}(h))above $\bar{V}\sim\mu$, where $\mu$ is here the chemical potential for unitary gas in the reservoirs. This was also the point were the transport properties entered the strongly disordered regime where they are equal to that of the weakly interacting Fermi gas. This threshold actually corresponds to the percolation threshold for {\it paris} composed to two particles, thus seeing twice stronger disorder. These pictures were analysed in terms of a continuous percolation of the density distribution, forming paths connecting the two reservoirs at lower densities with increasing disorder. 

Systems of superfluid Fermions in the presence of disorder have been thoroughly investigated in condensed matter physics, in particular since the 90s in thin superconductors where a superconductor to insulator transition was observed \cite{Goldman:1998aa,Gantmakher:2010aa}. The situation of superfluid Fermion pairs in the presence of disorder with a thin film configuration has a lot in common with superconducting films, with a few important differences. First the finite correlation length of the disorder introduces a new length scale in the problem, such that the results obtained have a lower degree of universality than homogeneously disordered films. In that respect, the situation is more similar to granular systems. Second, since atoms are neutral particles no mechanism analogous to Coulomb repulsion can lead to pair breaking, and nothing prevents pairs to form more and more localised compounds with a strong overlap between the two constituents. This was indeed the interpretation proposed in \cite{Krinner:2015ac} for Fermi superfluids forming a percolated phase of pairs in the two terminal measurements, even though the pairs in the clean system extend over distances larger than the correlation length of the disorder. 

A last difference, which is the focus of a lot of attention in the recent years, is the isolated nature of the atomic gases. This raises the question of the mechanisms by which a system reaches thermal equilibrium. In a situation where single particle states are localized over a large range of energies, like in strongly disordered systems, the phenomenon of many-body localization can localize the system in the Hilbert space, where even in a steady state some observable will strongly differ from the expected values at thermal equilibrium \cite{Basko:2006ac,Aleiner:2010kx,Altman:2014aa,Nandkishore:2015aa}. First evidences of non thermal states emerging in strongly disordered atomic gases have been reported very recently \cite{Schreiber:2015aa}. It should however be emphasized that slow equilibration or hysteric behavior is a general feature of systems having bad transport properties, even for purely classical reasons as is well known for glassy systems. Distinguishing such behavior from genuine many-body effects is a very challenging question that remains to a large extent open, both from experimental and theoretical point of views.

\subsection{Josephson oscillations in Fermionic superfluids}

\begin{figure}
    \centering
    \includegraphics[width=0.4\textwidth]{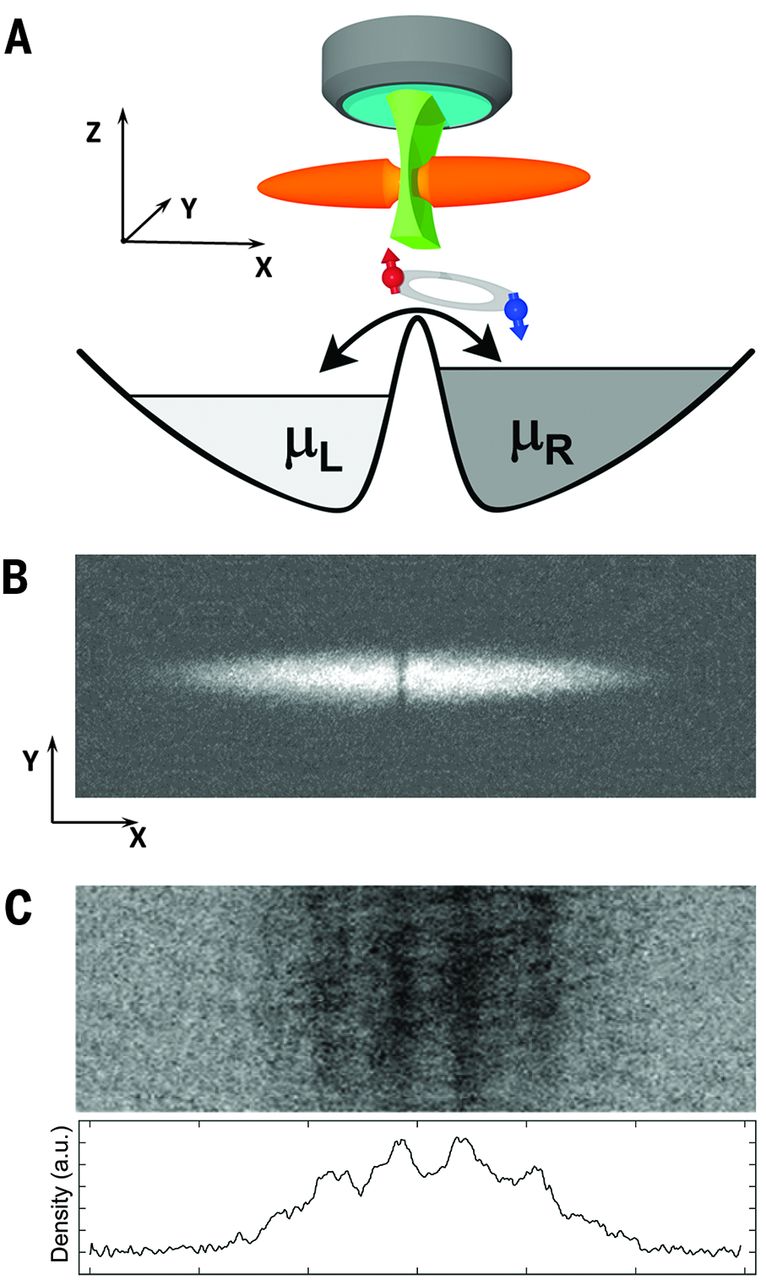}
    \caption{{\bf Josephson junction for superfluid Fermi gases}. A: schematic of the experimental configuration. B: typical absorption picture of a cloud in the presence of the thin barrier. C: fringe pattern observed after a rapid ramp and a time-of-flight, showing the phase difference between the two sides and allowing for the determination of the relative phase. Figure from \cite{Valtolina:2015ab}.
    }
    \label{fig:roati1}
\end{figure}

The relation between the superfluid phase and the current, as well as the Josephson-Anderson equation, is a feature that is independent of the Bosonic or Fermionic nature of the superfluid. In a two reservoirs situation similar to that of the two-modes BEC, Fermionic superfluids thus should exhibit plasma oscillations like their Bosonic counterpart, whenever the channel operates in a configuration close to the tunneling regime. Such an experiment was performed recently in Florence in the group of Giacomo Roati \cite{Valtolina:2015ab}, using a thin multimode channel.

The system studied consists of an elongated Fermi gas produced in a crossed optical dipole trap, with an elongated shape with an aspect ratio of $10$. The Fermi gas is rendered superfluid by working in the vicinity of the broad Feshbach resonance, realising the BEC-BCS crossover. Rather than imprinting a two dimensional channel, a very thin barrier with a short waist of $2\,\mu$m and controlled height was introduced at the center of the cloud, using a highly anisotropic laser beam. While this barrier is thicker than the coherence length of the superfluid in the BEC-BCS crossover regime, it was thin enough that no significant superfluid dynamics takes place in the transverse directions inside the barrier, realizing thus a Josephson junction, as shown in figure \ref{fig:roati1}A.

The experiments were initiated by introducing an atom number imbalance $z = \Delta N/N$ between the two sides of the junction. The key to the measurement was the use of a very low atom number imbalance of $z= 0.03(1)$, such that currents are low enough for the phases in the reservoirs to adjust and a single mode description remains valid. A second very important tool was the use of the rapid ramp technique to measure the relative phase of the two sides using atomic interferences after time of flight, as shown in figure \ref{fig:roati1}C. Without this technique, interference fringes cannot be observed away from the BEC regime, as observed in earlier attempts to observe Josephson dynamics with strongly interacting Fermi gases \cite{Kohstall:2011aa}.

\begin{figure}
    \centering
    \includegraphics[width=0.65\textwidth]{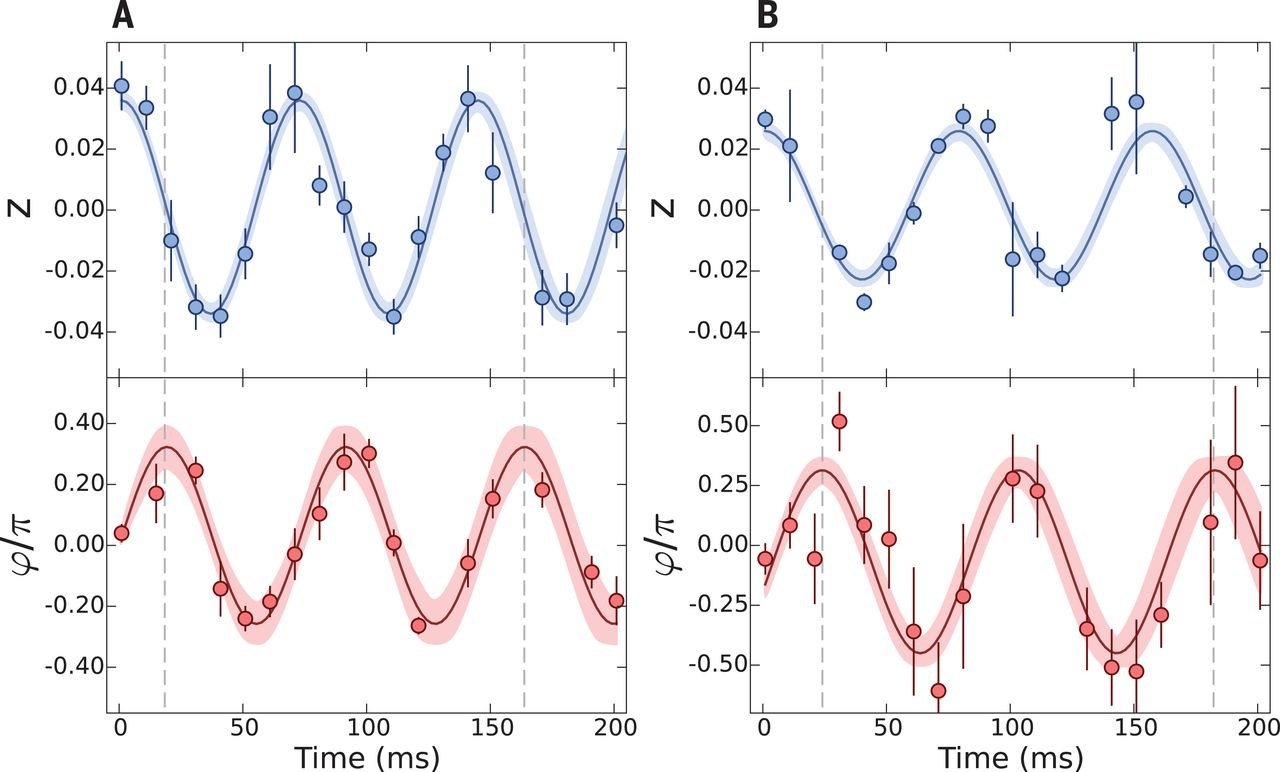}
    \caption{{\bf Plasma oscillations in a superfluid Fermi gases}. Population imbalance oscillations (top row) and phase difference (bottom row) oscillations in quadrature, on the BEC side of the resonance $1/k_Fa=4.3$ (A) and at unitarity (B). The height of the thin barriers are equal to $\mu$ and $1.1\,\mu$ respectively. Extracted from \cite{Valtolina:2015ab}.
    }
    \label{fig:roati2}
\end{figure}

The dynamics of both atom number difference and phase difference is shown in figure \ref{fig:roati2}. It is very similar to that observed with Bose-Einstein condensates in a double well configuration shown in figure \ref{fig:Oberthaler1}, and in the case of the molecular regime shown in figure \ref{fig:roati2}A, these are strictly identical. Qualitatively, this effect was seen to persist over the whole BEC-BCS crossover, highlighting the universality of the Josephson-Anderson equation. Following the model derived for Bosons, the observed oscillation frequency is given by $\omega_J = \sqrt{E_C E_J}$, with the charging energy $E_C$ arising from the reservoirs compressibility and the Josephson coupling strength $E_J$. Calculating the thermodynamic properties of the reservoirs using the measured trap frequencies, and extracting $E_C$, it was possible to directly extract $E_J$ across the BEC-BCS crossover. It was found that $E_J$ presents a broad maximum at unitarity when measured as function of scattering length at fixed total atom number and fixed height of the barrier. It arises due to the competition between an increase of the chemical potential when going from the BEC to the BCS side, which decreases the relative strength of the barrier, and the reduction of the paired fraction associated with the reduction of the strength of attractive interactions when going towards the BCS regime. 

\begin{figure}
    \centering
    \includegraphics[width=0.65\textwidth]{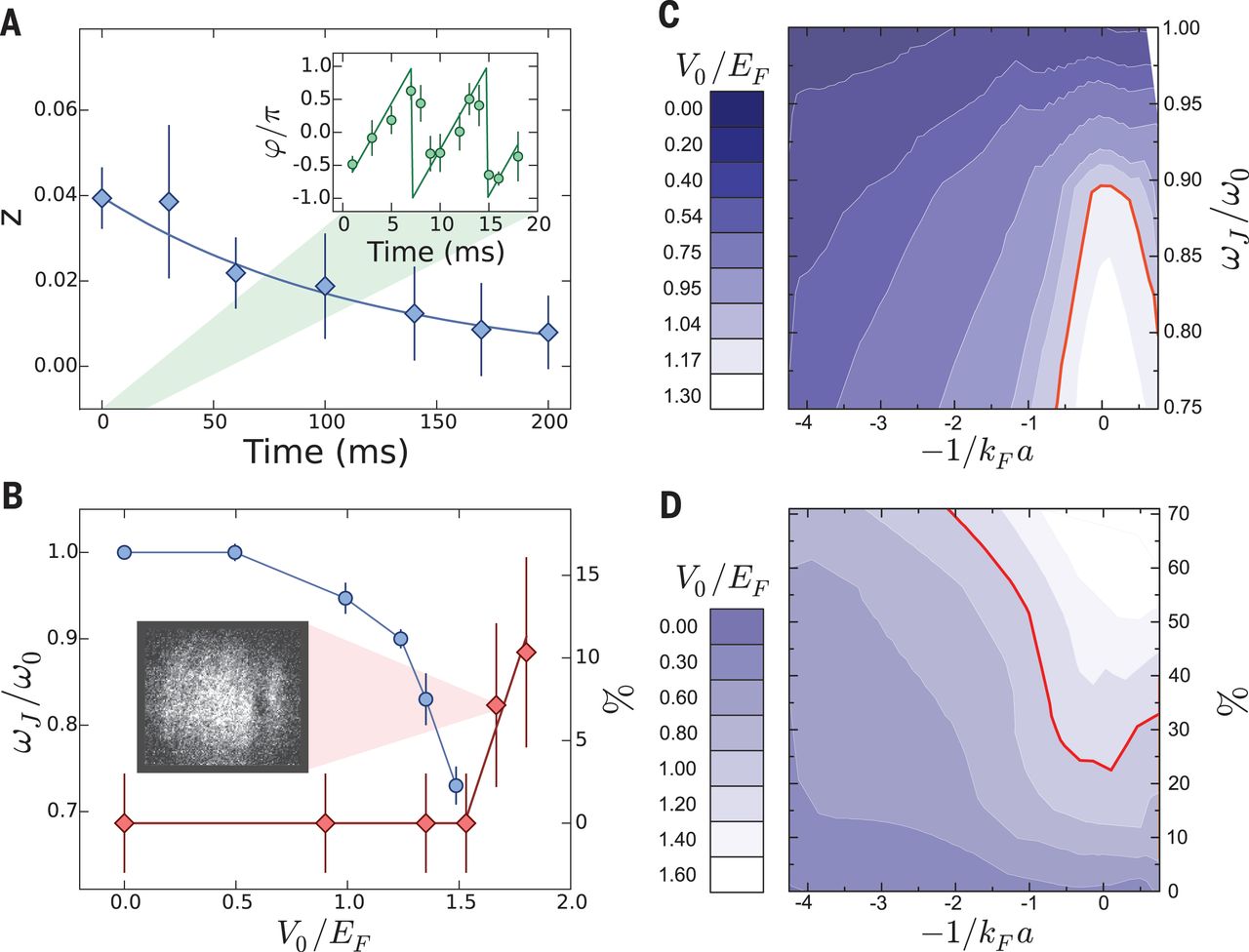}
    \caption{{\bf Onset of dissipation}. (A) Time evolution of the population imbalance for a large initial imbalance (blue diamonds) with an exponential fit to the data. The inset shows the time evolution of the phase for the same initial parameters. (B) Vortex occurrence probability, evaluated over a collection of 40 independent time-of-flight images (red diamonds, right axis), and normalized plasma frequency (blue circles, left axis) as a function of relative barrier height at unitarity, for an initial imbalance of 0.04. The inset shows a typical image of one vortex, recorded after a 10-ms time-of-flight expansion. (C) Contour plot of the relative plasma frequency versus interaction strength for different barrier heights. (D) As in (C), but for the probability of vortex occurrence. Extracted from  \cite{Valtolina:2015ab}.
    }
    \label{fig:roati4}
\end{figure}

When using larger initial imbalances or higher barriers, the system was seen to enter a regime of exponential decay of the initial imbalance, qualitatively similar to that of \cite{Stadler:2012kx} and presented in section \ref{section:superflow}. In this regime, phase coherence was maintained inside the reservoirs, as proved by the survival of the interference contrast observed in time of flight, as shown in figure \ref{fig:roati4}A. The onset of resistive flow leading to the exponential regime was attributed to vortex nucleation. Direct evidences for vortices inside the reservoirs where observed on some experimental pictures in this regime. The probability of vortex occurrences in the measurements was measured as a function of the height of the barrier in the unitary regime, showing a threshold above a critical height of about $1.5 E_F$ with $E_F= \hbar \bar{\omega}(6N)^{1/3}$ the Fermi energy of the trapped gas. The findings in this regime are presented in figure \ref{fig:roati4}B. A full map of the onset of dissipation in the BEC-BCS crossover was obtained as a function of both barrier height and interaction strength, and presented in figure \ref{fig:roati4} D.

The dissipative regime observed in this experiment is the counterpart of the self trapping regime observed in the case of a small Bose-Einstein condensate in the double well configuration. The qualitative difference between the two experiments is the geometry used, with much larger reservoirs and weaker trapping in the case of the Fermi gas. This allows for a breakdown of the two mode approximation, with the possibility of a phase dynamics to take place inside the junction, where vortices are nucleated and later move inside the fluid, generating dissipation.

Despite the wide differences in geometry, in interaction strength and even in the quantum statistical nature of the particles, the similarities between the Bose-Einstein condensates superfluid circuits in the ring geometry and the double well, multimode configuration are very striking. The unifying concept of phase coherence and the existence of the superfluid order parameter covering both Bose and Fermi superfluids, together with the fundamental Josephson-Anderson equation are very powerful tools to describe complex geometries and non equilibrium dynamics. This adds to the very large body of evidences for the universality of these concepts already collected in the context of superfluid Helium \cite{Packard:1998aa,RevModPhys.87.803}. 

\subsection{Superfluid Fermions in the single mode regime}

\begin{figure}
    \centering
    \includegraphics[width=0.6\textwidth]{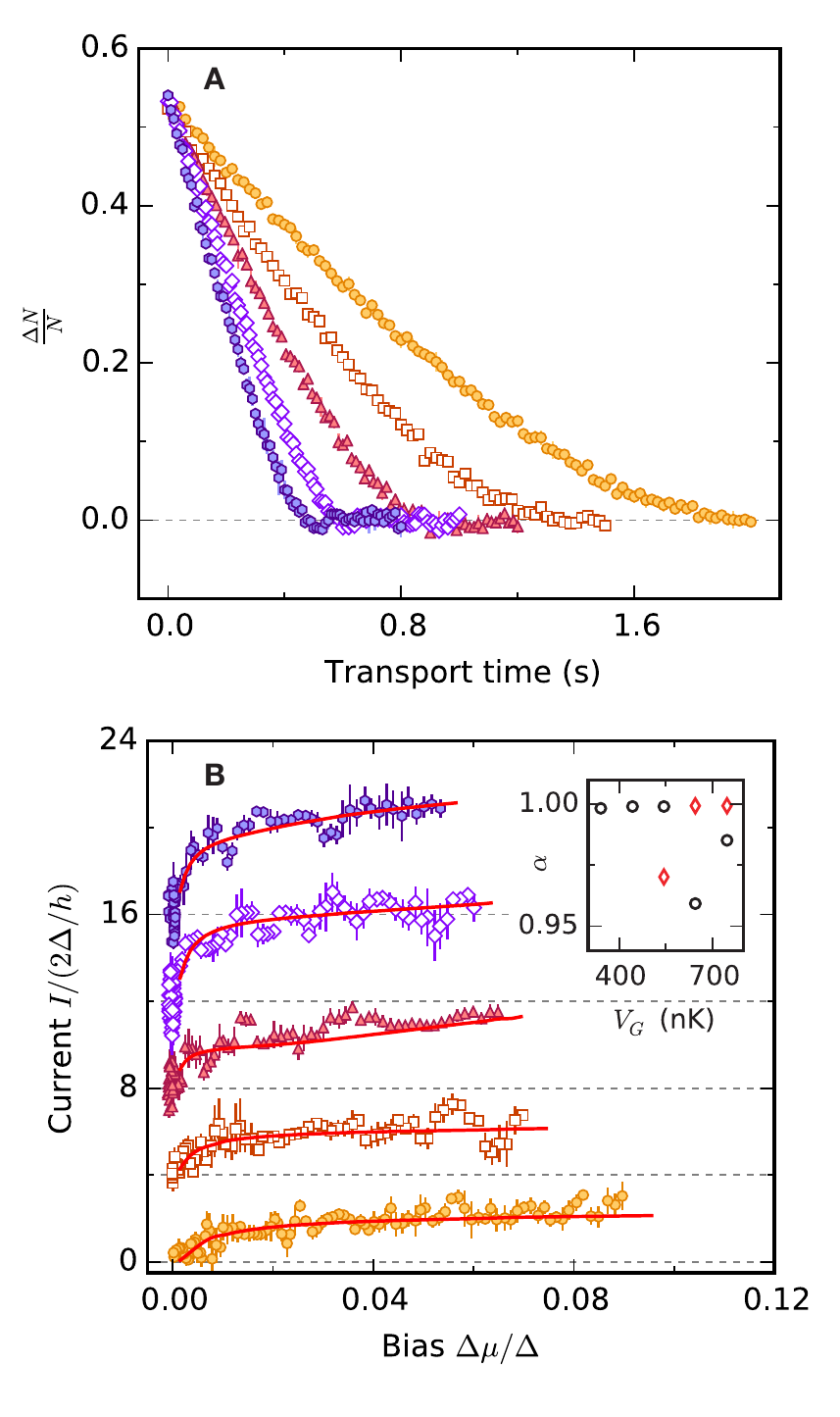}\caption{{\bf Non-linear characteristic of the QPC}. (A) Time evolution of the particle imbalance $\Delta N
/ N$ for gate potentials $V_{\mathrm{g}}=$341~nK (filled circle),
443~nK (open square), 544~nK (filled triangle), 645~nK (open diamond)
and 747~nK (filled hexagon). (B) Current-bias characteristics
normalized with respect to the superfluid pairing gap $\Delta$. The
error bars represent the variation of three averaged data
sets. Negative values of the current are artifacts from the numerical
derivation process. The red lines show the result of Keldysh
calculations with the transparency $\alpha$ of the QPC as the only free parameter.
% The values are given in \cite{materialsandmethods}. 
For clarity the curves are shifted vertically by 4 units. Temperature is $100(4)$\,nK for all data sets. Inset: fitted transparency $\alpha$ for the various gate potentials. Black circles refer to the lowest mode, red diamonds refer to the next two degenerate modes, when present.} 
    \label{fig:nonLinear}
\end{figure}

    The fact that the phenomenology of superfluid weak links, such as the Josephson effect, does not depend on the Bosonic or Fermionic nature of the underlying particles is a remarkable feature. This has actually played an important role in the history of the understanding of the physics of superfluidity \cite{Balibar:2007ab,Yang:1962aa}. Except for the quantitative evolution of the coupling constants, the observations of the Florence group \cite{Valtolina:2015ab} indeed demonstrate the universal character of superfluid dynamics from the Bose-Einstein condensate regime to the BCS regime. 
    
 An important question is therefore whether some transport phenomena could reveal the Fermionic or Bosonic character of superfluid, beyond the Josephson-Anderson relation. For BCS superfluids, pair breaking excitations have no Bosonic counterparts, and play a central role in situations where part of the system is in the normal state and is exchanging particles with the superfluid. The case of metallic weak links in superconductors is emblematic of this situation \cite{Post:1994aa,Goffman:2000aa} (see \cite{Bretheau:2013aa} for a recent review). We now present an experiment where such effects have been probed using superfluid Fermi gases contacted via a quantum point contact \cite{Husmann:2015ab}.

This experiment uses the setup realizing a quantum point contact (QPC) used for the observation of quantized conductance presented in \ref{section:QPC}, but uses atoms in the unitary regime. In particular, the use of an attractive gate potential (with strength $V_g$) in the QPC region allows for an independent control over the chemical potential relevant to transport, at the fixed temperature imposed by the reservoirs. The unitary gas is a particularly simple situation, despite the large interactions, since all the energy scales are universal functions of the fugacity, and at low temperature are just proportional to the Fermi energy. This allows in particular for reliable thermometry thanks to the detailed knowledge of the temperature dependent equation of state \cite{Luo:2009aa,Nascimbene:2010ys,Ku:2012aa}. 

The main finding is that in the single mode or few modes configuration of the QPC, the linear response breaks down and a non-linear relation between current and bias emerges. This is presented in figure \ref{fig:nonLinear}, where the time evolution of the atom number imbalance is presented as a function of time, showing an obvious non-exponential decay of imbalance for all strength of the gate potential. Differentiating these data, the current to bias relation is obtained confirming the non linearity. In order to facilitate the comparison with a microscopic model, the current to bias relation is normalized to the estimated value of the pairing gap $\Delta$ at the center of the QPC \cite{Schirotzek:2008aa}, based on the potential landscape and neglecting quantum corrections due to the tight confinement inside the QPC and in the confined region. This model describes multiple Andreev reflections in the QPC leading to a finite DC current, that drops sharply at very low bias, signaling the crossover to the Josephson regime, which was not observed in this experiment. The theory (solid lines in figure \ref{fig:nonLinear}) reproduces the data with the transmission coefficient for each channel as free parameter.

The physics of multiple Andreev reflections in a QPC can be understood as the coherent tunneling of several pairs across the quantum point contact. In the presence of a finite bias $\Delta \mu$, small compared to the pairing gap, the energy acquired by a pair upon crossing the contact is $2 \Delta \mu$. Due to the gap, this energy cannot be released into the bulk of the reservoirs, and thus unitary evolution drives the system back to its initial state after a time $\sim \hbar/\Delta \mu$, i.e. the system performs Josephson oscillations. However, coherently combining $n/2$ pair tunneling events yields an energy gain of $n \Delta \mu$. For any finite bias, there exists an $n$ for the $n$th order process to yield enough energy to create a pair breaking excitation above the gas. This process is called multiple Andreev reflection of order $n$, and it can yield a DC current through the QPC, as opposed to Josephson oscillations where the current averages to zero over one cycle. It is conceptually similar to Doppleron resonances in the physics of laser cooling \cite{Kyrola:1977aa}, or to Wannier-Stark resonances in optical lattices \cite{Gluck:2002aa}. Whenever the transmission probability $\alpha$ is smaller than one, the $n$th order process has a probability of order $\alpha^n$, thus the DC current through the QPC is exponentially suppressed at low bias. The order of Andreev reflections at which the DC drops gives indications on the value of $\alpha$. In the experiment, this is the way $\alpha$ is fitted on each curve using the theory. The fitted values are presented as insets in figure \ref{fig:nonLinear}, showing very large values, compatible with the observation of quantized conductance in the non interacting case. 

The agreement with the data provides indirect evidences of multiple Andreev reflections in the cold atom context, where interactions are very strong \footnote{Andreev bound states were invoked to explain measurements on solitons in the BEC-BCS crossover in \cite{Yefsah:2013aa}, but dismissed in later measurements \cite{Ku:2014aa}.}. More direct observations of multiple Andreev reflections have been observed in solid state superconductors, in the form of resonances at biases equal to integer fractions of the pairing gap \cite{Kleinsasser:1994aa,Goffman:2000aa}. This structure is not visible in the data on cold atoms, because the large transmission probabilities make it very weak, and the signal to noise ratio for current measurements is too low. 

The indications of Andreev states in cold atomic gases is an important step since they are one among the simplest non trivial bound state that appear at interfaces between normal and superfluid phases. For more complex structures such as spin dependent or spin orbit coupled contacts, more exotic states can appear such as Shiba \cite{Shiba:1968aa} or Majorana states \cite{Kitaev:2001aa,Jiang:2011aa}. 
%Importantly, it is also an example of dissipation induced by pair-breaking mechanism where the Fermionic nature of the underlying gas is fundamental, in contrast to the multimode situations where phase slips and vortices where likely responsible for the dissipation which occurred as well for Bose-Einstein condensates.

\subsection{Superfluid transition and the intermediate interaction regime}

Besides the two special cases of non-interacting particles and unitary limited scattering, which have been described above, the general question of the interplay of quantized conductance with many-body correlations can be approached using the control offered by the cold atoms settings. The cases of intermediate interactions and finite temperature are more difficult both from the conceptual and experimental point of view, because the addition of energy scales or length scales in the problem reduces the degree of universality, making it more difficult to draw general conclusions from the particular experimental conditions. Nevertheless, starting the well understood situations of universal conductance quantization or non-linear superfluid flow described by multiple Andreev reflections, one can introduce deviations in a systematic way and observe the outcomes on the transport properties. This was performed using again the single mode QPC.

\begin{figure}
    \centering
     \includegraphics[width=0.55\textwidth]{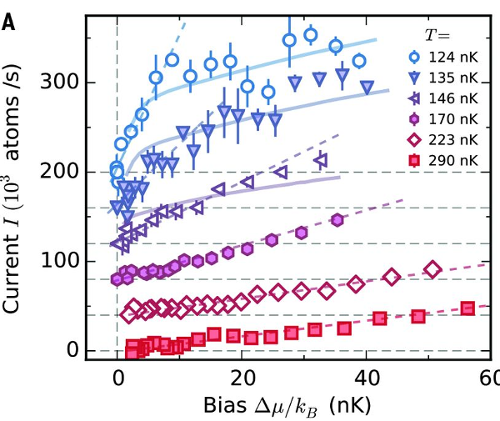}
        \caption{{\bf Non-linear current-bias relation for increasing temperatures}. Temperature is increased from $124\,$nK up to $290\,$nK. The non-linearity, evident at low temperature disappears with increasing temperatures. The multiple Andreev reflection theory at finite temperature is presented in solid lines, showing strong deviations already at $146\,$nK. The dashed line is a fit to the initial part of the curve, to guide the eye. Figure extracted from \cite{Husmann:2015ab}. }
    \label{fig:nonlinTemp}
\end{figure}

A first set of measurements was using the unitary Fermi gas where the multiple Andreev reflection picture holds at low temperature, and temperature was increased up to the expected critical temperature \cite{Husmann:2015ab}. The results are presented in figure \ref{fig:nonlinTemp}. Upon increasing temperature, the non-linear current to bias relation was observed to disappear and a linear regime was recovered. The measurements were however found to deviate strongly from the finite temperature mean-field theory successfully used to reproduce the low temperature, superfluid data. In particular, even in the linear regime, the conductance of the channel was much larger than expected from the mean-field theory i.e. the predictions of the Landauer formula for non interacting particles. On the one hand, this is not surprising since even in the non-superfluid phase, the system remains strongly interacting. On the other hand, it gives indications that in the normal phase, the correlations remain strong enough to prevent a description in terms of non interacting Fermionic quasiparticles. 

\begin{figure}
    \centering
    \includegraphics[width=0.8\textwidth]{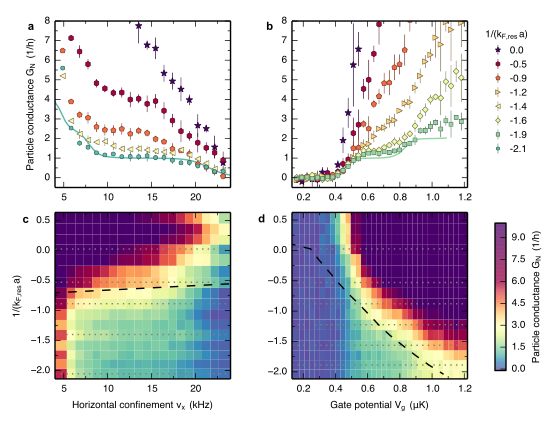}
        \caption{{\bf Particle conductance of the attractively interacting Fermi gas.} {\bf a,} Particle conductance $G_N$ as a function of the horizontal confinement frequency $\nu_x$ of the QPC, at fixed gate potential $V_g = 0.42\,\mu$K; and {\bf b,} as a function of the gate potential $V_g$ at fixed confinement frequency $\nu_x = 23.2$kHz, for different interaction strengths $1/(k_{\textup{F,res}} a)$ in the reservoirs. The solid lines are theoretical predictions for $1/(k_{\textup{F,res}} a) = 2.1$ and 1.9 respectively, based on the Landauer formula including mean-field attraction (Methods). Each data point represents the mean over 5 measurements and error bars indicate one standard deviation. {\bf c,} and {\bf d,} Two-dimensional colour plot of $G_N$ as a function of interaction strength $1/(k_{\textup{F,res}} a)$ and horizontal confinement (c) or gate potential (d). Both plots contain the cuts of (a) and (b) (grey dotted lines), and an estimation of the local superfluid transition at the QPC exits (black dashed line). Figure extracted from \cite{Krinner:2016ab}. }
    \label{fig:maps}
\end{figure}

A second set of experiments were performed at low temperature as a function of interaction strength, covering the regime from unitarity to weak attraction, where the low but finite temperature leads to a normal phase \cite{Krinner:2016ab}. For the weakest attractions, quantized conductance is observed with a plateau at $1/h$, in agreement with the expectations of the Landauer formula, including a mean-field effect reducing the length of the plateau. Upon increasing the strength of attractive interactions, the plateau was preserved but it was observed to continuously shift to higher conductances. These observations are presented in figure \ref{fig:maps}. For large densities or stronger interactions, the plateau fully disappeared and a non-linear current-bias relation was recovered, as expected from multiple Andreev reflections in the contact.

\begin{figure}
    \centering
    \includegraphics[width=0.65\textwidth]{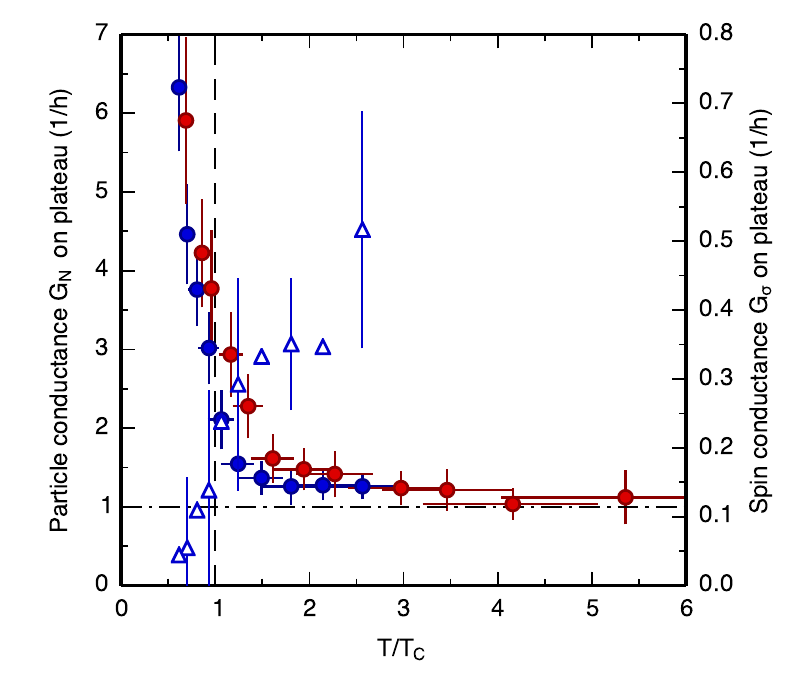}
 \caption{{\bf Particle and spin conductances in the single mode regime}.
    $G_N$ (closed circles) and $G_\sigma$ (open triangles, every second error bar displayed) for various interaction strengths are presented as a function of the reduced temperature $T/T_c$, which varies due to the dependence of $T_c$ on density and scattering length.
    Blue data points are obtained from the measurements shown in Fig. \ref{fig:maps}d and Figs. \ref{fig:spinCond}, for $V_g=0.64\mu$K and $\nu_x=23.2$\,kHz.
    Red data points are obtained from the measurements shown in Figs. \ref{fig:maps}a, for $V_g=0.42\mu$K and $\nu_x=14.5$\,kHz.
    $G_N$ tends to the conductance quantum $1/h$ (horizontal dash-dotted line) for weak interactions ($T/T_c\gg 1$). Error bars contain statistical and systematic errors. Figure extracted from \cite{Krinner:2016ab}.}
    \label{fig:singleModeSF}
\end{figure}

A difficulty in interpreting these data is the fact that the equation of state changes as interactions are varied across the BEC-BCS crossover, in particular the finite temperature effects have to be evaluated in order to obtain a faithful estimate of the critical temperature and the part of the parameter space where the system is expected to be superfluid. This was done using the state of the art theoretical calculations for the critical point \cite{Haussmann:2007aa}, the measured ground state equation of state in the BEC-BCS crossover \cite{Navon:2010ab} and the finite temperature corrections to first order in the interactions, extrapolated to the strongly interacting case where perturbation theory was breaking down. Applying this procedure allowed for the extraction of the conductance in the single mode regime (the center of the plateau in the weakly interacting case) as a function of $T/T_c$, where $T_c$ is evaluated from the theory \cite{Haussmann:2007aa} at the entrance and exit of the QPC where the density is the highest, using the equation of state \cite{Navon:2010ab}. The result is presented in figure \ref{fig:singleModeSF}, where the systematic deviation of the conductance from $1/h$ is obvious even way above the expected transition temperature.

Two interpretations of these deviations from the conductance quantum predicted by the Landauer formula have been formulated. 
\begin{enumerate}
\item A first explanation put forward in \cite{PhysRevLett.118.105303} focuses on the physics of the reservoirs, pointing out that even above the superfluid transition, the strongly attractive Fermi gas can still show deviations from the conventional Fermi liquid picture, as seen in previous experiments in the BEC BCS crossover \cite{Gaebler:2010zr,Sagi:2015aa}. Some superfluid correlations can persist over finite but long range, in particular in the vicinity of the transition. The critical regime, with slow power law decay of the superfluid correlations extends up to $2 T_c$ \cite{Taylor:2009aa,PhysRevA.93.023642}. To explain why such slowly decaying (power law) superfluid correlations could influence the conductance, one can compare them with another system having such power law correlations, namely the Luttinger liquid in one dimensions, where the physics giving rise to power law superfluid correlations also gives rise to enhanced conductance compared to $1/h$ \cite{Giamarchi:2003aa}.

\item The second explanation, described in \cite{PhysRevLett.117.255302}, focuses on the physics of the channel, in particular the fact that confinement enhances the attraction inside the channel compared to the reservoirs, which leads at finite temperature to a situation where the reservoirs are in the normal state but superfluidity persists inside the QPC. Then, Andreev reflection processes take place at the entrance and exit of the QPC, leading to enhanced conductance up to a factor of two. The key to explain the continuous increase of the conductance up to higher value is to realise that the position at which the superfluid to normal interface exists can be outside the single mode region, and thus Andreev reflections take place in several channels in parallel. 
\end{enumerate}

The very different nature of these explanations illustrates the richness of the physics of cold gases in mesoscopic structures. Whilst the signal to noise ratio is smaller than in state of the art condensed matter physics experiments, the microscopic Hamiltonian is known a priori and unexpected features can arise due to the interplay of strong interactions with finite size and non-equilibrium.

\subsection{Spin transport in interacting Fermi gases}

The spin degree of freedom plays a fundamental role in the physics of Fermions, since it is the degree of freedom that allows for interactions between particles. There is a fundamental difference between spin and particle currents regarding the effects of interactions: since scattering conserves the total momentum, the total current of particles is not expected to be affected by interactions. In contrast, scattering does not conserve spin currents and is thus damped by interactions, as observed already in early experiments on Fermi gases \cite{DeMarco:2001aa,PhysRevLett.87.173201}.

\begin{figure}
    \centering
    \includegraphics[width=0.6\textwidth]{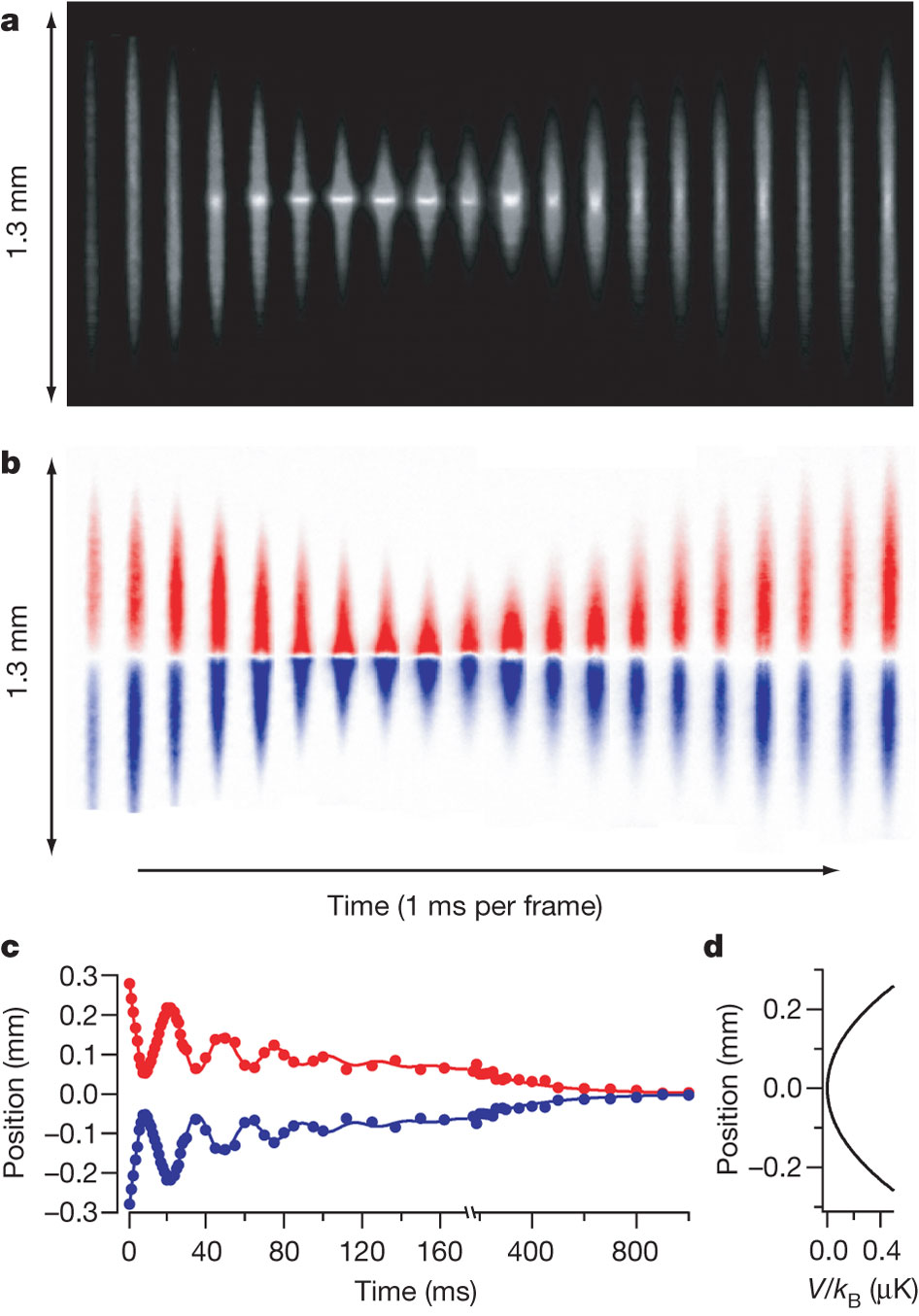}
 \caption{{\bf Observation of spin transport in the collision of two spin polarized clouds.} Total column density {\bf a} and difference in column density {\bf b} as a function of time. {\bf c}, Time evolution of the average position of the spin up and spin down atoms, showing a transcient oscillating regime regime followed by a slow diffusion. {\bf d} Trapping potential.  Extracted from \cite{Sommer:2011uq}.}
    \label{fig:zwierlein1}
\end{figure}

Several experiments have investigated spin transport in interacting Fermi gases, using configurations close to the two terminal setup. In particular, in the seminal experiment \cite{Sommer:2011uq} a two component Fermi gas is first cooled to low temperatures, then the two spin components are separated in the absence of interactions, using a dipole oscillations in a spin dependent harmonic trap produced at low magnetic field. When the two parts oscillate in phase opposition, interactions are restored by ramping up the magnetic field in the vicinity of the Feshbach resonance. The relative motion of one component with respect to the other is monitored as a function of time. For short times the two spin components keep oscillating and bounce on each other. In this regime, the dynamics is dominated by the oscillations in the reservoirs, and interactions prevent the two spin to diffuse through each other. At longer time scales, the oscillations damp, and an exponential decay of the spin-dipole is observed, corresponding to the slow diffusion of one component through the other. This is illustrated in figure \ref{fig:zwierlein1}.

\begin{figure}
    \centering
    \includegraphics[width=0.65\textwidth]{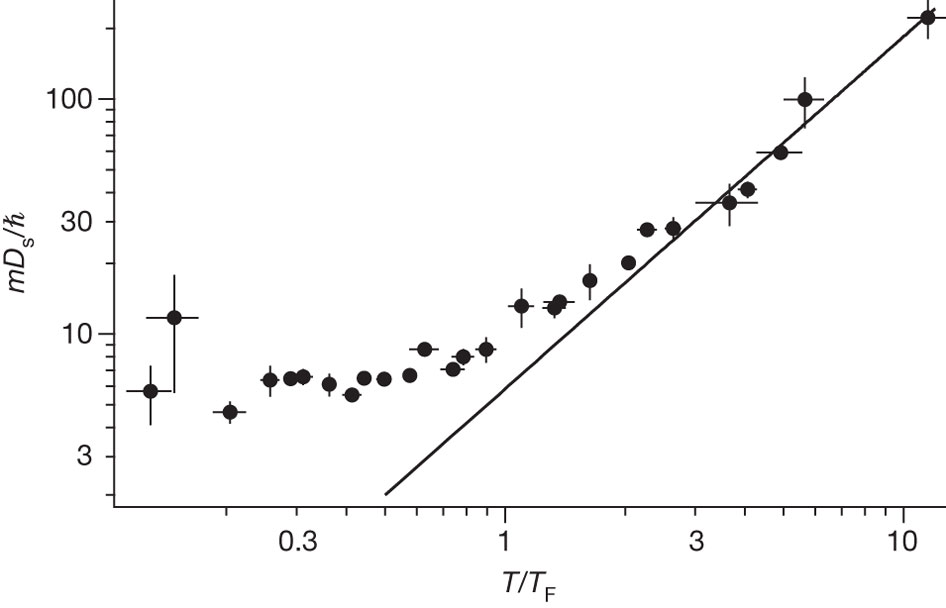}
 \caption{{\bf Spin diffusion coefficient for the trapped gas}, as a function of temperature, for the unitary Fermi gas. The solid line shows a $T^{3/2}$ trend expected at high temperature. For low temperature, the diffusion coefficient approaches a constant limit. Extracted from \cite{Sommer:2011uq}.}
    \label{fig:zwierlein2}
\end{figure}

Using the average spin current extracted from the dynamics of the center of mass of the clouds, divided by the measured spin gradient in the region where the two components merge, a spin diffusion coefficient was estimated for the unitary Fermi gas as a function of temperature. These results are presented in figure \ref{fig:zwierlein2}. In the high temperature regime, the evolution of the diffusion coefficient is controlled by the velocity dependence of the scattering cross-section, which is inversely proportional to the square of the relative momentum in the unitary limit, leading to a dependence like $T^{2/3}$. In the degenerate regime, the diffusion coefficient is saturating to $1.11\hbar/m$, as Pauli suppression of collisions gradually cancels the increase of scattering cross section. 

In the homogeneous unitary Fermi gas, dimensional analysis implies that the diffusion coefficient is a universal number multiplied by $\hbar / m$. Indeed, the only length scale is $1/k_F$ so the mean free path has to be a number divided by $k_F$, and the only velocity scale is the Fermi velocity. However, the measured number is likely to depend on the experimental situation, in particular since the density variations across the trap lead to a position dependent diffusion coefficient that cannot be recovered based on a local density approximation only (since the direction of the current density cannot be inferred from the distribution only). Direct experimental evidences of the effect of the trapping potential was found when these experiments where reproduced in the presence of a spin imbalance, where a paired core forms at the center of the trap with the excess particles repelled at the side where they form a ballistic shell. The paired core was observed to move through the unpaired atoms \cite{1367-2630-13-5-055009}.

These experiments raised two important questions regarding spin transport. The first one is the long standing question of the possible ferromagnetic instability on the molecular side of the Feshbach resonance. Indeed, as the two clouds collide, atoms explore the upper branch of the Feshbach resonance leading to repulsive interactions before three-body recombinations allow for the formation of bound molecules. Very similar spin transport measurements were performed recently with the goal to study the possible metastable ferromagnetic state emerging in the molecular regime of the Feshbach resonance. Several evidences were gathered, in particular a softening of the spin dipole mode, even though the dynamics remains dominated by the underlying molecule formation \cite{Valtolina:2016aa}.

\begin{figure}
    \centering
    \includegraphics[width=0.65\textwidth]{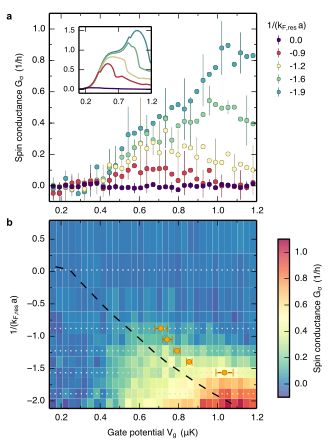}
    \caption{{\bf Spin conductance of the attractively interacting Fermi gas.} {\bf a,} 
    Spin conductance $G_\sigma$ as a function of the gate potential $V_g$ for different interaction strengths $1/(k_{\textup{F,res}} a)$ in the reservoirs. Each data point represents the mean over 9 measurements and error bars indicate one standard deviation plotted for every third point. Inset: $G_\sigma$ obtained from a mean-field phenomenological model, reproducing the non-monotonic behaviour of the experimental data.
    {\bf b,} Two-dimensional color plot of $G_\sigma$ as a function of $1/(k_{\textup{F,res}} a)$, with cuts of (a) indicated as grey dotted lines.
    The points where $G_\sigma$ is maximum, obtained from a parabolic fit along $V_g$, are displayed as orange circles for comparison. The black dashed line represents the superfluid critical line estimated at the entrance and exit regions of the QPC, using the results of \cite{Haussmann:2007aa}. Figure extracted from \cite{Krinner:2016ab}. }
    \label{fig:spinCond}
\end{figure}

The second question is the effect of pairing on spin transport. These were indirectly observed in \cite{1367-2630-13-5-055009}, where the spin separation followed by relaxation was reproduced in the presence of a partial spin polarization, and at low temperatures. Due to the inhomogeneous nature of the system, the spin currents consisted of unpaired particles expelled from the center of the trap \cite{Partridge:2006fk,Shin:2006aa} and flowing in the wings of the cloud. 

These effects have been investigated using the two terminal configuration with the quantum point contact in \cite{Krinner:2016ab}, exploring the low temperature regime with varying interaction strength in conditions identical to that of the conductance at the emergence of superfluidity in the previous section. A partial spin separation is produced following a procedure similar to that of \cite{Sommer:2011uq}, yielding reservoirs with identical chemical potential but opposite spin polarisation. The crucial difference with \cite{Sommer:2011uq,1367-2630-13-5-055009} is the fact that in the presence of superfluidity, the point contact geometry forces the spin current to occur at the center of the cloud, along the longitudinal direction only: it forces the spin current {\it through} the superfluid. 

The analysis follows the linear response model of section \ref{section:RCmodel}. The spin current is written within linear response as $I_{\sigma} =G_{\sigma} \Delta b$, with $\Delta b = (\Delta\mu_\uparrow-\Delta\mu_\downarrow)/2$ the difference between the chemical potential bias for the two spin components. In the absence of net magnetization, symmetry imposes that the spin bias does not couple to particle currents, and the response is only described by the spin conductance $G_\sigma$. The evolution of $G_\sigma$ is presented in figure \ref{fig:spinCond} as a function of gate potential and interactions. For the weakest interactions, spin conductance is growing with increasing gate potential, but with stronger attraction a broad maximum is developing, and for the largest gate potentials $G_\sigma$ decreases with gate potential, indicating a spin insulating behavior. With increasing interactions $G_\sigma$ decreases earlier, and is entirely suppressed upon approaching unitarity. For the fixed temperature of the reservoirs, increasing the gate potential increases the density, which brings the system into the superfluid phase beyond a critical value. This transition line was estimated using theory like for particle transport, with the boundary shown in figure \ref{fig:spinCond}b. The regime of vanishing spin conductance coincides with the expected superfluid regime, where very large particle transport is observed (see figure \ref{fig:maps}). 

The spin insulating character of BCS superfluid is expected as a result of the spin-singlet character of the pairs. The pairing gap acts as a spin gap separating singlet from triplet excitations. In that sense, the QPC in the superfluid regime acts as an efficient singlet filter. As a consequence, within the BCS framework, the magnetization is exclusively carried by excitations above the gap, which are the Bogolyubov quasiparticles. They form a ballistic, free Fermi gas, for which a Landauer formula can be written. In a semi-classical approximation, the large pairing gap in the QPC region acts like a repulsive barrier for the excitations, effectively closing the contact and thus suppressing spin transport. Based on such a model, the spin conductance was estimated (see inset in figure \ref{fig:spinCond}a), with a qualitative agreement with the data. A better agreement was found using the sophisticated model of \cite{PhysRevLett.117.255302} which also reproduces the anomalous conductance plateau for particle transport. 

These spin transport experiments are focused on magnetisation transport, i.e. longitudinal spin transport. the case of transverse spin diffusion focusing on decoherence of spin textures imprinted on atomic cloud is different, and is not studied using the two terminal configuration but rather as a quench experiment \cite{Koschorreck:2013aa,Bardon:2014aa,Trotzky:2015aa}.

\section{Perspectives}
The investigations of mesoscopic transport phenomena using cold atoms is only at its infancy. The experiments are so far at the status of proofs of principle. Nevertheless, several new insights have already been gathered about transport physics, including a few novel features that were not anticipated based on the existing knowledge of the condensed matter physics community. The extension of cold atoms into the field of mesoscopic transport is a very exciting perspective, and we expect that the interaction between the two fields will lead to important progress in the next years. We now present a list of challenges and perspectives for future developments. This is by no means exhaustive, and reflects our main interests at the time of the writing. 

\subsection{Technical developments}

The experimental techniques are improving fast in particular in the variety of mesoscopic structures that can be produced and investigated. In that respect, the growing use of holographic projection techniques in cold atoms experiments opens perspectives for fully programmable potential landscapes, implemented using computer programmed holograms that even provide dynamical control \cite{Zupancic:2016ab}. This will allow in the near future for the projection of tunnel barrier, and short scale potentials with large interference effects, such as quantum dots or mesoscopic lattices. The progresses in detecting atoms at the single-atom and single-site level in optical lattices could also be used in order to investigate the details of atomic distributions inside driven mesoscopic structures \cite{PhysRevLett.114.193001,PhysRevLett.114.213002,Yamamoto:2015aa,Haller:2015aa,PhysRevA.92.063406,PhysRevLett.115.263001}. Combining these high resolution techniques with automated processes for the optical control of potential landscapes could lead to fully integrated simulation machines, taking as input Hamiltonian parameters and allowing to read out transport coefficients, similar to a high performance simulation facility. This would indeed come close to a practical implementation of a quantum simulator as envisionned Richard Feynman \cite{Feynman:1982aa,Cirac:2012aa}.

A major limitation of the currently used systems is the difficulty in measuring very low currents, below $1000$ atoms per second, expected when the atoms in the channel are in an insulating state. The main problem is the reproducibility of the preparation of the system: first, atom number fluctuations are of the order of a few percent, then the separation in two reservoirs also introduces extra noise. Therefore, the outcome of atom number measurements in the reservoirs fluctuates by typically thousand atoms from one realization to the next. This yields upon averaging to the typical error bars shown in figure \ref{fig:confinementScan}. In addition, drifts inherent to long-term, systematic measurements limit the number of consecutive data sets that can be acquired and compared with each other before recalibration is required. In our setup, measurements over five consecutive days could be performed for the spin conductance measurements, with realignments every hours approximately. Improving on the preparation techniques could reduce the errors but eventually the technical stability of the setup is likely to be insufficient. Remarkably, the preparation of initial bias in small ensemble of atoms out of a Mott insulator using quantum gas microscopes reaches limitations in the reproducibility of the same order \cite{Choi:2016ab}.

There, techniques based on quantum non-demolition measurements could track the population inside the reservoirs in real time with very high precision, using for example a high finesse cavity as detector \cite{Lye:2003aa,Ottl:2005aa}. This opens the perspective of an atom-by-atom measurement of currents, allowing for a full counting analysis of currents in mesoscopic structures \cite{Albert:2012aa,Haack:2014aa}. This would also allow to perform two terminal measurements with optical lattices in the Hubbard regime. Indeed, placing the atoms in a lattice increases the effective mass of the particles, slowing down the dynamics of the reservoirs and thus the time scale where meaningful conductance measurements can be performed. In addition, the current is limited by the band width which is exponentially reduced in the Hubbard regime \cite{Esslinger:2010aa}. Because of the relevance of lattice systems to condensed matter physics, and the ability they provide to operate in the strongly repulsive regime, the extension of conductance measurements to lattices would be an important breakthrough. 

\subsection{Transport at quantum phase transitions}

Thanks to the control of the experimental parameters, cold atoms can be precisely tuned at quantum critical points, such as the superfluid to Mott-insulator transition \cite{Greiner:2002aa}. On the one hand, the universal character of the physics in the vicinity of such a transition makes it of interest in particular for comparison with theory and across different fields of physics  but on the other hand the lack of a quasi-particle description of the excitations makes the system challenging to describe (see \cite{Sachdev:2007aa} and reference therein).

A two-terminal configuration allows for the direct measurement of conductances for a mesoscopic system tuned at a critical point, such as a two-dimensional system of Bosons close to the superfluid to Mott insulator transition \cite{Chen:2014aa,Witczak-Krempa:2014aa}. Connecting this system to large reservoirs could be done as described in section \ref{section:QPC}, and the conductance in the critical regime could be directly accessed, providing a direct test-bed for theoretical predictions. Of particular interest would be the ability to make quantitative predictions, controlling the effects of finite size. This could be performed using programmed, controlled potentials allowing for a finite size scaling analysis, extracted the intrinsic length scales from the size dependence of the conductance, and extrapolating to the infinite size limit. 

\subsection{Emergent states at interfaces}

The interplay of the connection to reservoirs with interactions inside the contact, as provided by mesoscopic cold atomic gases, is a natural environment to study the effects of interfaces. An archetype of such state is the Andreev bound state inside a quantum point contact connected to superfluids (see Appendix for more details). Such a system is the subject of ongoing research in condensed matter physics, in particular as a resource for quantum information processing \cite{Janvier:2015aa}. For more complex structures, which could include spin orbit coupling or artificial gauge fields in the contact for instance, the emerging states can have non-trivial topological properties \cite{Jiang:2011aa}. The use of a contact with large reservoirs may be useful in mitigating some of the heating effects in existing protocols for the implementation of gauge structures with atoms. 

Emergent quantum states have been invoked to explain some features observed in electronic systems known as the 0.7 anomaly \cite{Thomas:1996aa,Rejec:2006aa,Bauer:2013aa,Iqbal:2013aa}. A cold atomic gas could be used as quantum simulator with tunable interactions in order to investigate these phenomena. In particular, the use of charge neutral particles with short range interactions, the ability to measure spin conductances, spin drag or thermopower, and the use of attractive or repulsive interactions could be an interesting test-bed for existing theories. 

At the same time, the emergent state in the contact could provide cold atoms physicists with an access to strongly correlated physics like the Kondo effect or quantum magnetism, forming a starting point for the engineering of complex quantum systems with cold atoms.

\section*{Acknowledgements}

We thank the present and past members of the quantum transport group at ETHZ, Henning Moritz, Bruno Zimmermann, Torben M\"uller, Jakob Meineke, David Stadler, Dominik Husmann, Martin Lebrat, Charles Grenier, Samuel Hausler, Shuta Nakajima and Laura Corman. We would like to thank Antoine Georges, Thierry Giamarchi, Corinna Kollath, Shun Uchino, Pjotrs Grisins, Johann Blatter, Thomas Ihn, Klaus Ensslin, Lei Wang, Joseph Imry, Vincent Josse, Wilhelm Zwerger, Paivi T\"orm\"a, Christian Glattli, Wolfgang Belzig, Christoph Bruder, for numerous discussions, and Giacomo Roati for discussions and a careful reading of the manuscript. 

We acknowledge financing from NCCR QSIT, the ERC project SQMS, the FP7 project SIQS, the Horizon2020 project QUIC, Swiss NSF under division II. JPB is supported by the Ambizione program of the Swiss NSF and by the Sandoz Family Foundation-Monique de Meuron program for Academic Promotion.

\section*{Appendix}
\subsection*{Andreev reflections}
The case of a single mode point contact in between two weakly correlated superfluids is a textbook example of mesoscopic physics, and provides an elegant link between the macroscopic Josephson effect, described by the Josephson-Anderson equation, and the microscopic (even though effective) description of the contact \cite{Beenakker:1991aa} (see \cite{Bretheau:2013aa} for a recent overview).

\begin{figure}
    \centering
    \includegraphics[width=0.6\textwidth]{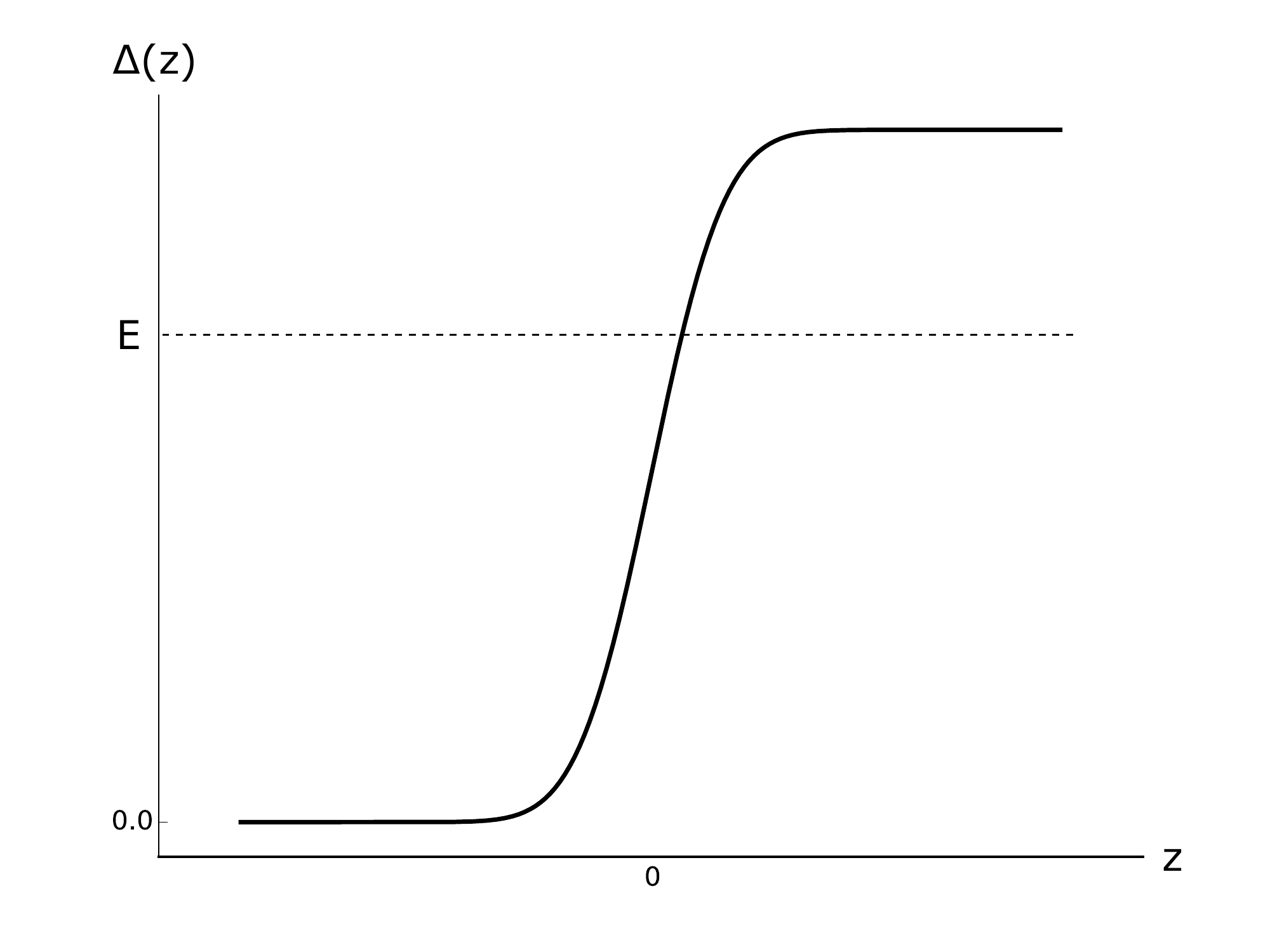}
    \includegraphics[width=0.6\textwidth]{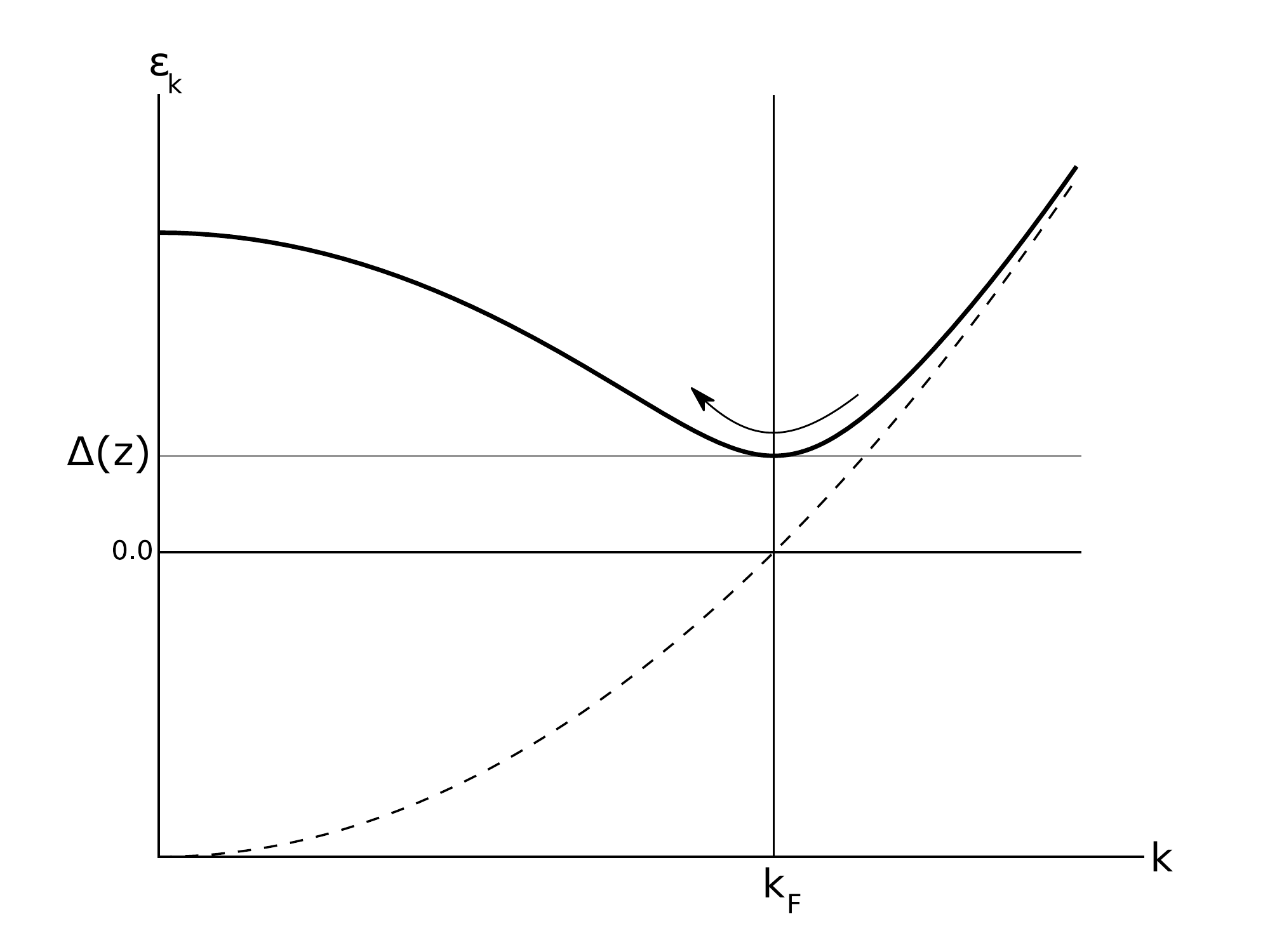}
    \caption{{\bf Schematic picture of Andreev reflection}. Up: real space variations of the gap with a normal-superfluid interface at $z=0$. Down: BCS dispersion relation of the quasi-particles at position $z$ for which the gap is $\Delta(z)$.
    }
    \label{fig:Andreev}
\end{figure}

Consider first a one dimensional situation with an interface between a superfluid and normal region located around $z=0$. Consider now a particle wave packet incident from $-\infty$ with energy $\epsilon_k$ above the Fermi energy $E_F$, such that to first order $\epsilon_k=\hbar k v_F$, with $v_F = \hbar k_F/m$, $k_F$ is the Fermi momentum and the total momentum is $k_F+k$. The total energy of the incident particle is supposed to be lower than $\Delta(+\infty)$, as illustrated in figure \ref{fig:Andreev}. We suppose that the gap $\Delta(z)$ varies with position much slower than $k_F^{-1}$, consistently with the BCS framework where the fastest variations of $\Delta$ are of order $\xi = \hbar \Delta /v_F$. Thus, we describe the motion of the incident particle semiclassically with a local momentum $k_F + k(z)$ and a local energy $E(z) = \sqrt{\epsilon_k^2 + \Delta(z)^2}$. 

The equations of motion for the wave packet are readily obtained:
\begin{eqnarray}
\frac{dz(t)}{dt} &=& \frac{1}{\hbar}\frac{\partial E}{\partial k}  = \frac{\hbar v_F^2}{E}k(t) \\
\frac{dk(t)}{dt} &=& -\frac{1}{\hbar}\frac{\partial E}{\partial z} = -\frac{\Delta}{\hbar}\frac{\partial \Delta}{\partial z}
\end{eqnarray}

The incident wave packet is moving at the local group velocity, with a momentum constrained by energy conservation. Since $\Delta \geq 0$  and $\frac{\partial \Delta}{\partial z} \geq 0$ everywhere, momentum is monotonically decreasing during the whole evolution. The group velocity, positive when the particle is at $z \ll 0$, cancels at the "turning point" set by $E(z) = \Delta(z)$. At this point, the momentum of the wave packet is exactly the Fermi momentum and the quasi particle is an equal mixture of a particle and a hole. The momentum then keeps decreasing and the group velocity becomes negative. The corresponding wave packet still has an overall {\it positive} momentum (slightly below $k_F$) and a negative group velocity, thus a hole like character. In the limit $t \rightarrow \infty$, the position of the wave packet goes to $-\infty$ and the wave packet turns into a pure hole-like excitation. 

In the frame of the wave packet, the Andreev reflection is equivalent to a Landau-Zener transition well known in quantum optics. The wave packet sees a growing "driving", represented by $\Delta$ which couples coherently particles and holes of a given momentum, with a resonance at $k_F$. The semiclassical motion of the wave packet amounts to a slow sweep of momentum across $k_F$. 

It is important to realise that in this process, no actual backscattering takes place: all processes happen in a thin momentum window around $+k_F$, while actual backscattering would connect $+k_F$ to $-k_F$. Since the order parameter cannot vary on the length scale of $k_F^{-1}$ such processes have to originate from another mechanism, such as a sharp jump in the electron density at an oxide interface in condensed matter. Of course the reverse process also takes place, by which a hole is coherently converted into a particle. As usual, the equilibrium situation represents the equilibrium between particle and hole reflection processes. Accounting for the spin degree of freedom, the hole is reflected with the opposite spin. 

Within this description, the total number of particles is only conserved up to the addition or subtraction of any number of pairs in ground state, consistently with the usual mean-field BCS theory. The conversion of a particle into a hole by Andreev reflection is accompanied by the creation of one pair in the condensate in the superfluid part, in the same way as a Landau-Zener transition is accompanied by the absorption of one photon. This also means that an incident pair from the superfluid side is actually coherently converted into a particle-hole pair with conjugated momenta and opposite spin. 

This picture does not depend on the details of the connection between the normal part and the superfluid. The normal state does not actually need to be normal, it could just be a region where the gap is lower, and we consider an excitation that has enough energy to live in the low gap region but not in the other. 

\subsection*{Andreev bound state}

Let's now consider the situation of a weak link between two superfluids. Inside the weak link, a particle incident onto a superfluid contact is coherently converted into a hole, which is then coherently converted back into a particle at the other interface. Each such cycle coherently adds a pair in one contact and destroys another in the other contact, a process sometimes referred to as the 'Andreev pump'. There is however a fundamental extra ingredient, namely the relative phase between the two superfluids. 

\begin{figure}
    \centering
    \includegraphics[width=0.6\textwidth]{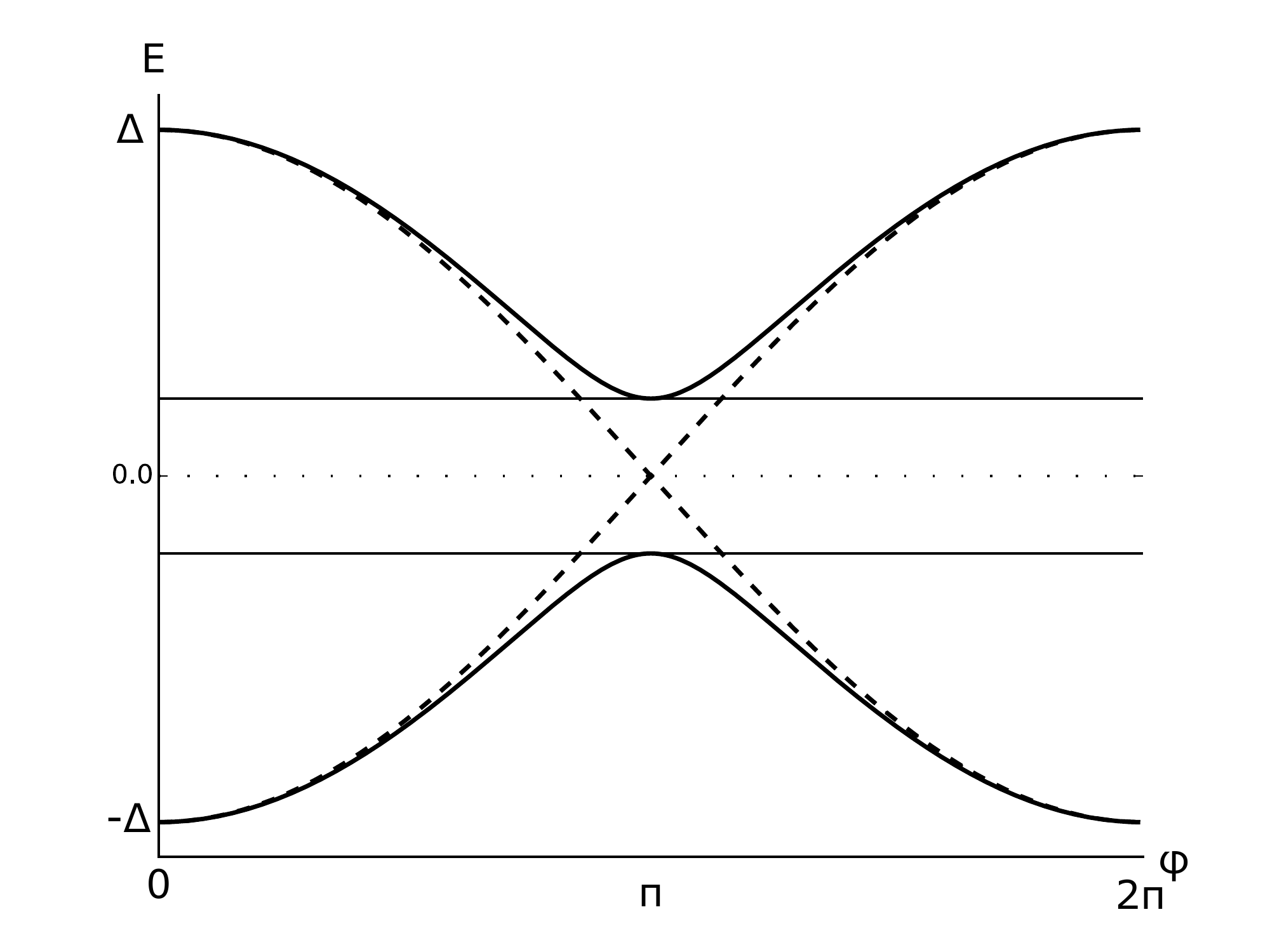}
    \caption{{\bf Andreev Bound states}. Energy of the Andreev bound states as a function of phase difference, without (dashed line) and with (solid line) backscattering mechanisms.
    }
    \label{fig:ABS}
\end{figure}

Indeed, following the analogy with a Landau-Zener process, the phase of the superfluid is imprinted onto the outgoing hole upon Andreev reflection. When the hole in converted back into a particle upon Andreev reflection on the second contact, the phase difference is imprinted onto the excitation, and interference with the original state takes place. 

Writing down the condition for constructive interferences inside the weak link, one obtains a pair of eigenenergies $E_{\rightarrow}(\phi)$ and $E_{\leftarrow}(\phi)$ lying below the pairing gap, corresponding to two bound states $\ket{\rightarrow}$ and $\ket{\leftarrow}$ which depend on $\phi$, the phase difference between the two superfluids $E_{\rightarrow}(\phi) =  -\Delta \cos{\frac \phi 2}$, and $E_{\leftarrow}(\phi) = - E_{\rightarrow}(\phi)$. Here we have supposed that the length of the weak link is shorter than the coherence length of the superfluid, such that the phase accumulated due to free space propagation inside the link is negligible. These carry a current $I_j = -\frac 2 \hbar \frac{\partial E_j}{\partial \phi}$, $j = \leftarrow, \rightarrow$. For $\phi < \pi$, the ground state is $\ket{\rightarrow}$, which carries a positive current and for $\phi > \pi$ the ground state is $\ket{\leftarrow}$ carrying a negative current. This corresponds to the idea that the current is directed in the direction of the phase gradient. The two states $\ket{\rightarrow}$ and $\ket{\leftarrow}$ are associated to wave packets performing oscillations around $+k_F$ and $-k_F$ respectively. For $\phi=\pi$, the two states are degenerate. 

In the presence of a backscattering mechanism, yielding a transparency for the weak link $\alpha<1$, the two states $\ket{\rightarrow}$ and $\ket{\leftarrow}$ get coupled. Inside the weak link, an left propagating particle can be coherently backscattered into a right propagating particle with probability $1-\alpha$. This coherent coupling lifts the degeneracy of the $\ket{\rightarrow}$ and $\ket{\leftarrow}$ states around $\phi=\pi$. The new eigenstates are $\ket{+}$ and $\ket{-}$ with energies $E_{\pm} = \pm \Delta \sqrt{1-\alpha \sin^2(\frac \phi 2)}$. 

In the presence of a weak tunnel barrier, $\alpha \ll 1$, this model reproduces the Josephson effect, where to first order the current exhibits sinusoidal oscillations. In the presence of a bias between the two reservoirs, the phase difference increases linearly in time. When the bias is smaller than the gap, the systems evolves adiabatically along the state $\ket{-}$, the ground state of the system. This yields an oscillating current as the admixture of $\ket{\rightarrow}$ and $\ket{\leftarrow}$ oscillates in time. Using the relation $I = -\frac 2 \hbar \frac{\partial E}{\partial \phi}$ we obtain the Ambegaokar-Baratoff formula for the maximum current sustained by the junction $I_c = \frac{\pi \Delta}{2} G_n $, with $G_n = \frac 2 h \alpha$ the conductance in the normal state given by the Landauer formula \footnote{Note the factor of $2$ due to spin degeneracy, which follows the condensed matter convention in contrast to the discussions of the Landauer formula in the main text.}. Remarkably, adding several channels in parallel with difference transmissions do not alter the phenomena, since the phase difference and the frequency is identical for all channels. This explains the robustness of the Josephson oscillations to trap averaging and other imperfections. 

\subsection*{Multiple Andreev reflections and dissipative flow}

In the absence of dissipation, i.e. in the fully adiabatic limit, applying a bias to the junction yields Josephson oscillations that average out to zero in the DC limit, leading to zero DC conductance. However, with a finite bias, there is a non zero probability to perform a Landau-Zener transition from the ground Andreev state to the excited one, which then leads to a finite current in the DC limit. Upon sweeping the phase, one single particle excitation is taken at $-\Delta$ and "lifted up" above the gap via the Andreev bound states and the Landau-Zener transition, creating one excitation. The probability of a non-adiabatic transition is given by the Landau-Zener formula $p = e^{-\frac{\pi (1-\alpha) \Delta}{\Delta \mu}}$, where $\Delta \mu$ is the bias. A crossover from adiabatic Josephson oscillations and dissipative current occurs when $\Delta \mu \sim \pi (1-\alpha) \Delta$. For very good contacts with $\alpha \leq 0.95$ like in our experiment, this crossover takes place for very low bias. 

Dissipation and DC currents also occur via a resonance mechanism named multiple Andreev reflections. This can be understood by considering the energy needed to create one single particle excitation. When $\Delta \mu < 2\Delta$, the bias is not large enough to directly break pairs. However, upon transferring one pair from the high chemical potential lead to the low chemical potential one, an energy $2 \Delta \mu$ is acquired. Coherently adding $n/2$ pair transfer events can create a single particle excitation provided $\Delta \mu \leq \frac{2 \Delta}{n}$, where $n$ here is the number of Andreev reflections involved in the process, or the order of the process. Since each Andreev reflection involves the transmission of a particle or a hole through the weak link, an $n$th order process occurs with probability $\alpha^n$. Upon reducing the bias, the current drops sharply at $\Delta \mu < 2 \Delta /n$ for each integer number $n$, corresponding to the closing of  the Andreev reflections of order $n$, leading to a cusp in the current-bias relation, as observed in superconducting point contacts. 

This process is formally analogous to Wannier-Stark resonances in optical lattices. Consider the conjugated variable to $\phi$, namely the pair population difference $N$. States with a given $N$ are analogous to sites of an optical lattice, and $\phi$ is the counterpart of the quasi-momentum measuring the phase difference between neighbouring sites. The two Andreev bound states then are the counterparts of two bands of the lattice. Introducing a bias $\Delta \mu$, leads to an offset $\Delta \mu$ between neighbouring sites. When the bias is equal to a fraction of the gap between the two states $2\Delta$, resonant tunnelling can take place from the ground band to the excited band across several sites. 

In contrast to the Josephson oscillations, multiple Andreev reflections or the adiabatic to dissipative crossover described by the Landau-Zener formula is specific to single mode junction. In the multimode case, different order of reflections take place in the different channels, and the crossover to dissipative currents takes place for a different bias in each channel, depending on its transmission.

\bibliographystyle{unsrt}
\bibliography{review}
\end{document}